\documentclass[11pt,a4paper]{article}
\usepackage[authoryear]{natbib}
\usepackage{times}
\usepackage{graphics}
\usepackage{ifpdf}
\usepackage{graphicx}
\usepackage{pgf}
\usepackage{epsfig}
\usepackage{amssymb,amstext}
\usepackage{amsmath,latexsym,xspace}
\usepackage{array}
\usepackage{setspace} 
\usepackage{comment}
\usepackage{authblk}
\usepackage{caption}
\usepackage{hyperref}
\usepackage{verbatim}
\usepackage{hyperref}
\usepackage[scale={0.7,0.85},centering,includeheadfoot]{geometry}

\def\E{\text{E}}
\def\mm#1{\ensuremath{\boldsymbol{#1}}} 

\begin{document}
\begin{titlepage}
\title{Model-based bias correction for short AR(1) and AR(2) processes}

\author{Sigrunn H. S{\o}rbye\thanks{\textit{Address for correspondence}: Sigrunn Holbek S\o rbye, Department of Mathematics and Statistics, Faculty of Science, UiT The Arctic University of Norway, 9037 Troms{\o}, Norway. \E-mail: sigrunn.sorbye@uit.no} $^1$, Pedro G. Nicolau$^1$ and H\aa vard Rue$^2$}
\affil{$^1$UiT The Arctic University of Norway, Troms{\o}, Norway}
\affil{$^2$King Abdullah University of Science and Technology, Thuwal, Saudi Arabia}

\end{titlepage}
\maketitle
\begin{abstract}
The class of autoregressive (AR) processes is extensively used to model temporal dependence in observed time series. Such models are easily available and routinely fitted using freely available statistical software like \texttt{R}. A potential caveat in analyzing short time series is that commonly applied estimators for the  coefficients of AR processes are severely biased.  This paper suggests a model-based approach for bias correction of well-known estimators for the coefficients of first and second-order stationary AR processes, taking the sampling distribution of the original estimator into account. This is achieved by modeling the relationship between the true and estimated AR coefficients using weighted orthogonal polynomial regression, fitted to a huge number of simulations. The finite-sample distributions of the new estimators are approximated using transformations of skew-normal densities and their properties are demonstrated by simulations and in the analysis of a real ecological data set. The new estimators are easily available in our accompanying \texttt{R}-package \texttt{ARbiascorrect} for time series of length $n=10, 11, \ldots , 50$, where original estimates are found using exact or conditional maximum likelihood, Burg's method or the Yule-Walker equations.
\end{abstract}


\section{Introduction} \label{sec:introduction}
The class of autoregressive (AR) processes is one of the most central and widely applied time series models. These processes are simple and intuitive, expressing the current value of a time series as a linear combination of previous values and additional random noise. Specifically, a $p$th order autoregressive process (AR($p$))  can be defined by 
\begin{equation}
x_t -\mu=  \sum_{j=1}^p \phi_j (x_{t-j}-\mu)+w_t  \label{eq:arp}
\end{equation}
where $\mu=\E(x_t)$, the set $\{\phi_j\}_{j=1}^p$ are fixed coefficients and $\{w_t\}$ is a Gaussian white noise process, $w_t\sim N(0,\sigma^2)$.  Originated by the famous Yule-Walker equations \citep{yule:27, walker:31}, the statistical properties of AR models have been thoroughly studied and established,  see e.g. \cite{brockwell:02}, \cite{box:08} and \cite{shumway:17} for comprehensive introductions.  However, one problem still remains. Commonly used estimators for the AR coefficients are severely biased for small sample sizes, i.e.~having less than 50 observations \citep{shaman:88, huitema:91, decarlo:93, cheang:00}. This is problematic in several fields of applications in which realistic time series lengths are inherently short, e.g. in  behavioral  \citep{arnau:01} and 
ecological  research studies  \citep{bissonette:99, ives:10}. 

This paper studies  the finite-sample properties of commonly applied estimators for the coefficients of stationary AR(1) and AR(2) processes and provides bias-corrected versions of such estimators.  The AR(1) and AR(2) processes do have considerable practical importance \citep[p.53]{box:08} and various estimators for the AR coefficients are easily available. The popular \texttt{stats:::ar}-function in \texttt{R} \citep{rteam:20} implements the closed form solution given by the Yule-Walker equations as its default. The \texttt{ar}-function also provides estimates using  Burg's method \citep{burg:67} and a conditional maximum likelihood estimator (MLE),  maximizing the likelihood given initial values of $x_0,\ldots , x_{p-1}$ in \eqref{eq:arp}. The \texttt{R}-package \texttt{FitAR:::FitAR} \citep{mcleod:08b} provides the exact maximum likelihood estimator, where initial values are set using Burg's method.  All of the mentioned estimators give very similar results for large sample sizes while the exact MLE  has been claimed to usually perform better than alternatives for short time series \citep{mcleod:06, box:97}. 

\begin{figure}[h]
\centering
\includegraphics[scale=0.5]{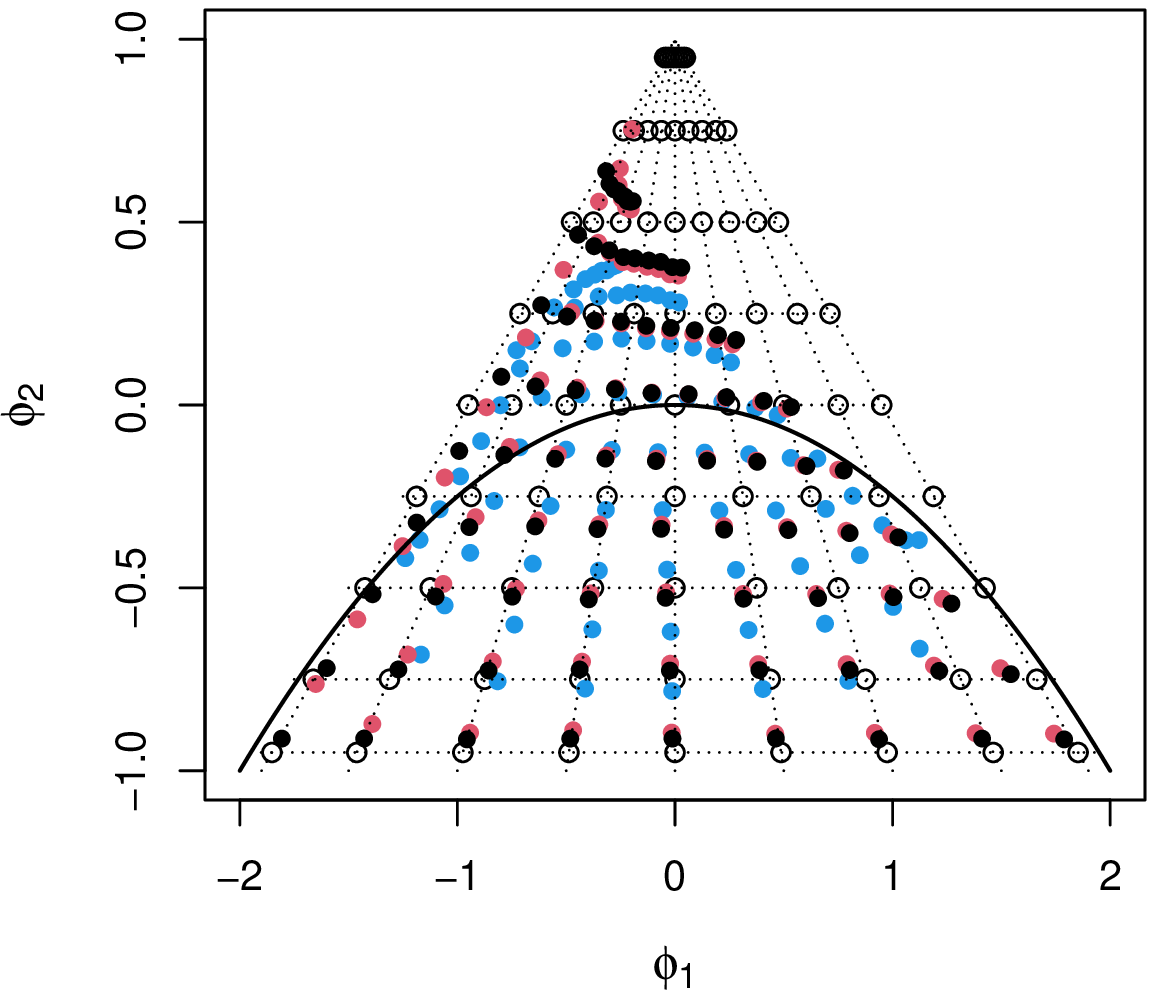}
\includegraphics[scale=0.5]{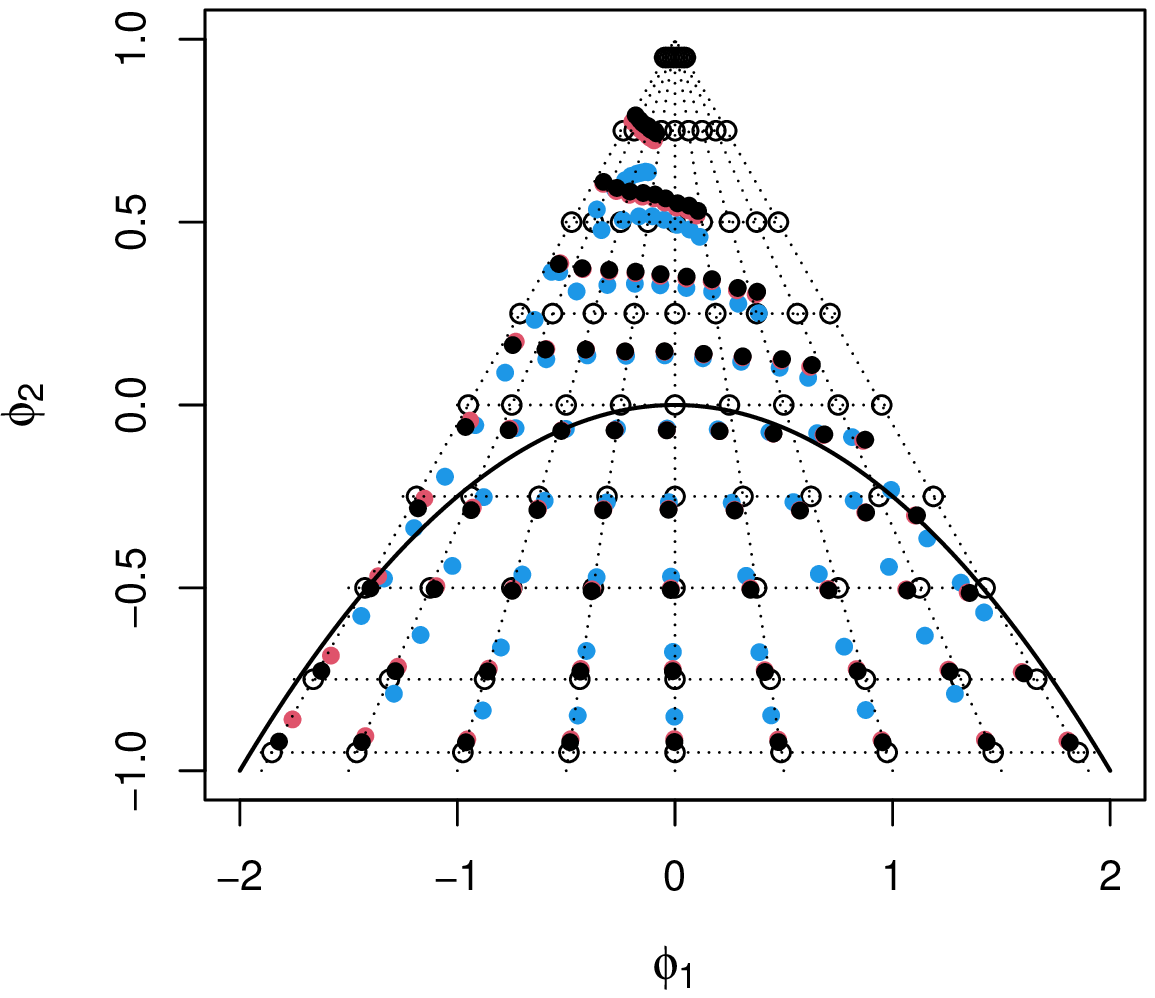}\\
\includegraphics[scale=0.5]{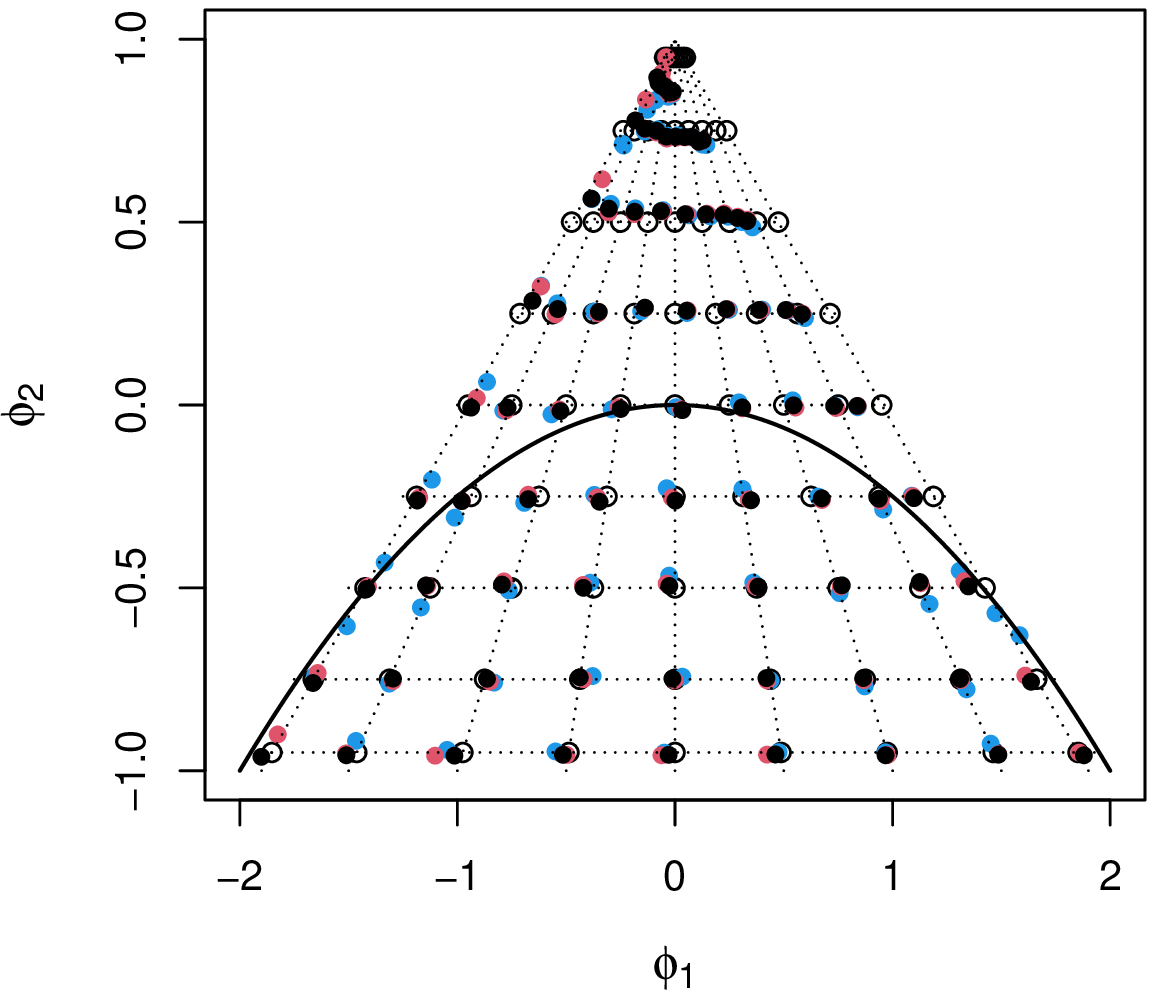}
\includegraphics[scale=0.5]{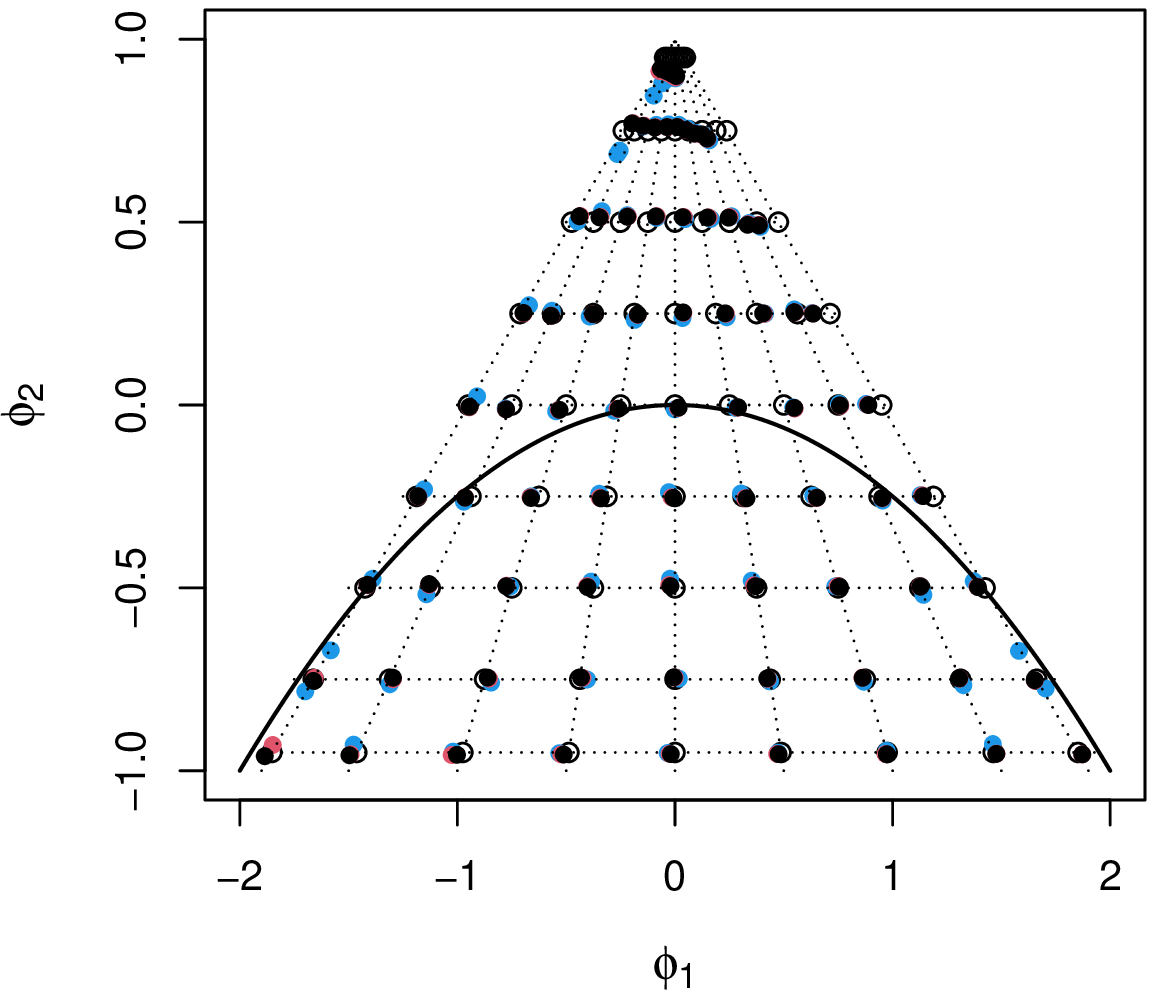}
\caption{Upper panels: Mean estimates of selected pairs $(\phi_1,\phi_2)$ based on 10000 simulations, using the exact MLE (black), Burg's algorithm (red) and the Yule-Walker solution (blue) when $n=15$ (left) and $n=30$ (right).  Lower panels: The corresponding corrected mean estimates found by our proposed simulation-based approach.}
\label{fig:phis}
\end{figure}

The upper panels of Figure \ref{fig:phis} illustrate the average coefficient estimates of AR(2) processes of length $n=15$ (left) and $n=30$ (right), using the exact MLE, Burg's method and the Yule-Walker estimator. The results are based on generating 10000 time series for each selected combination of $(\phi_1,\phi_2)$ within the triangular stationary area of such processes. The AR(2) process has pseudo-periodic behavior for pairs of coefficients below the given parabolic curve. In the lower part of this region, the bias of the exact MLE and Burg's method is not too severe but it increases with increasing values of $\phi_2$. The Yule-Walker approach clearly gives the most biased results for both sample sizes. The average estimates using the conditional MLE are not illustrated as these were not visually distinguishable from the exact MLE.

A number of methods to provide bias-corrected estimators for the AR coefficients have been proposed in literature. These range from asymptotic-based formulas for the bias  \citep{marriott:54, kendall:54, shaman:88, tanaka:84, cordeiro:94} to methods using restricted maximum likelihood  \citep{cheang:00} and bootstrapping \\ \citep{thombs:90, kim:03}. 
\cite{andrews:93} introduced a median-unbiased correction for the least squares estimator of the AR(1) coefficient which was generalized to give an approximate median-unbiased estimator of AR($p$)  processes in \cite{andrews:94}.  This estimator is implemented in the \texttt{R}-package \texttt{BootPR} \citep{bootpr:14}, also including the  estimators of \cite{shaman:88} and \cite{roy:01}. The resulting estimates are not constrained to fall within the stationary area of the processes. 

Our bias-correcting approach differs from previous suggestions in the sense that we model the true AR coefficients as a function of original estimates,  accounting for the sampling distribution of the original estimator.  This is achieved by a brute-force simulation approach where we generate  AR(1) and AR(2) processes for a fine grid of underlying true values of the coefficients. The relationship between the true and estimated coefficients is  then described using a weighted orthogonal polynomial regression model. In the AR(2) case, the regression model is fitted based on a total of more than 59 million time series for a given sample size $n$.    The resulting bias-corrected average estimates are illustrated in the lower panels of Figure \ref{fig:phis}, again using the exact MLE, Burg's method and the Yule-Walker solution as original estimators.  We do see a clear improvement in the bias properties for all of the methods.  Admittedly, the bias-corrected estimators will not be completely unbiased as the relationship between the true and originally estimated coefficients cannot be modeled perfectly. Especially this is the case for coefficient combinations along the borders of the triangular area. 

The given brute-force simulation approach makes it possible to also derive sampling distributions for both the original and bias-corrected estimators. Specifically, we have fitted skew-normal distributions to transformations of the originally estimated coefficients of AR(1) and AR(2) processes.  The parameters of the skew-normal distribution are then modeled in terms of the true underlying AR coefficients, again using orthogonal polynomial regression. In the AR(1) case, it is  straightforward to obtain confidence intervals for the original and bias-corrected estimators using Monte Carlo sampling. In the AR(2) case, confidence intervals are found by combining Monte Carlo sampling and a Gaussian copula representation to preserve correlation between the estimated AR coefficients. 
 
This paper is structured as follows. Section~\ref{sec:ar1} outlines our modeling approach giving bias-corrected estimators of the first-lag autocorrelation coefficient of AR(1) processes and derive their  sampling distributions. In Section~\ref{sec:ar2}, the suggested approach is extended to give bias correction and approximate sampling distributions in estimating the coefficients of AR(2) processes. Section~\ref{sec:example} illustrates bias correction for a real ecological data set, where the autoregressive coefficients are used to characterize density dependence and population dynamics. Concluding remarks are given in Section~\ref{sec:conclusions}. The appendix~\ref{sec:appendix} describes our accompanying \texttt{R}-package which can be used to obtain the bias-corrected estimates and $95\%$ confidence intervals for time series of length $n=10,11,\ldots , 50$, using the four mentioned original estimators.

\section{Bias correction and finite-sample properties for AR(1) models}\label{sec:ar1}
\begin{figure}[h]
\centering
\includegraphics[scale=0.5]{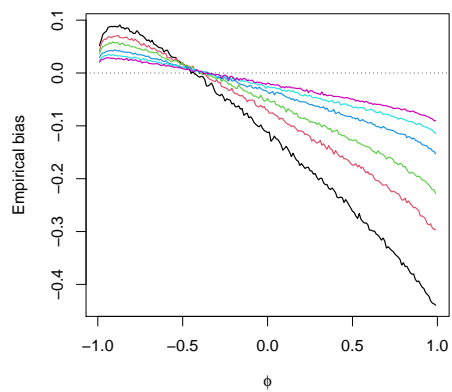}
\caption{The estimated empirical bias $\hat \phi-\phi$ of the exact MLE for AR(1) processes with length $n=10$ (black), $n=15$ (red), $n=20$ (green), $n=30$ (blue), $n=40$ (light blue) and $n=50$ (pink).}
\label{fig:ar1-bias}
\end{figure}

The AR(1) model has a simple one-step dependence as defined by \eqref{eq:arp} where $p=1$. 
The bias in estimating the first-order autocorrelation coefficient $\phi_1=\phi$ has been studied by several authors,   see e.g  \cite{krone:17} for a recent comparative simulation study of AR(1) estimators in short time series. The bias of the exact MLE is illustrated for different sample sizes in Figure~\ref{fig:ar1-bias}, giving empirical averages for 10000 simulations for a fine grid of $\phi$-values in the stationary area $(-1,1)$. 

A common way to explicitly construct an unbiased estimate for a parameter $\phi$ is simply to subtract an estimate of the bias from the original estimator, providing a bias-corrected estimator of the form  
$$ \hat \phi_c = \hat \phi - \hat{E}(\hat \phi-\phi).$$
For example, the exact MLE can be corrected using the  asymptotic bias $-(1+3\phi)/n$ \citep{tanaka:84, cordeiro:94}. Naturally, such a linear correction would not be accurate enough for small sample sizes. 
In \cite{arnau:01}, the  ordinary autocorrelation estimator for $\phi$ is bias-corrected by adding the absolute value of a polynomial fit to the empirical bias.  This has a slight resemblance to our approach, but an important difference is that we do not try to estimate or model the bias. We model  the true parameter value $\phi$ directly as a function of original estimates based on a huge number of simulations.
Similarly, we provide approximate sampling distributions by modelling parameters of  skew-normal approximations as functions of the true values of $\phi$.

\subsection{Deriving  bias-corrected estimators by simulation}
Let $\hat \phi$ denote an original estimator for the first-lag autocorrelation coefficient of AR(1) processes. Our goal is to construct a bias-corrected estimator for $\phi$ such that  $\E(\hat \phi_c)=\phi$ for all values of $\phi$. To achieve this we model the relationship between the true and estimated parameter values using a weighted orthogonal polynomial regression model. The coefficients of the regression model are derived by minimizing the weighted squared error between the corrected estimate and true parameter values. The specific steps in constructing the bias-corrected estimator can be summarized as follows:

\begin{enumerate}
\item To avoid constraints on the support of $\phi$, we first introduce a monotonic transformation
\begin{equation}
g(\phi)=\mbox{logit}\left(\frac{\phi+1}{2}\right), \label{eq:logit}
\end{equation}
which has infinite support. This facilitates optimization and implies that our inverse transformed bias-corrected estimate will always be within the stationary area  of the AR(1) process.   
\item Let $\hat \phi$ denote an original estimator of $\phi$. We model the true AR coefficient using an orthogonal polynomial model
\begin{eqnarray}
\phi =  f(\hat \phi,\mm{\beta}) &= & g^{-1}\left(\sum_{k=0}^{K} \beta_k h_k(g(\hat \phi))\right), \quad \hat\phi \in (-1,1)\label{eq:correct-est}
\end{eqnarray}
where $\mm{\beta}=\{\beta_k\}_{k=0}^K$ denotes a fixed set of  regression coefficients while $\{h_k(.)\}_{k=0}^K$ represents a set of orthogonal polynomials of order $k$. Here, we choose to use the probabilists' Hermite polynomials which are orthogonal with respect to the standard normal density. These polynomials are defined by
\begin{equation}
h_0(x)=1,\quad 
h_1(x)  =  x, \quad
h_{k+1}(x)  =  x h_{k}(x)-k h_{k-1}(x),\quad k\geq 1.
\end{equation}   
\item To estimate $\mm{\beta}$ for a given sample size $n$, we generate a total of $m =10000$ time series  for a fine grid of $\phi$-values. Note that the suggested estimator is a non-linear function of $\hat \phi$ implying 
$$\E(f(\hat \phi,\mm{\beta})) \neq f(\E(\hat \phi),\mm{\beta}). $$
 This means that the optimization takes the estimated value for each time series into account not just the average estimate of the $m$ simulations for each $\phi$. The regression coefficients  are thus found by solving the  optimization problem
\begin{eqnarray}
\hat{\mm{\beta}} & = & \mbox{arg}\min_{\mm{\beta}} \sum_{r=1}^l \frac{1}{s^2_r}\left(\frac{1}{m}\sum_{j=1}^m g^{-1}\left(\sum_{k=0}^{K} \beta_k h_k(g(\hat \phi_{rj}))\right)-\phi_r\right)^2\nonumber \\
& = & \mbox{arg}\min_{\mm{\beta}} \sum_{r=1}^l \frac{1}{s^2_r}\left(\frac{1}{m}\sum_{j=1}^m f(\hat \phi_{rj},\mm{\beta})-\phi_r\right)^2.\label{eq:optim-ar1}
\end{eqnarray}
The quantity  $\hat{\phi}_{rj}$ represents the estimate of $\phi_{r}$ in simulation $j$. Specifically, we choose the grid $\phi_r  \in (-0.95,-0.94,\ldots , 0.95)$, implying that  $l=191$.  The ordinary sample variances $s^2_r$ are used as weights. This gives a unique set of regression coefficients $\hat{\mm{\beta}}$ for each sample size $n$ and for each original estimator $\hat \phi$, implying that $\hat{\mm{\beta}}$ accounts for the sampling distribution of the estimator $\hat\phi$. The bias-corrected estimator is then given by  $\hat \phi_c =  f(\hat \phi,\hat{\mm{\beta}})$ which is used to predict the true value $\phi$.
 \end{enumerate}  
 In minimizing \eqref{eq:optim-ar1}, we have chosen to exclude values of $\phi$ close to the edges of the stationary interval. This is to avoid a severe inflation of the variance caused by forcing the estimator to give unbiased estimates at the edges of the interval. 
In practice, the given approach can be used to find corrected estimates for any estimator $\hat \phi$ giving values within the stationary range. The corrected estimates might be equal to $\pm 1$, but will never fall outside of the interval $[-1,1]$. Our implementation includes the exact and conditional MLEs, Burg's algorithm and the Yule-Walker solution. We have stored all the sets of regression coefficients for these four estimators for AR(1) series  of length $n=10,11, \ldots , 50$, and the resulting bias-correction is available from our \texttt{R}-package, see Section~\ref{sec:appendix}.

\subsection{Bias-correcting curves}
\begin{figure}[h]
\centering
\includegraphics[scale=0.5]{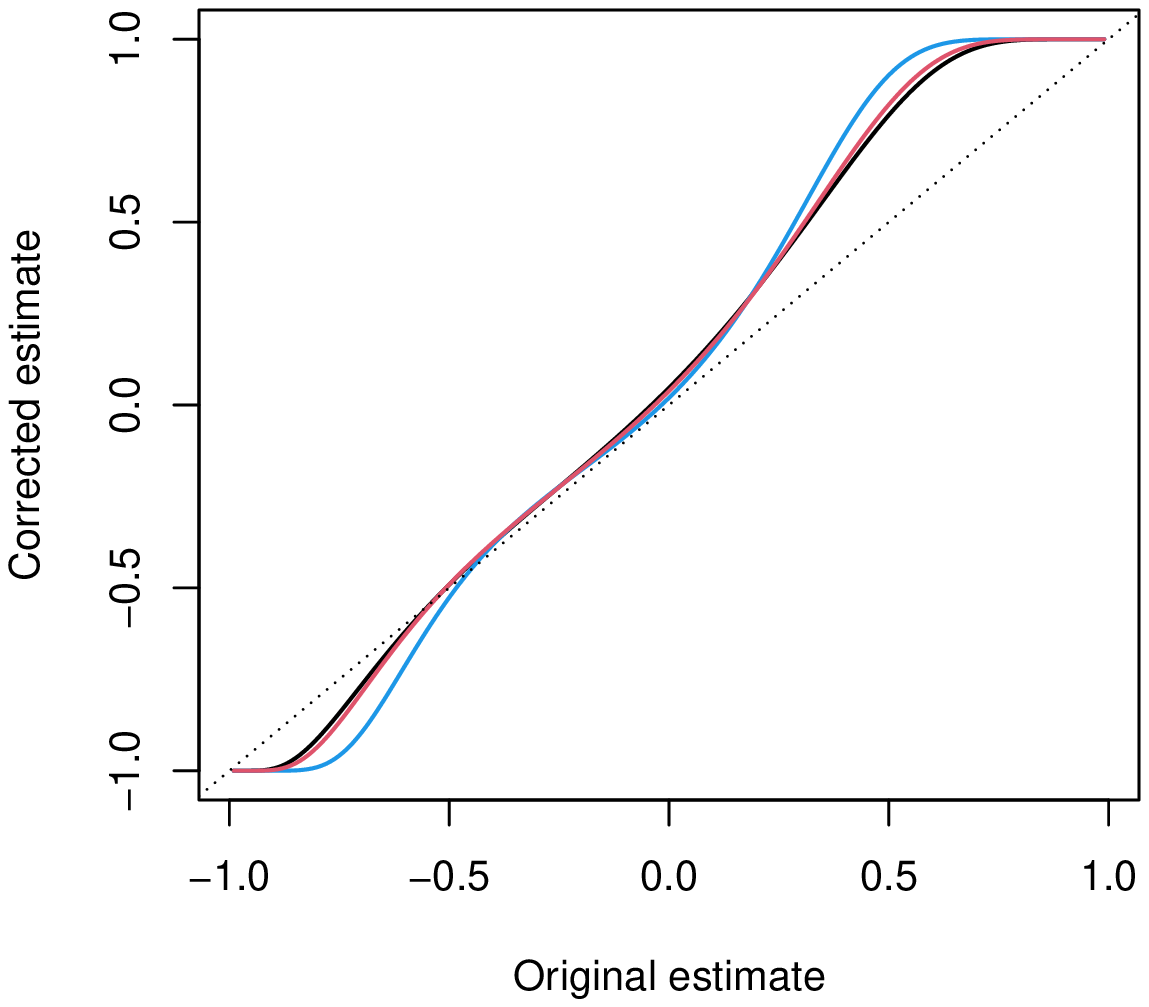}
\includegraphics[scale=0.5]{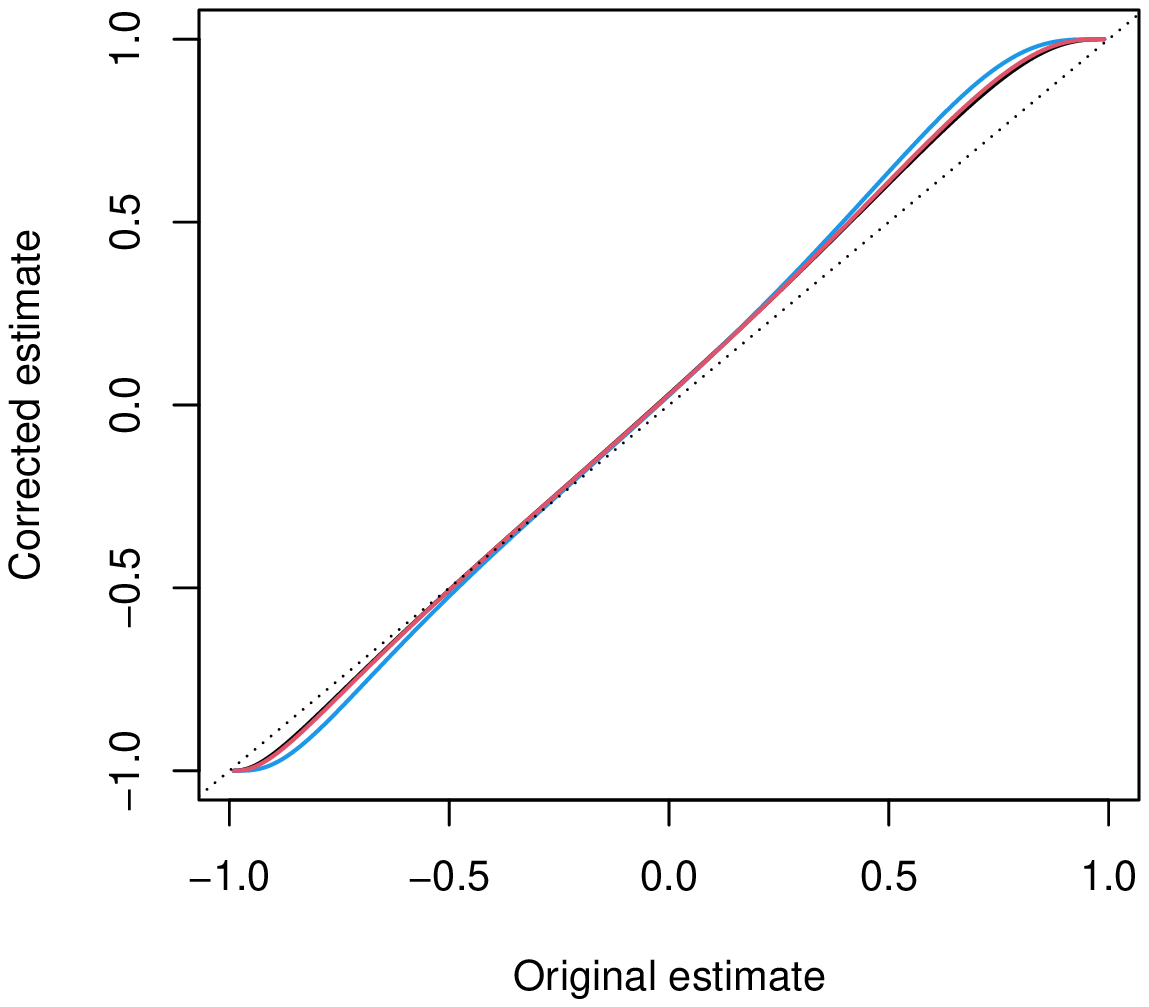}
\caption{The computed correction curves when $n=15$ (left) and $n=30$ (right) for the exact MLE (black), Burg's method (red) and the Yule-Walker solution (blue).}
\label{fig:correction-curve}
\end{figure}

Figure~\ref{fig:correction-curve} illustrates how the correction works for  AR(1) processes of length $n=15$ and $n=30$ for a fine grid of estimated $\phi$ values in the interval (-1,1).  The given curves correspond to using the exact MLE, Burg's algorithm and the Yule-Walker solution where the corrected estimators are calculated using up to cubic Hermite polynomials ($K=3$).   For an original estimate  $\hat\phi$ given at the horizontal axis, we calculate the corresponding corrected estimate $\hat \phi_c$ at the vertical axis. Naturally, this gives a quite large bias-correction when $n=15$, where original estimates above 0.5 are corrected to a value close to 1. In the case of $n=30$, we notice that the correction curves are close to linear for an internal subset of the interval.

\begin{figure}[h]
\centering
\includegraphics[scale=0.5]{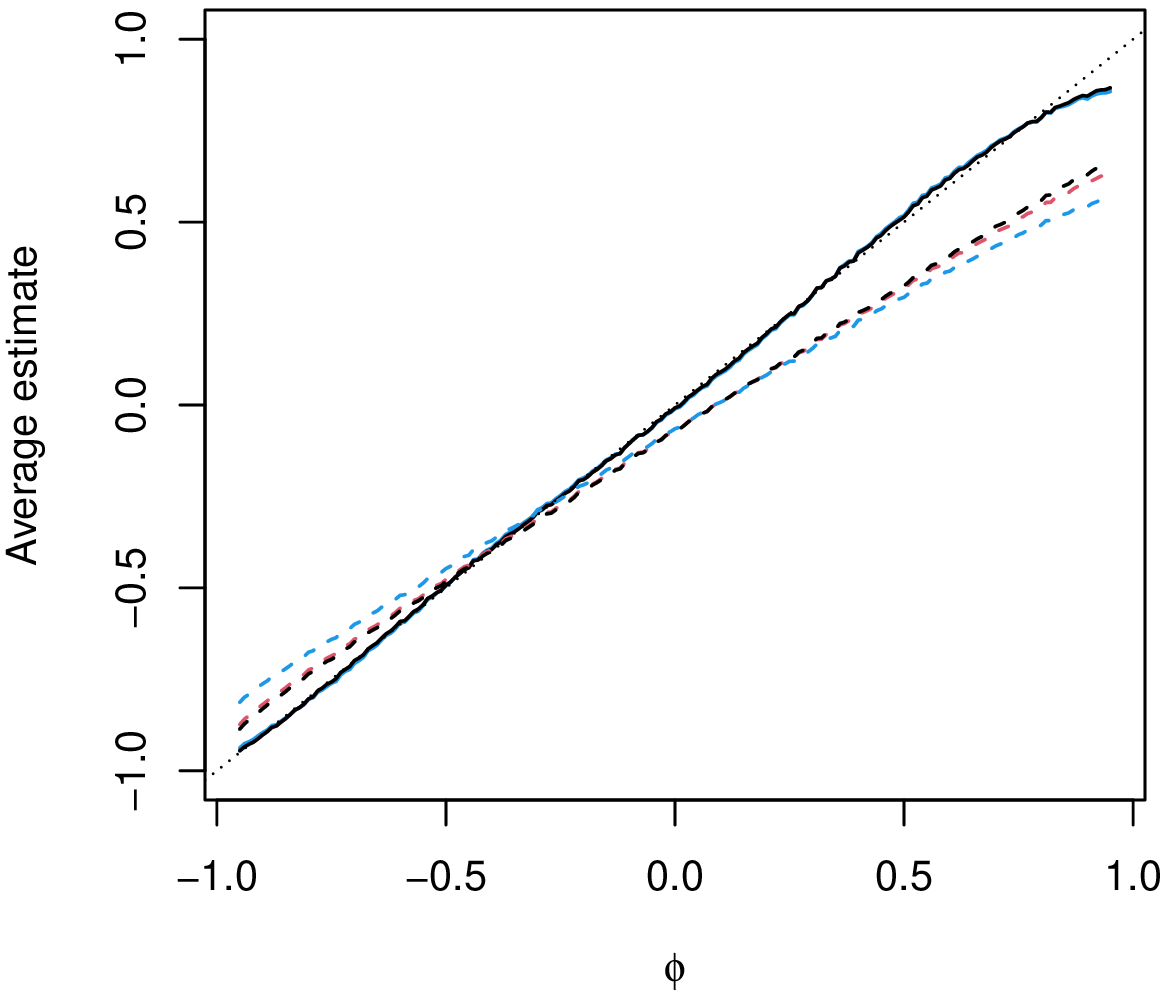}
\includegraphics[scale=0.5]{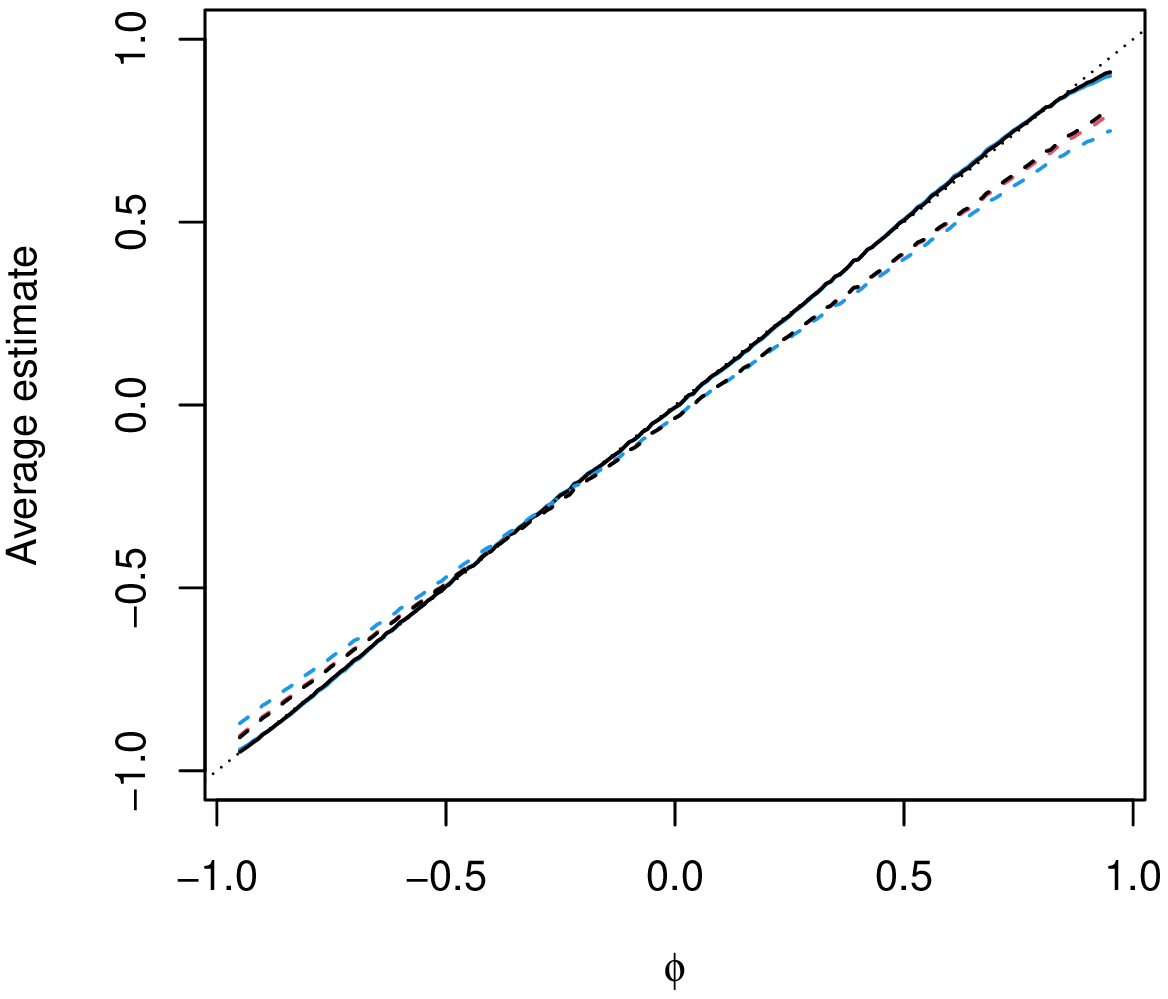}
\caption{The average original (dashed) and resulting corrected average estimates (solid) for 10000 simulations, generating time series of length $n=15$ (left) and $n=30$ (right). The estimators used include the exact MLE (black), Burg's method (red) and the  Yule-Walker solution (blue).}
\label{fig:correction-est}
\end{figure}

The original  and the corresponding corrected average estimates for  $m=10000$ simulations are displayed in  Figure~\ref{fig:correction-est}. Using the correction,  we do get close to unbiased results both when $n=15$ and $n=30$. The overall average bias and sample variance for the estimators are given in Table~\ref{tab:ar1-rmse}.  We also compute the overall root mean squared error (RMSE), which in the case of the corrected estimator is defined by
\begin{equation*}
\mbox{RMSE}(\hat {\phi}_c) = \sqrt{ \frac{1}{ml}\sum_{r=1}^l  \sum_{j=1}^m \left( \hat \phi_{c,rj}-\phi_r \right)^2}
\end{equation*}
where $\hat \phi_{c,rj}$ is the corrected estimate of $\phi_r$ in simulation $j$. The RMSE for the original estimator and the bias and variances are  computed correspondingly.  
The  results illustrate the well-known bias-variance trade-off showing that a decrease in the bias of an estimator will inherently cause an increase in the variance. For time series of length $n=15$, the choices we have made imply that RMSE is slightly larger for the corrected estimator versus the originals. When $n=30$,  we get approximately unbiased results only causing a negligible increase in RMSE.  In minimizing \eqref{eq:optim-ar1}, we could have chosen to  exclude more of the $\phi$-values at the ends of the unit interval, e.g. using $\phi \in (-0.9,0.9)$. This would reduce the variance but naturally also increase the bias close to the limits of the interval $(-1,1)$.

\begin{table}[h]
\begin{center}
\begin{tabular}{ll|rrr|rrr}
& & \multicolumn{3}{c}{Original estimator} & \multicolumn{3}{c}{Bias-corrected estimator} \\
&Method & Bias & Variance & RMSE & Bias & Variance & RMSE\\\hline 
n=15&Exact MLE& -0.080 & 0.055 & 0.263 & -0.0015 & 0.084 & 0.286 \\
&Conditional MLE & -0.080 & 0.055& 0.263 & -0.0015 & 0.084 & 0.286 \\
&Burg's method&  -0.081 & 0.052 & 0.263 & -0.0015 & 0.085 & 0.288 \\
&Yule-Walker& -0.079 & 0.046 & 0.265 & -0.0013 & 0.089 & 0.296 \\\hline
n=30&Exact MLE& -0.037 & 0.026 & 0.171 & -0.0004 & 0.031 & 0.174 \\
&Conditional MLE & -0.037 & 0.026 & 0.171 & -0.0004 & 0.031 & 0.174 \\
&Burg's method&  -0.038 & 0.025 & 0.172 & -0.0005 & 0.031 & 0.175 \\
&Yule-Walker& -0.038 & 0.024 & 0.175 & -0.0005 & 0.032 & 0.178 \\
 \end{tabular}
\caption{The average bias, variance and root mean square error for the original and bias-corrected estimators of $\phi$ when $n=15$ and $n=30$. The averages are computed based on  $m=10000$ simulations for each value of $\phi\in (-0.95,-0.94,\ldots , 0.95)$.}
\label{tab:ar1-rmse}
\end{center}
\end{table}

\subsection{Finite-sample distributions for the original and corrected estimators}
\begin{figure}[h]
\centering
\includegraphics[scale=0.28]{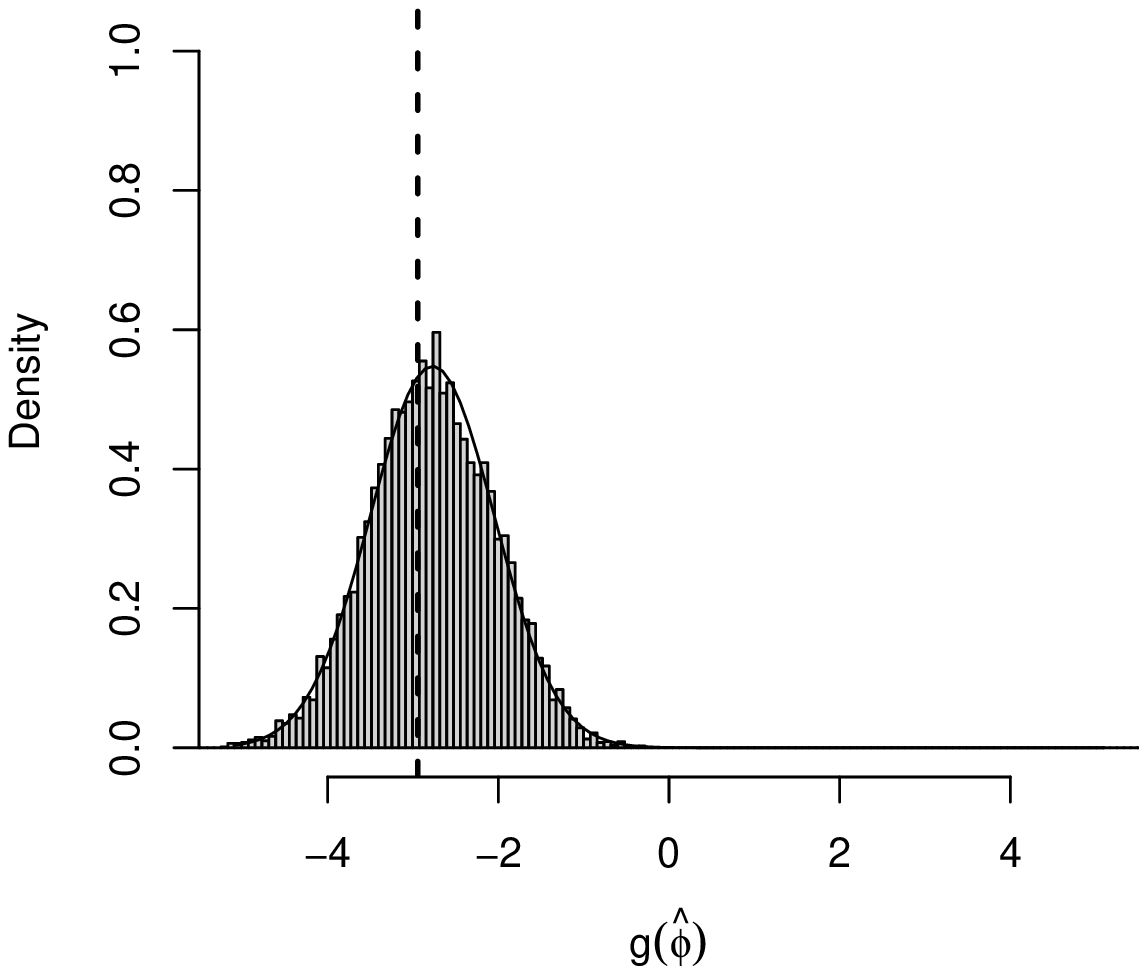}
\includegraphics[scale=0.28]{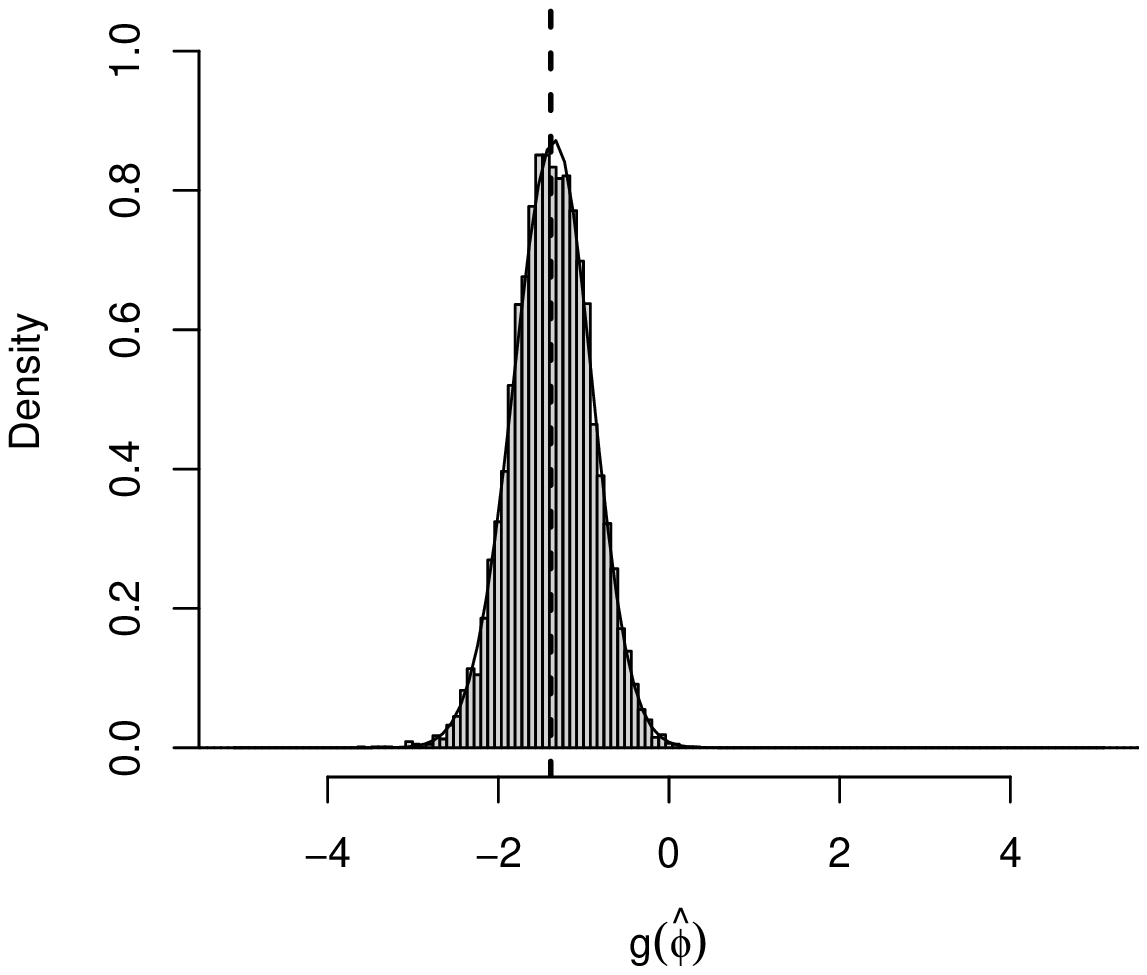}
\includegraphics[scale=0.28]{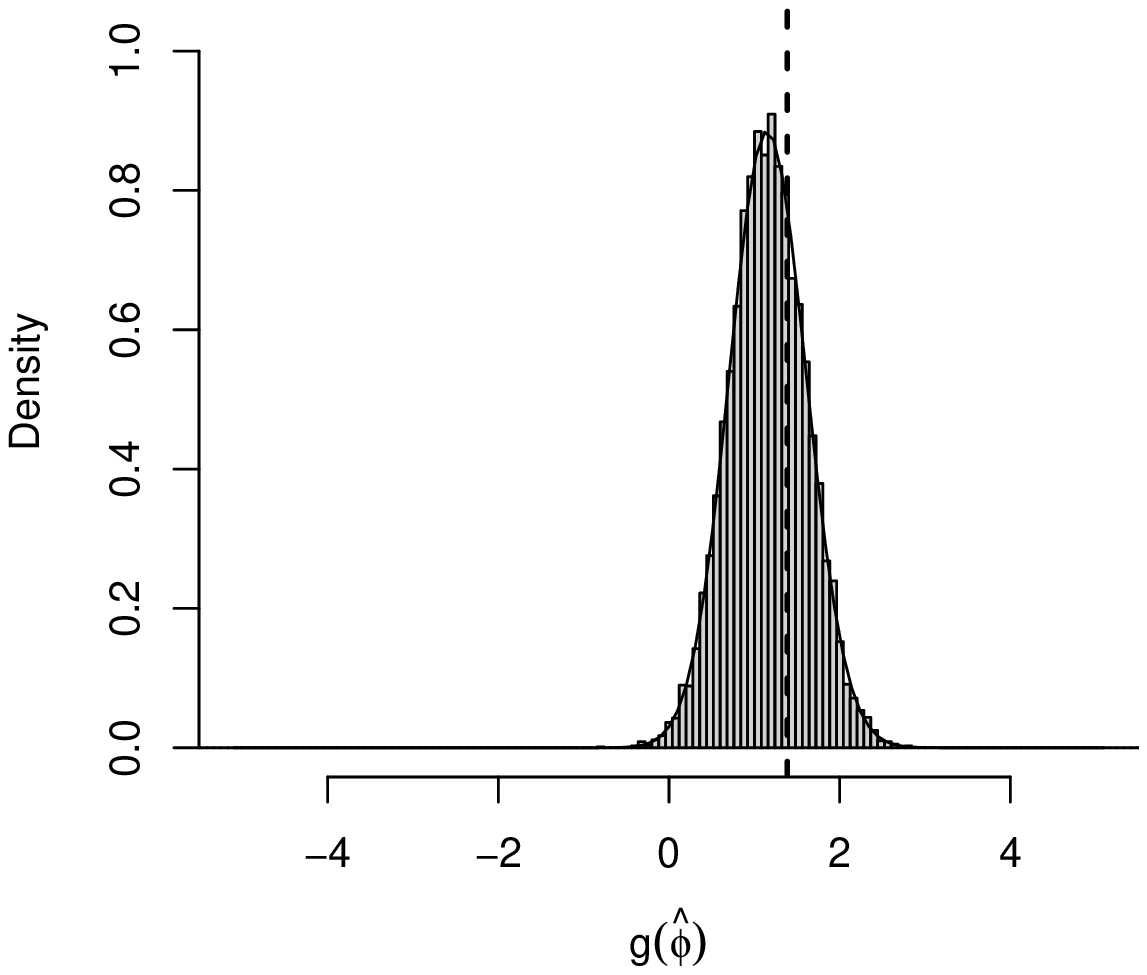}
\includegraphics[scale=0.28]{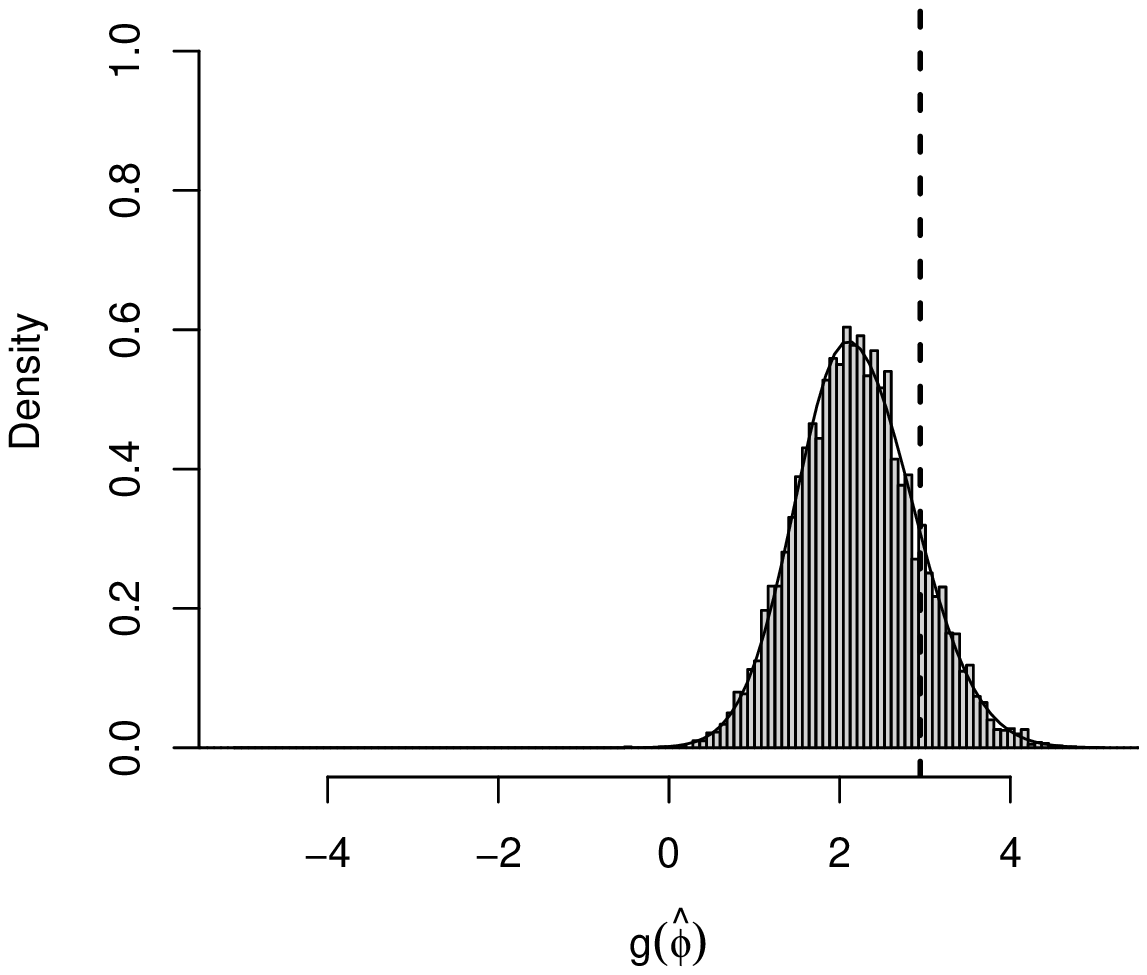}
\caption{Sampling distributions for $g(\hat \phi)$ where $\hat \phi$ is the exact MLE. The dashed lines represent the true values of $g(\phi)$ where $\phi\in (-0.9,-0.6,0.6,0.9)$. }
\label{fig:hist}
\end{figure}

 Commonly applied estimators for the coefficients of AR($p$) processes are asymptotically normal \citep{hannan:70} but the finite-sample distributions of these estimators have not been thoroughly studied. Figure~\ref{fig:hist} illustrates the sampling distribution for the logit-transformation $g(\hat \phi)$ where $\hat\phi$ is the exact MLE for AR(1) processes of length $n=30$ and where the underlying true values are $\phi \in (-0.9,-0.3,0,3, 0.9)$. The fitted curves are skew-normal densities which are seen to give very good approximations to the given sampling distributions. These have been fitted using the function \texttt{fGarch:::snormFit} in \texttt{R} \citep{fgarch:20}, implementing the skew-normal density as defined by \cite{fernandez:98}, i.e.
\begin{equation}
\pi_{\mbox{\scriptsize{sn}}}(x) =  \frac{2}{\xi+\frac{1}{\xi}}\left(\pi_G(x/\xi)H(x) + \pi_G(x\xi)H(-x)\right).
 \end{equation}
 The function $\pi_G(.)$ denotes the standard normal density while $H(x)$ is the ordinary Heaviside or unit step function.  In addition to the skewness parameter $\xi>0$, the skew-normal density is parameterized in terms of a mean $\mu$ and standard deviation $\sigma$, using an input argument $(x-\mu)/\sigma$.   
 
\begin{figure}[h]
\centering
\includegraphics[scale=0.28]{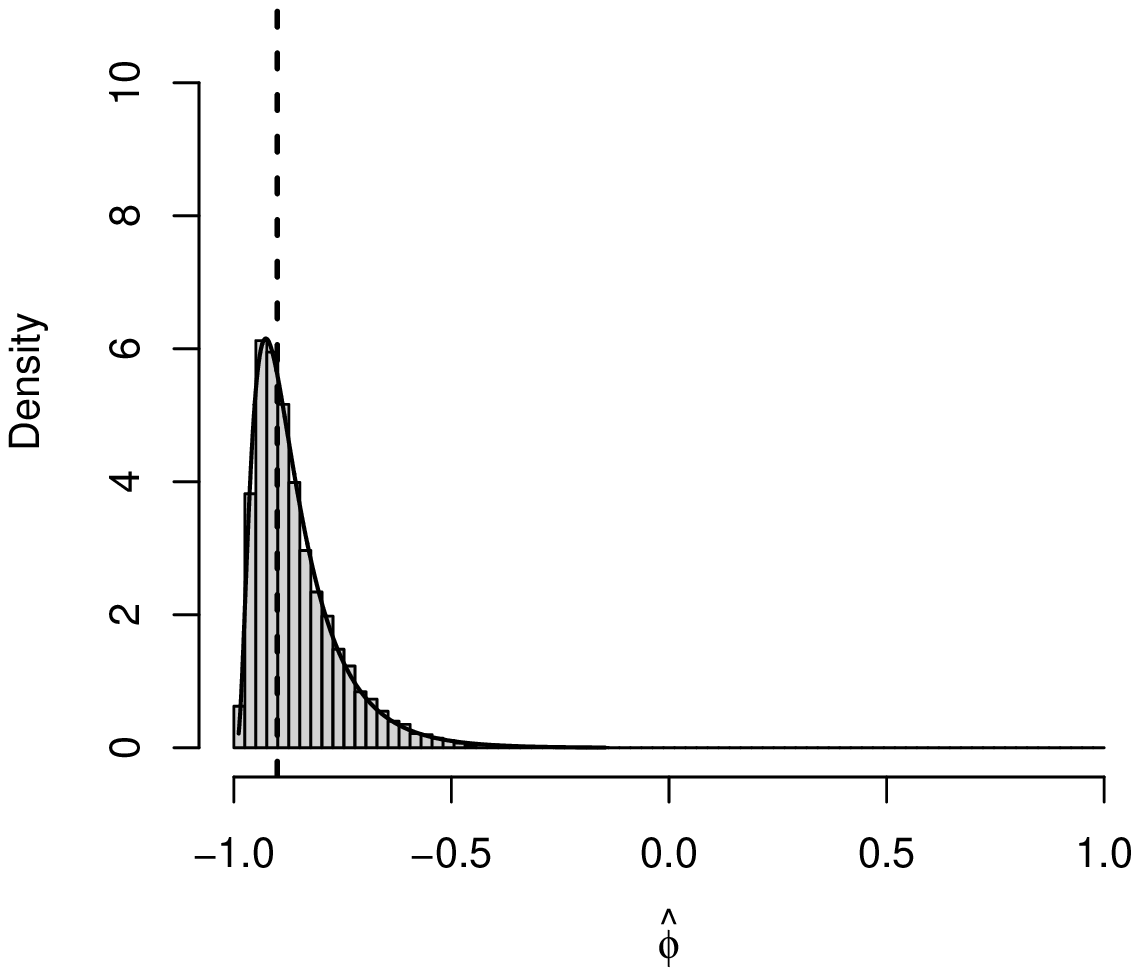}
\includegraphics[scale=0.28]{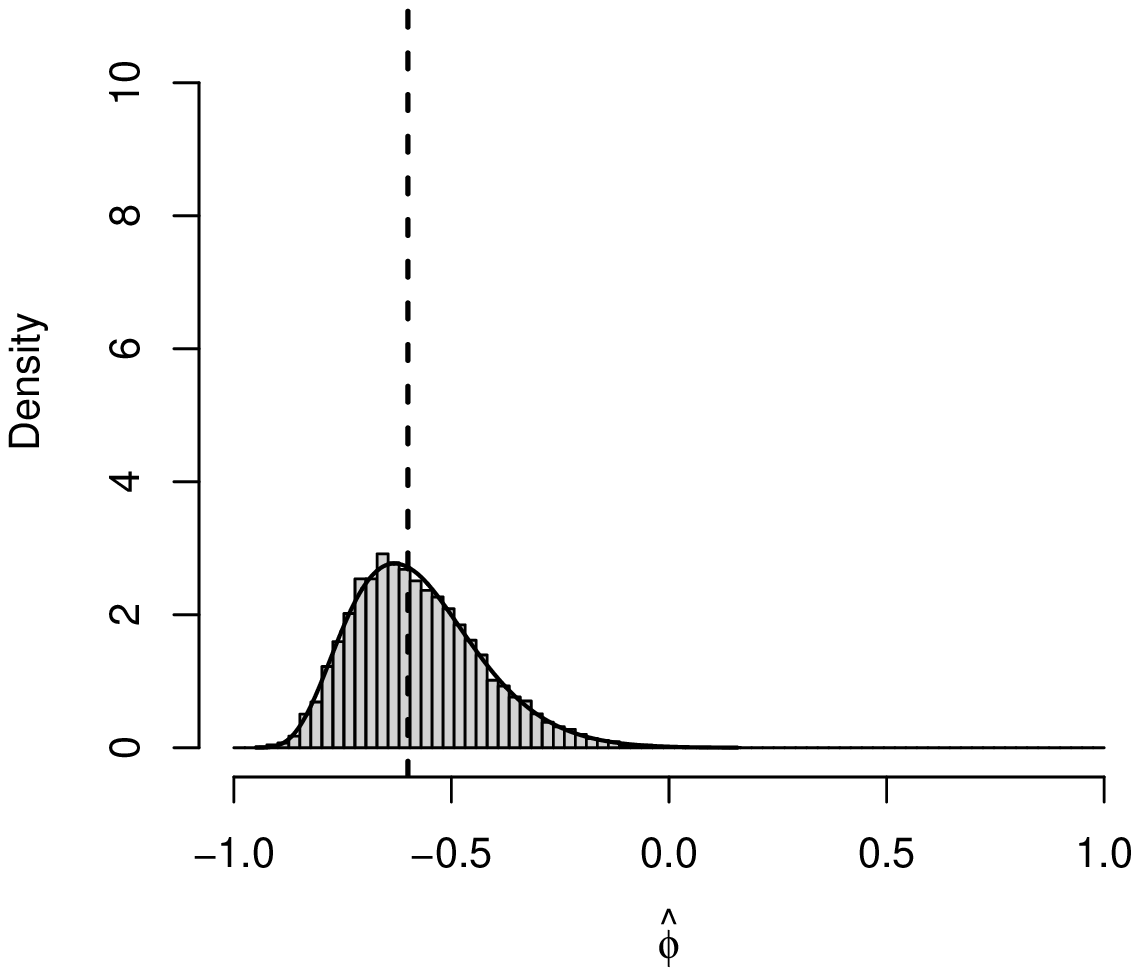}
\includegraphics[scale=0.28]{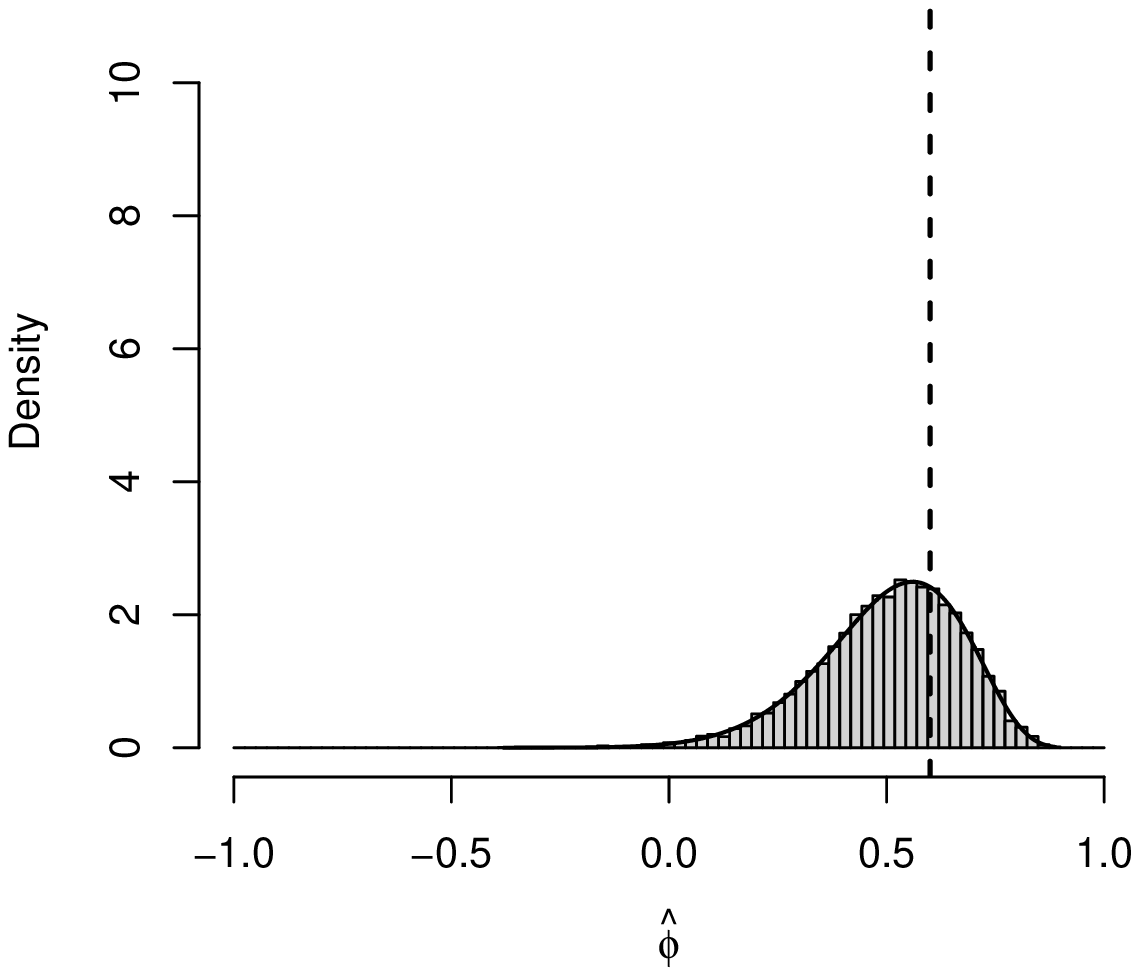}
\includegraphics[scale=0.28]{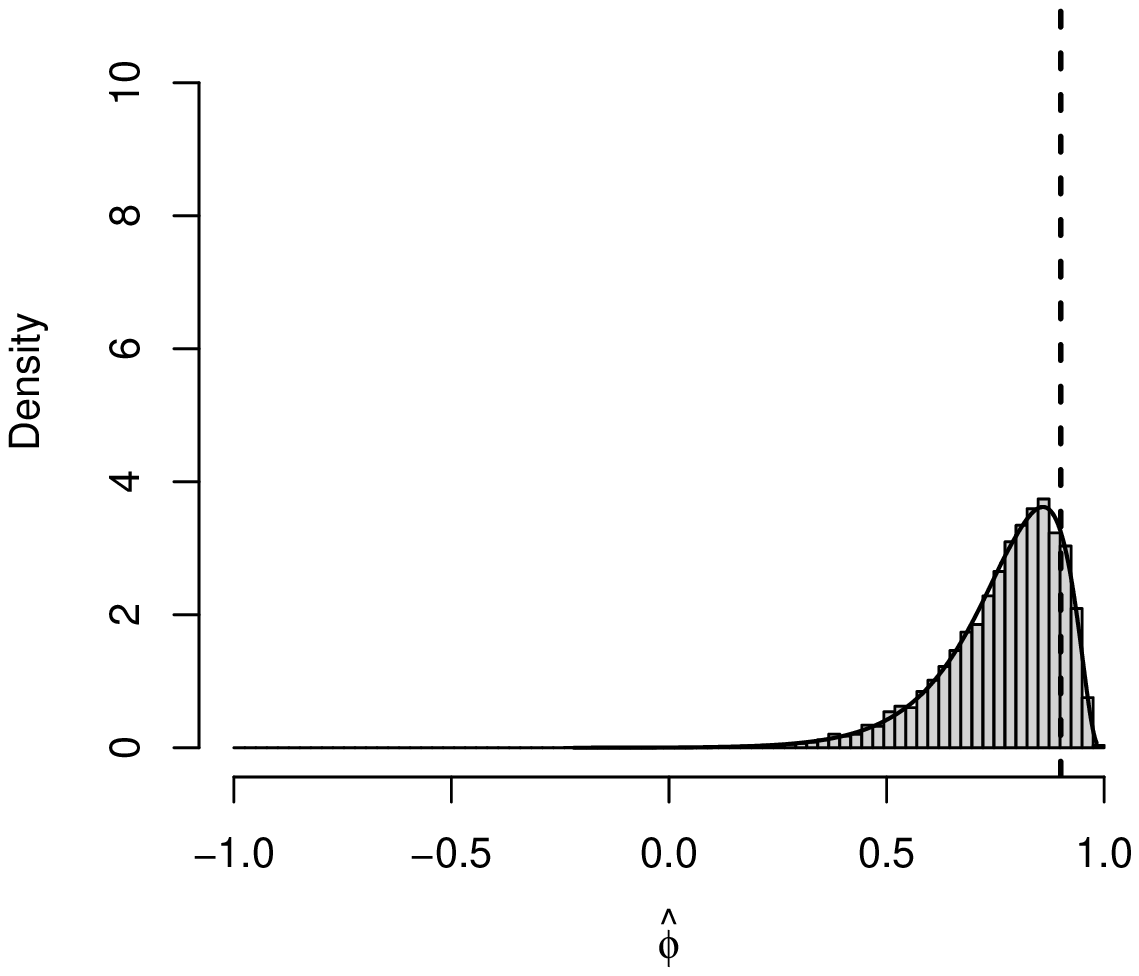}\\
\includegraphics[scale=0.28]{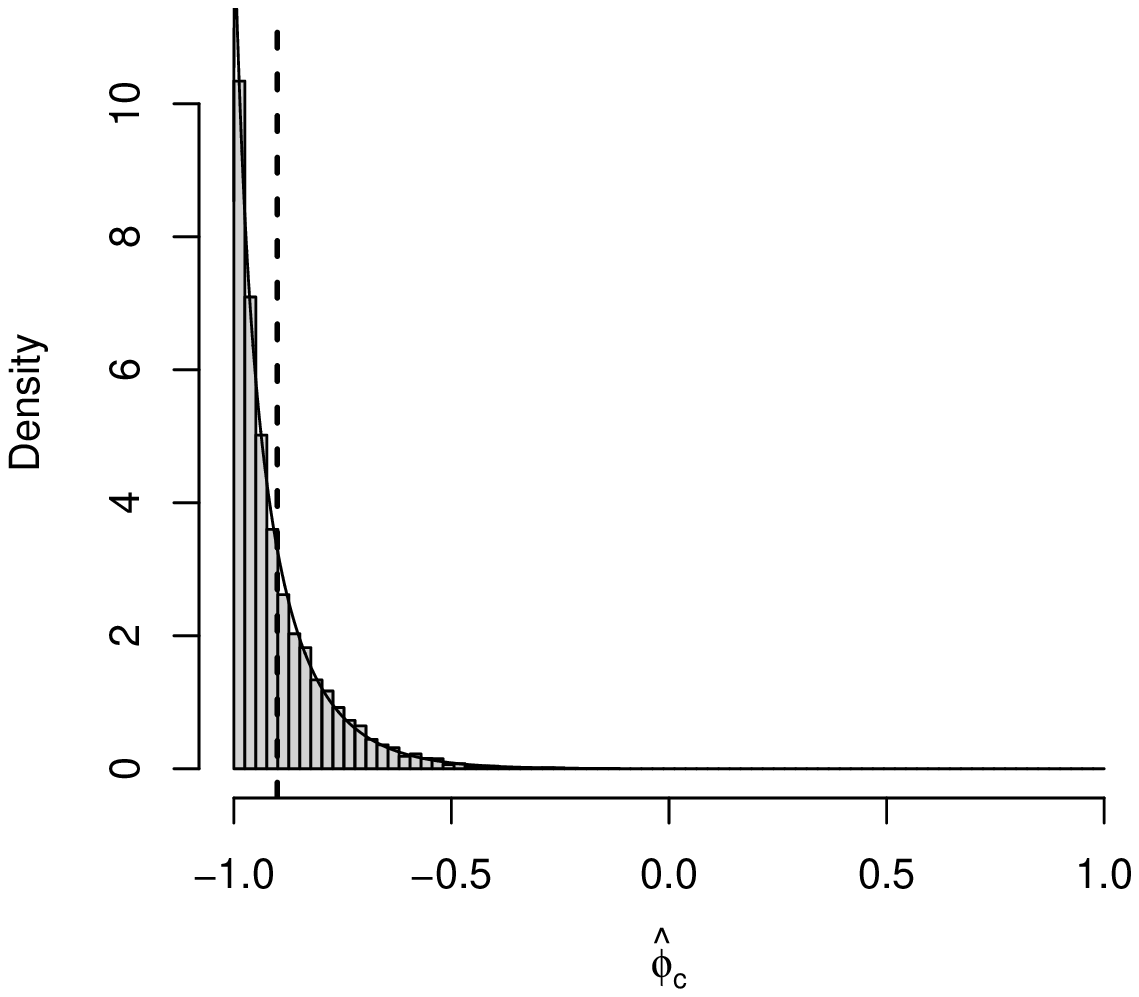}
\includegraphics[scale=0.28]{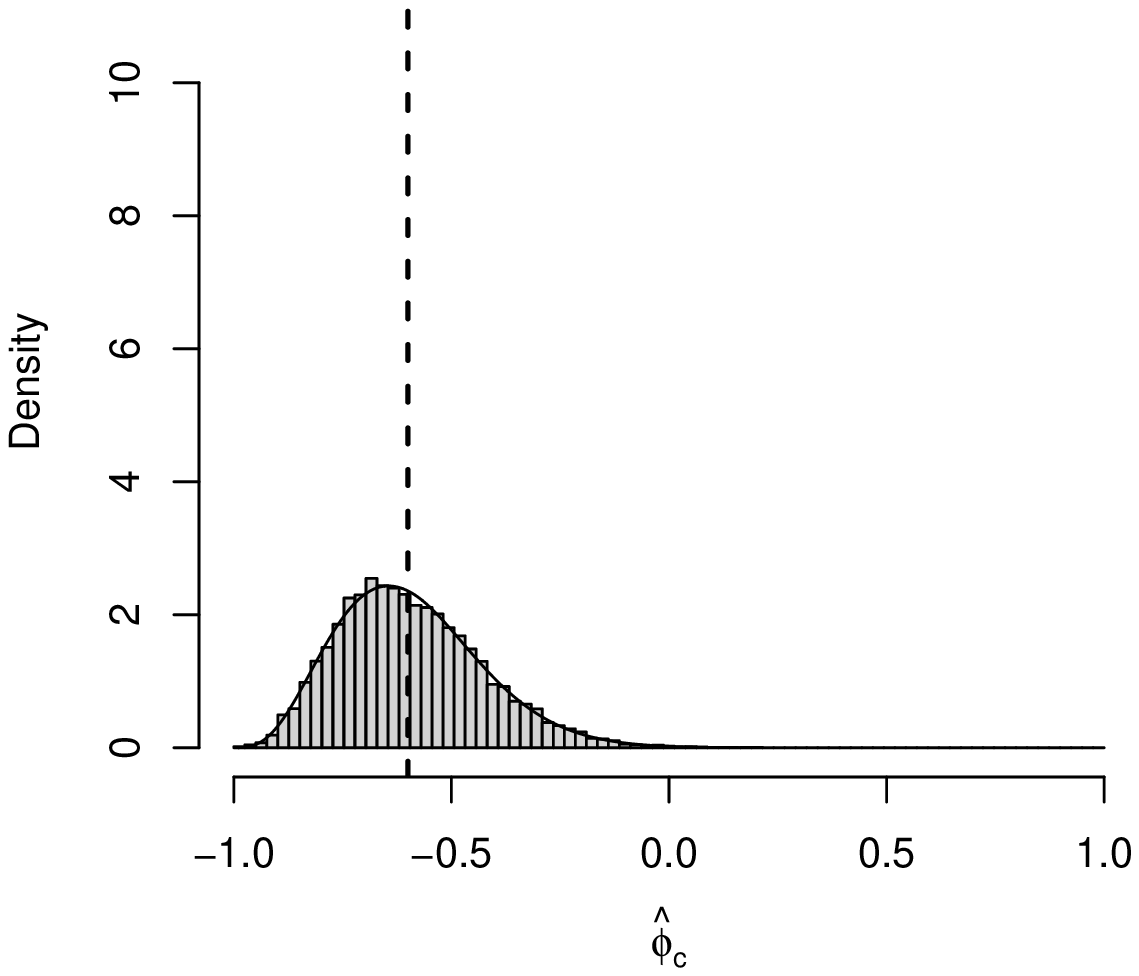}
\includegraphics[scale=0.28]{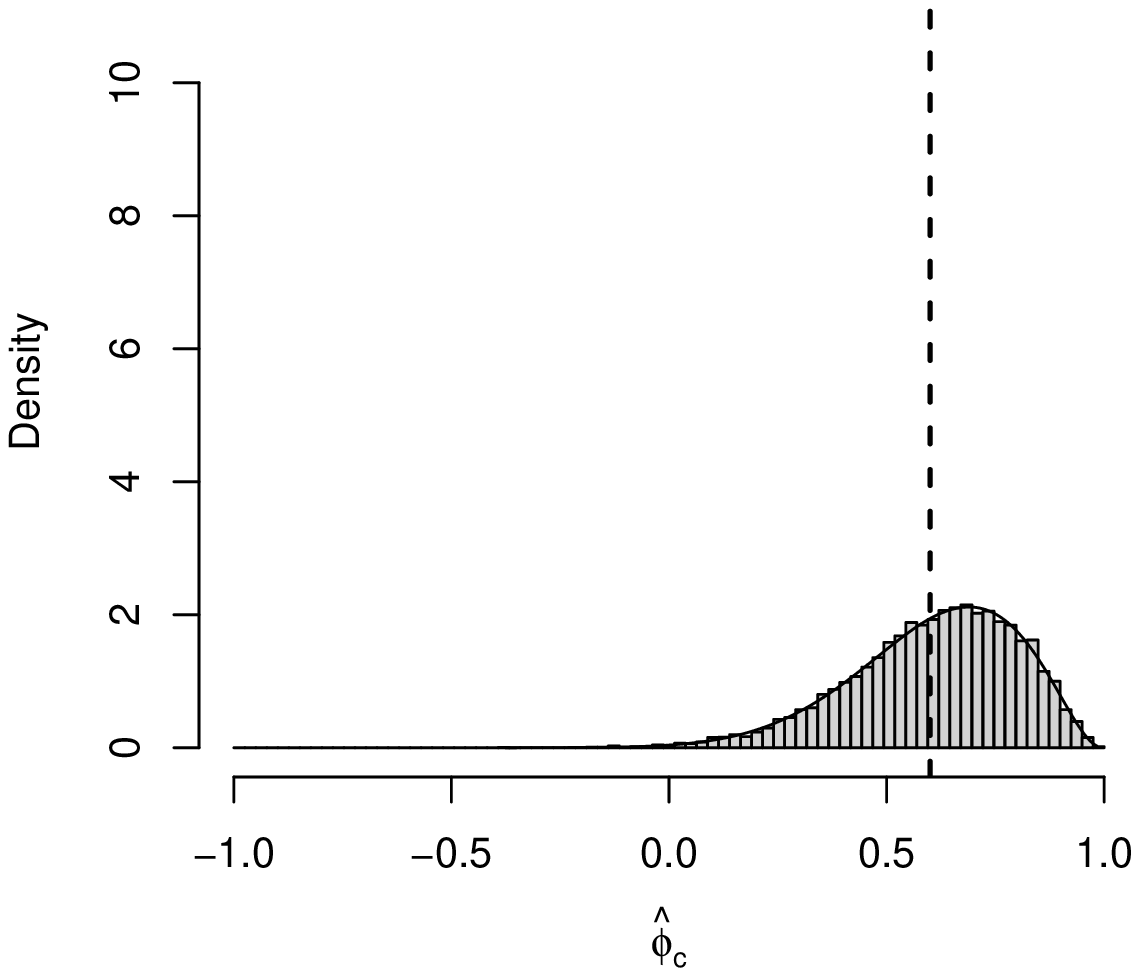}
\includegraphics[scale=0.28]{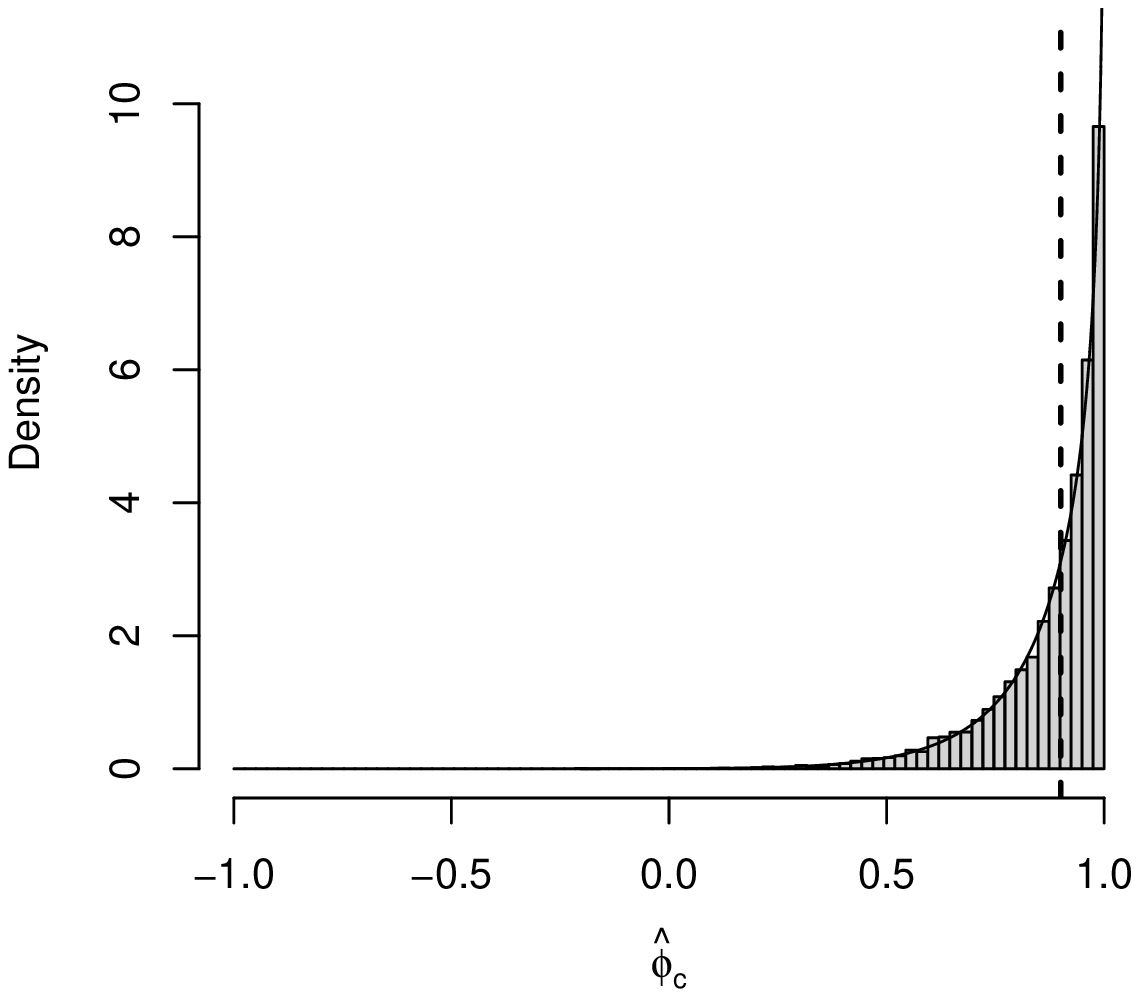}
\caption{Sampling distributions for $\hat \phi$ (upper panels) and  $\hat \phi_c$ (lower panels) where $\hat \phi$ is the exact MLE. The dashed lines give the true values of $\phi \in ( -0.9,-0.6,0.6,0.9)$.}
\label{fig:hist2}
\end{figure}

Applying the skew-normal approximation, we can 
derive finite-sample distributions for the estimators $\hat \phi$ and the corresponding bias-corrected estimators $\hat \phi_c$. Let $\tilde \pi_{\mbox{\scriptsize{sn}}}(.)$ denote the skew-normal approximation for the logit transformation $g(\hat \phi)$. The approximate sampling distribution for the original estimator is easily expressed analytically by the ordinary change of variable transformation
$$ \tilde \pi(\hat\phi) = \tilde \pi_{\mbox{\scriptsize{sn}}}(g(\hat \phi))\left|(1+\hat{\phi})^{-1}+(1-\hat{\phi})^{-1}\right|.$$
Likewise, an approximation to the sampling distribution for the corrected estimator can be derived numerically as 
$$ \tilde \pi(\hat{\phi}_c) = \tilde \pi_{\mbox{\scriptsize{sn}}} (s(\hat{\phi}_c))\left| \frac{ds(\hat{\phi}_c)}{d \hat{\phi}_c}\right |,$$
where $s(\hat \phi_c)$ represents a spline function approximating the monotonic relationship between $\hat \phi_c$ and $g(\hat\phi)$. The resulting sampling distributions for the original and corrected estimators are shown in Figure~\ref{fig:hist2}, where $n=30$ and $\phi \in (-0.9,-0.6,0,6, 0.9)$. 

\begin{figure}[h]
\centering
\includegraphics[scale=0.33]{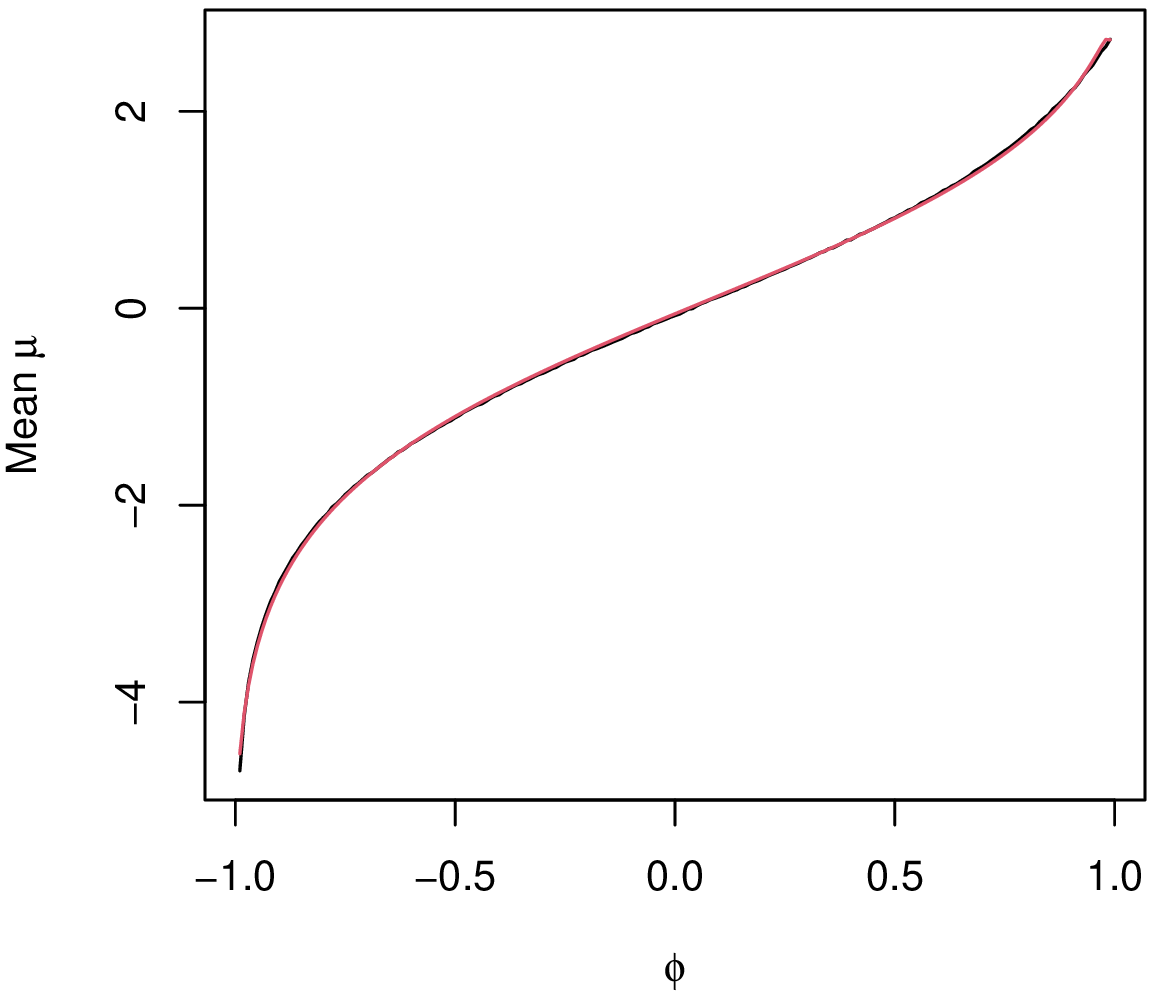}
\includegraphics[scale=0.33]{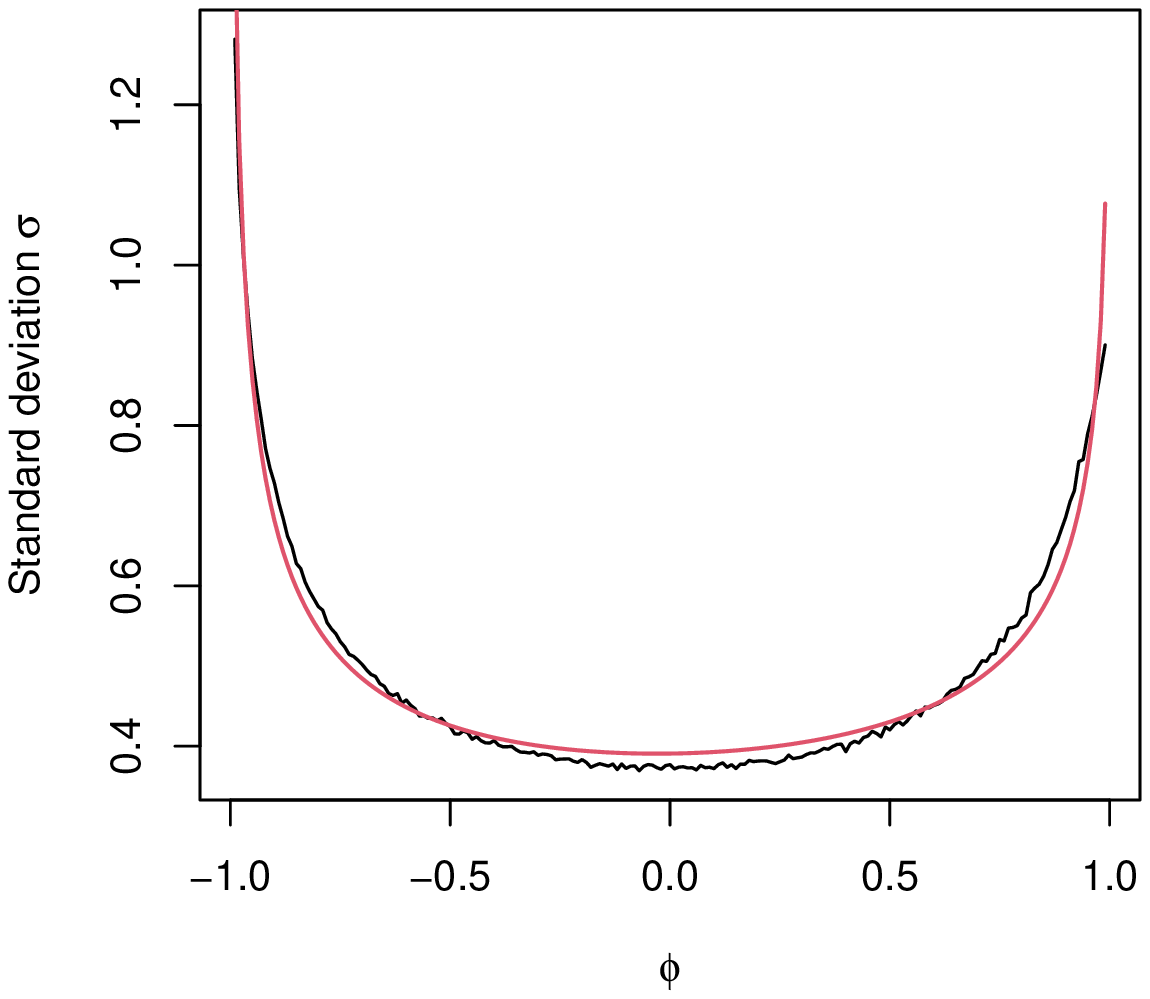}
\includegraphics[scale=0.33]{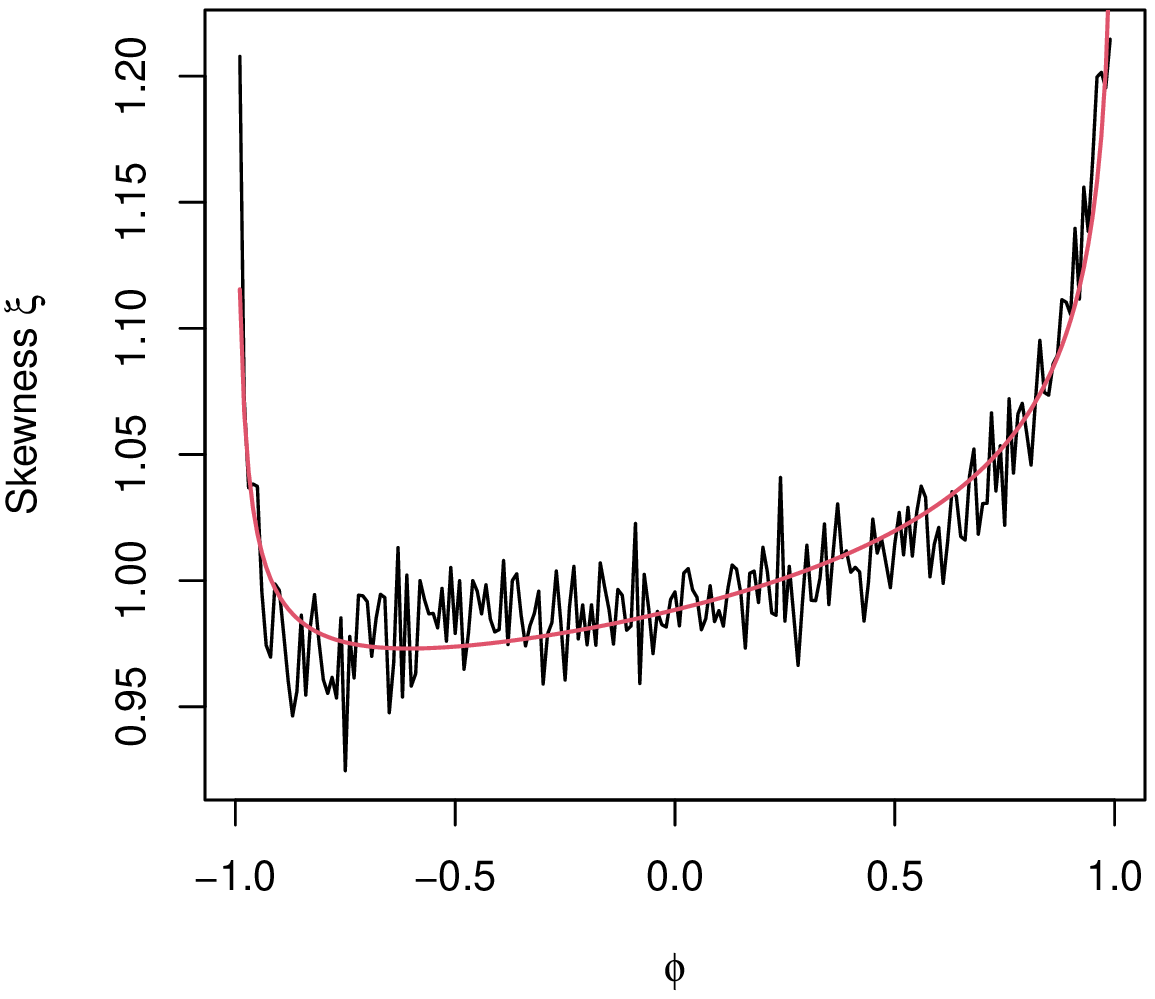}
\caption{The mean, standard deviation and skewness parameter of the skew-normal approximation for $g(\phi)$ as  smoothed functions (red) of $\phi$ using the exact MLE when $n=30$.}
\label{fig:sn-param}
\end{figure}

For each of the estimators and each $n$, we can now model the parameters of the skew-normal approximation for $g(\hat \phi)$ as a function of $\phi$. This is done using a third order orthogonal polynomial regression model similar to \eqref{eq:correct-est}, now using the parameters of the skew-normal approximation as the response variables. Specifically, let $\mm{\theta} = (\mu,\sigma,\xi)$. The predicted value of  each of the parameters is then defined by  
\begin{equation}
\hat \theta_ s= \sum_{k=0}^K \hat b_{k,s} h_k(g(\phi)),\quad s=1,2,3,\label{eq:ar1-fit-sn}
\end{equation}
where the coefficients $\{\hat b_{k,s} \}$ are found by the ordinary least squares approach using a polynomial order of  $K=3$. 
Figure~\ref{fig:sn-param} illustrates the relationships between each of the skew-normal parameters and $\phi$ using the exact MLE when $n=30$. The red curves illustrate the  fitted curves defined by \eqref{eq:ar1-fit-sn}. Especially, we notice that the skewness parameter is quite close to 1 for the interior of the given interval, implying that the given distributions for $g(\hat\phi)$ are not very far from being Gaussian. The skewness increases as $\phi$ increases towards the upper limit of the stationary area. 

To predict the skew-normal parameters for a new AR(1) series we use the estimates $\hat \phi$ as plug-in estimates for $\phi$ in \eqref{eq:ar1-fit-sn}. We then obtain confidence intervals for the estimates $\hat \phi$ and $\hat \phi_c$ by Monte Carlo simulations. To investigate the coverage probability of the resulting confidence intervals, we have performed a simulation study in which we generated 10000 AR(1) process with a uniformly drawn coefficient, i.e. $\phi\sim \mbox{Uniform}(-1,1)$. The simulation study was performed for sample sizes $n=10, 15, 20, 30, 40, 50$, in which the coefficient $\phi$ was estimated using each of the four original methods. We then calculated the bias-corrected estimates and found $95\%$ equi-tailed confidence intervals for all cases by Monte Carlo simulation.   The results demonstrate that the coverage probabilities for the original estimators are below the nominal level of 0.95 for all estimators and all sample sizes, see Table~\ref{tab:coverage}.  Especially, the nominal level using the Yule-Walker solution is below 0.90 for all sample sizes. Using the bias-corrected estimators, the coverage properties are clearly better being quite close to the nominal level of 0.95 in all cases. For the smaller sample sizes, this can partly be explained by the larger variance of the bias-corrected estimators giving wider confidence intervals. For the large sample sizes, the confidence lengths are approximately the same.

\begin{table}[h]
\begin{center}
\begin{tabular}{c|cc|cc|cc|cc|}
 & \multicolumn{2}{c|}{Exact MLE} & \multicolumn{2}{c|}{Conditional MLE} & \multicolumn{2}{c|}{Burg's method} &  \multicolumn{2}{c|}{Yule-Walker} \\
n &  $\hat \phi$ &  $\hat \phi_c$ &  $\hat \phi$ &  $\hat \phi_c$ &  $\hat \phi$ &  $\hat \phi_c$  &  $\hat \phi$ &  $\hat \phi_c$ \\\hline
10 & 0.8092 & 0.9771 & 0.8097 & 0.9768 & 0.7798 & 0.9829 & 0.6692 & 0.9942\\
15 & 0.8326 & 0.9741 & 0.8330 & 0.9736 & 0.8012 & 0.9754 & 0.7198 & 0.9791\\
20 & 0.8463 & 0.9644 & 0.8468 & 0.9635 & 0.8233 & 0.9620 & 0.7415 & 0.9624\\
30 & 0.8617 & 0.9527 & 0.8608 & 0.9532 & 0.8460 & 0.9508 & 0.7814 & 0.9476\\
40 & 0.8728 & 0.9506 & 0.8773 & 0.9502 & 0.8621 & 0.9488 & 0.8088 & 0.9437\\
50 & 0.8813 & 0.9474 & 0.8878 & 0.9457 & 0.8722 & 0.9456 & 0.8234 & 0.9396\\
\end{tabular}
\caption{Coverage probabilities of $95\%$ confidence intervals using the original estimators $\hat \phi$ and the respective bias-corrected versions $\hat \phi_c$ for a total of 10000 simulations where the true value $\phi$ is randomly generated from $(-1,1)$.  The confidence intervals are found by Monte Carlo sampling after fitting a skew-normal approximation to  $g(\hat\phi)$. }
\label{tab:coverage}
\end{center}
\end{table}

\section{Bias correction and sampling distribution in the AR(2) case}\label{sec:ar2}
In this section we extend the given model-based approach to construct bias-corrected estimators for the pair of coefficients ($\phi_1,\phi_2$) of an AR(2) process. This is far more computationally expensive than in the AR(1) case as we need to generate time series for a fine two-dimensional grid of the coefficients in the triangular stationary area.  In total we have fitted a weighted polynomial regression model to more than 2.43 billion time series, providing regression coefficients  which can be used to bias-correct estimators for sample sizes from $n=10,11,\ldots , 50$. As previously, the original estimators used include the exact and conditional MLE, Burg's algorithm and the Yule-Walker solution.  In addition, we obtain approximate sampling distributions by constructing a Gaussian copula where the  marginals are generated as transformations of skew-normal densities and combine this with Monte Carlo simulations. 

\subsection{Modeling approach in two dimensions}
The  AR(2) process is defined by \eqref{eq:arp} where $p=2$ and it is stationary within the triangular area constrained by $\phi_2+|\phi_1|<1$ where $ |\phi_2|<1$.  
A more appealing parameterization of this process is given by the partial autocorrelations, 
\begin{equation}
\psi_1 = \frac{\phi_1}{1-\phi_2}, \quad \psi_2=\phi_2, \label{eq:psi2}
\end{equation}
as  the stationary area of the AR(2) process is then defined by the square $\psi_i\in (-1,1)$,  $i=1,2$. The area in which the process has pseudo-periodic behavior is characterized by $\psi_1^2(1-\psi_2)^2+4\psi_2<0$.
 
We now extend  the algorithm in Section~\ref{sec:ar1} to construct bias-corrected estimators $(\hat{\phi}_{c,1},\hat{\phi}_{c,2})$, again taking the sampling distribution of the original pair of estimators $(\hat{\phi}_{1},\hat{\phi}_{2})$ into account. In this case we estimate the parameters of the regression model by minimizing the squared error between the  corrected and true partial autocorrelations. This is computationally beneficial to avoid the triangular constraints  on $\phi_1$ and $\phi_2$. Also, the correlation between $\hat \psi_1$ and $\hat \psi_2$ is much smaller than the corresponding correlation between the $\phi$-coefficients. Naturally, this only makes a difference for the first coefficient  as $\phi_2=\psi_2$. The algorithm extending the weighted polynomial regression model to two dimensions can be summarized as follows:

\begin{enumerate}
\item Using the logit-transformation in \eqref{eq:logit}, the underlying true partial autocorrelations are modeled by 
\begin{eqnarray}
\psi_{i} = f(\hat \psi_1,\hat \psi_2,\mm{\beta}_i) = g^{-1}\left(\sum_{k=0}^K \sum_{q=0}^{K-k}\beta_{k,q,i} h_{k,q}(g(\hat\psi_{1}),g(\hat\psi_{2}))\right),\quad i=1,2 \label{eq:polregr}
\end{eqnarray}
where 
$h_{k,q}(g(\hat\psi_1),g(\hat\psi_2)) =h_k(g(\hat\psi_1)) h_q(g(\hat\psi_2))$
 denotes the product of Hermite polynomials of order $k$ and $q$. Notice that the two partial autocorrelations are modeled separately, giving separate sets of regression coefficients 
 $\mm{\beta}_i=\{\beta_{k,q,i}\}$ for $i=1,2$. However, each of the true partial autocorrelations need to be modeled in terms of the estimated pair $(\hat\psi_1,\hat\psi_2)$ as these parameters are not independent. 
\item  Due to the dependence, the regression coefficients $\mm{\beta}=\{\mm{\beta}_1,\mm{\beta}_2\}$ of the given predictors for $\psi_1$ and $\psi_2$ are  found simultaneously. This is achieved by solving the following optimization problem:
\begin{eqnarray}
\hat{\mm{\beta}} & = & 
\arg\min_{\mm{\beta}} \sum_{r=1}^l 
\sum_{i=1}^2\frac{1}{s^2_{ri}}\left(
\frac{1}{m}\sum_{j=1}^m g^{-1}\left(\sum_{k=0}^{K}\sum_{q=0}^{K-k} \beta_{k,q,i} h_{k , q}(g(\hat{\psi}_{rj1}),g(\hat{\psi}_{rj2}))\right)-\psi_{ri}\right)^2 \nonumber \\
 & = & \arg\min_{\mm{\beta}} \sum_{r=1}^l 
\sum_{i=1}^2\frac{1}{s^2_{ri}}\left(
\frac{1}{m}\sum_{j=1}^m  f(\hat \psi_{rj1},\hat \psi_{rj2},\mm{\beta}_i) -\psi_{ri}\right)^2. \label{eq:optim-ar2}
\end{eqnarray}
The values $(\hat{\psi}_{rj1},\hat\psi_{rj2})$ denote the original estimates for the $r$th pair of the partial autocorrelations in simulation $j$ while $s^2_{ri}$ denotes the sample variances for the $m$ simulations in each case.  The value $l$ denotes the total number of pairs of the partial autocorrelations that are included in the minimization. 
\end{enumerate}
In solving \eqref{eq:optim-ar2}, we needed to generate time series for a fine two-dimensional grid of the parameter values $(\psi_1,\psi_2)$ within the square defining the stationary area. 
Specifically, our estimate $\hat{\mm{\beta}}$ is based on using the grid $\psi_i \in (-0.95, - 0.925,\ldots ,  0.95)$, $i=1,2$.  This gives a total of $l=77^2 = 5929$ different combinations of the partial autocorrelations. For each pair $(\psi_1,\psi_2)$, we generated $m=10000$ times series of  a specific length $n$ implying that the regression coefficients are estimated based on approximately 59 million time series. This was repeated for all sample sizes  $n=10,11,\ldots , 50$, such that the total number of generated time series is equal to $77^2 \cdot 41\cdot 10000 = 2430890000 $ or approximately 2.43 billion time series for a given value of $K$. We then saved the regression coefficients for each sample size and for each of the original estimation methods providing bias-corrected estimators
$$\hat \phi_{c,i} =  g^{-1}\left(f(\hat \psi_1,\hat \psi_2,\hat{\mm{\beta}}_i)\right),\quad i=1,2.$$
As in the AR(1) case, the estimated coefficients might fall at the border of the stationary area but not outside.  

If run sequentially on an ordinary single-core laptop, the given brute-force simulation approach to compute the bias correction would be
computationally infeasible as it would take approximately 5 years of CPU
time. The computations were therefore done on the Ibex cluster at KAUST (\href{https://www.hpc.kaust.edu.sa/ibex}{https://www.hpc.kaust.edu.sa/ibex}), which reduced the time down to
about 2 days.

\subsection{Properties of the bias-corrected estimators for AR(2)}
By using the partial autocorrelations in \eqref{eq:optim-ar2}, the resulting corrected estimators are not constructed to  be completely unbiased. However, in addition to the numerical advantages, we have  noticed that this gives a smaller variance and RMSE compared to performing the optimization with respect to the $\phi$-coefficients. In constructing the bias-corrected estimator it is important that we also consider the ordinary bias-variance trade-off. This is also relevant in determining the maximum order $K$ of the Hermite polynomials in \eqref{eq:optim-ar2} as increasing values of $K$ naturally give decreased bias but higher variance. 

\begin{figure}[h]
\centering
\includegraphics[scale=0.38]{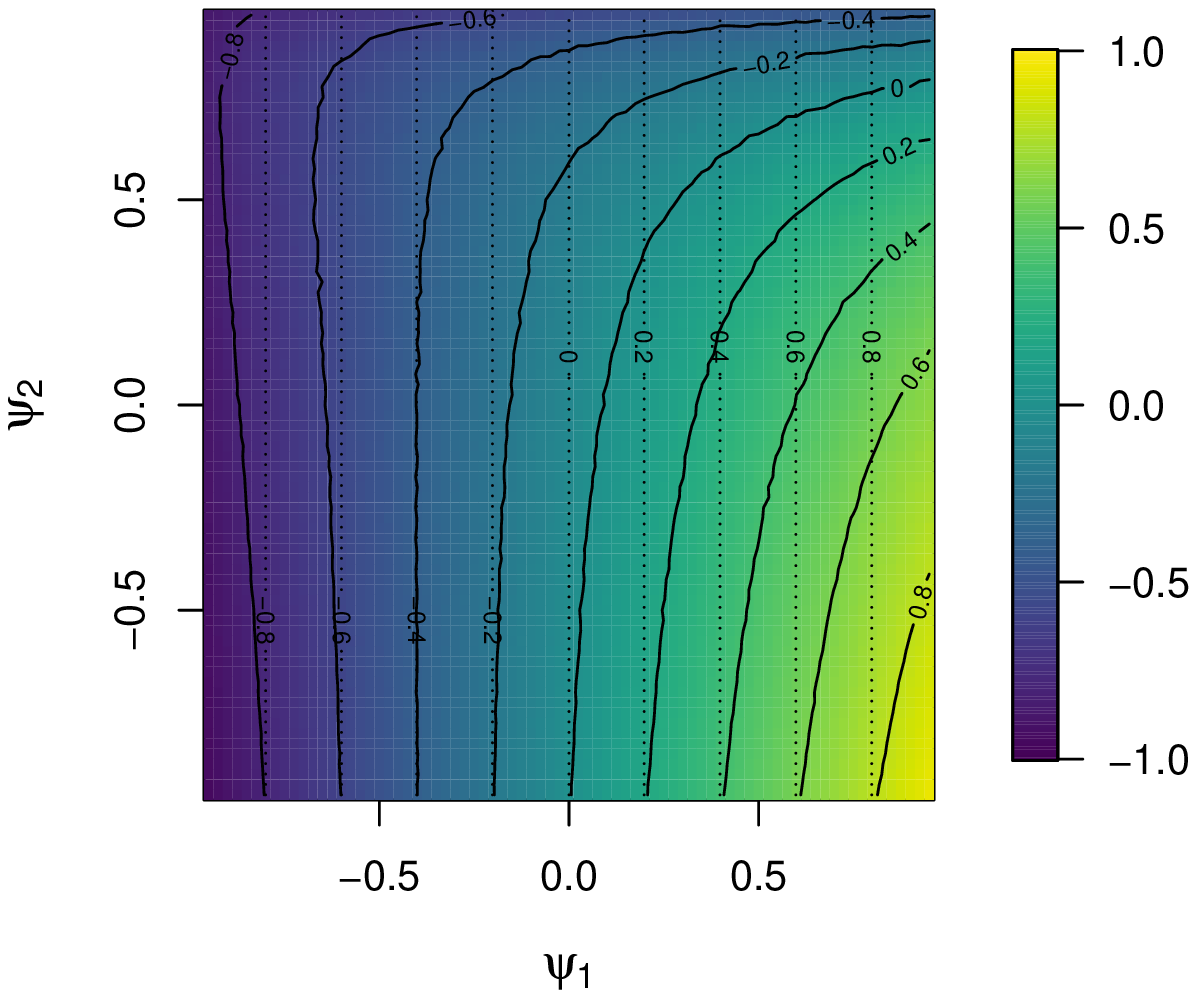}
\includegraphics[scale=0.38]{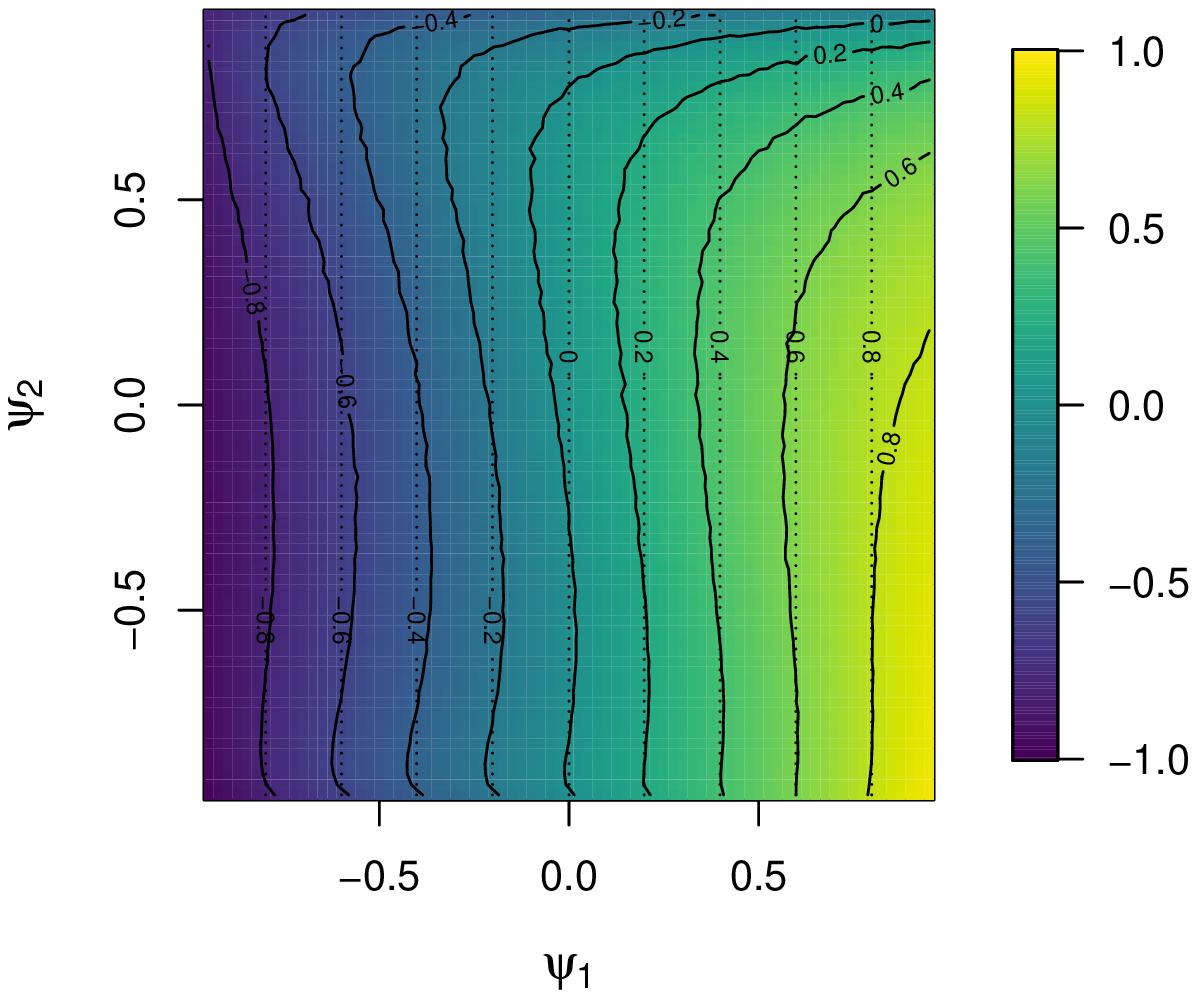}
\includegraphics[scale=0.38]{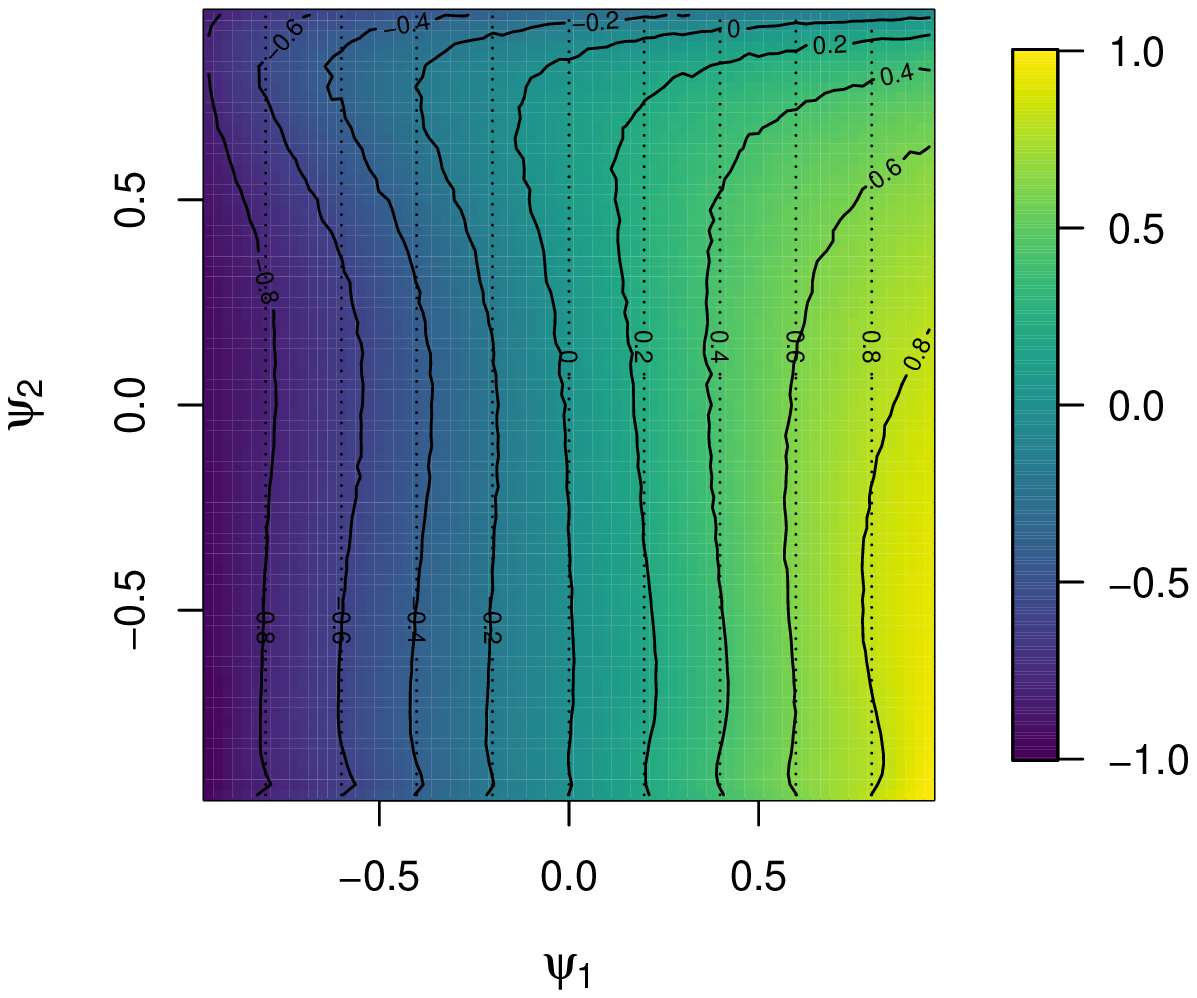}\\
\includegraphics[scale=0.38]{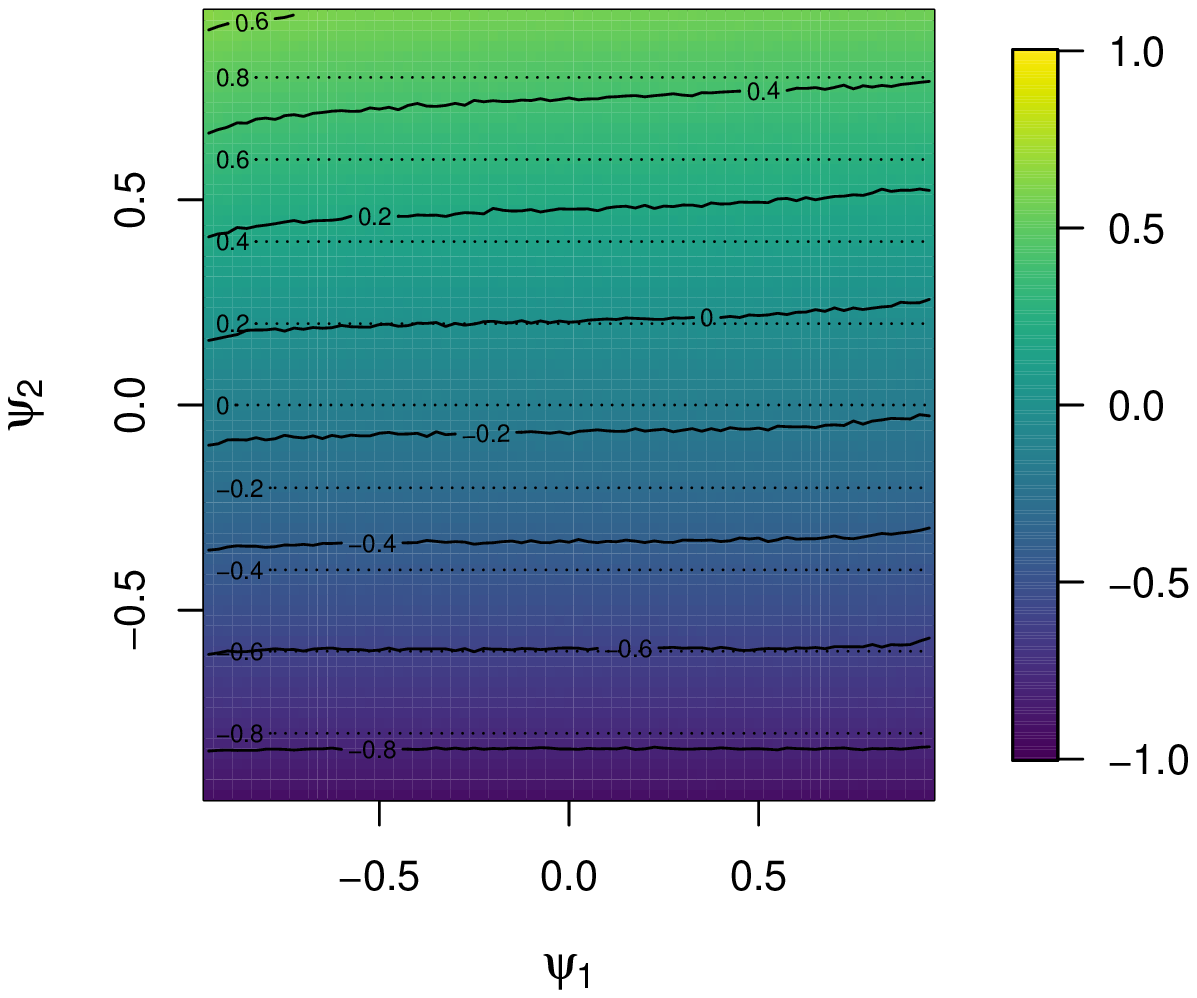}
\includegraphics[scale=0.38]{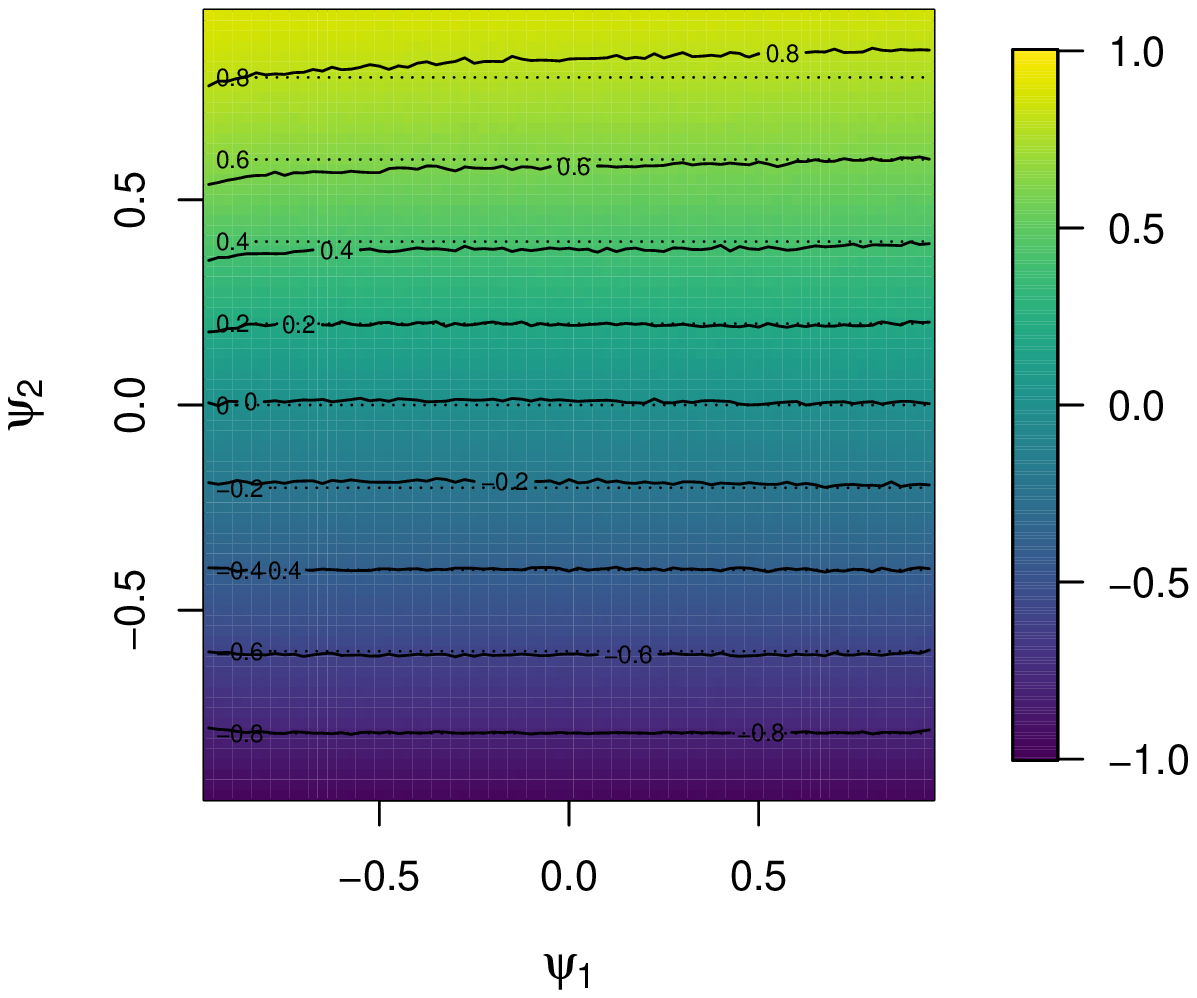}
\includegraphics[scale=0.38]{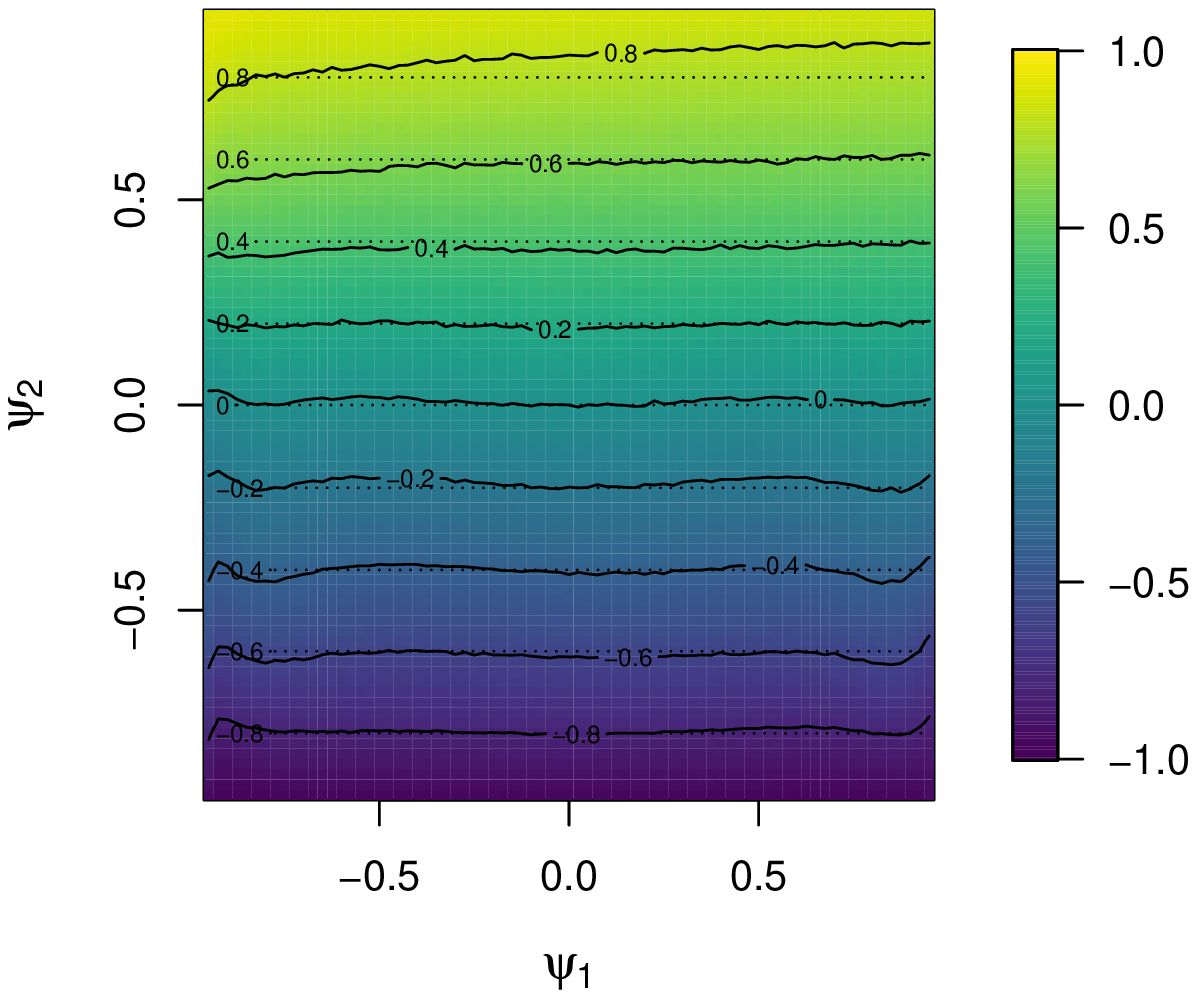}\\
\caption{Results for $n=15$. 
The mean estimated values of $\psi_1$ (upper) and $\psi_2$ (lower)  using the exact MLE (left), the bias-corrected estimator with $K=3$  (middle) and the bias-corrected estimator with $K=7$ (right).
}
\label{fig:ar2-image-n15}
\end{figure}

\begin{figure}[h]
\centering
\includegraphics[scale=0.38]{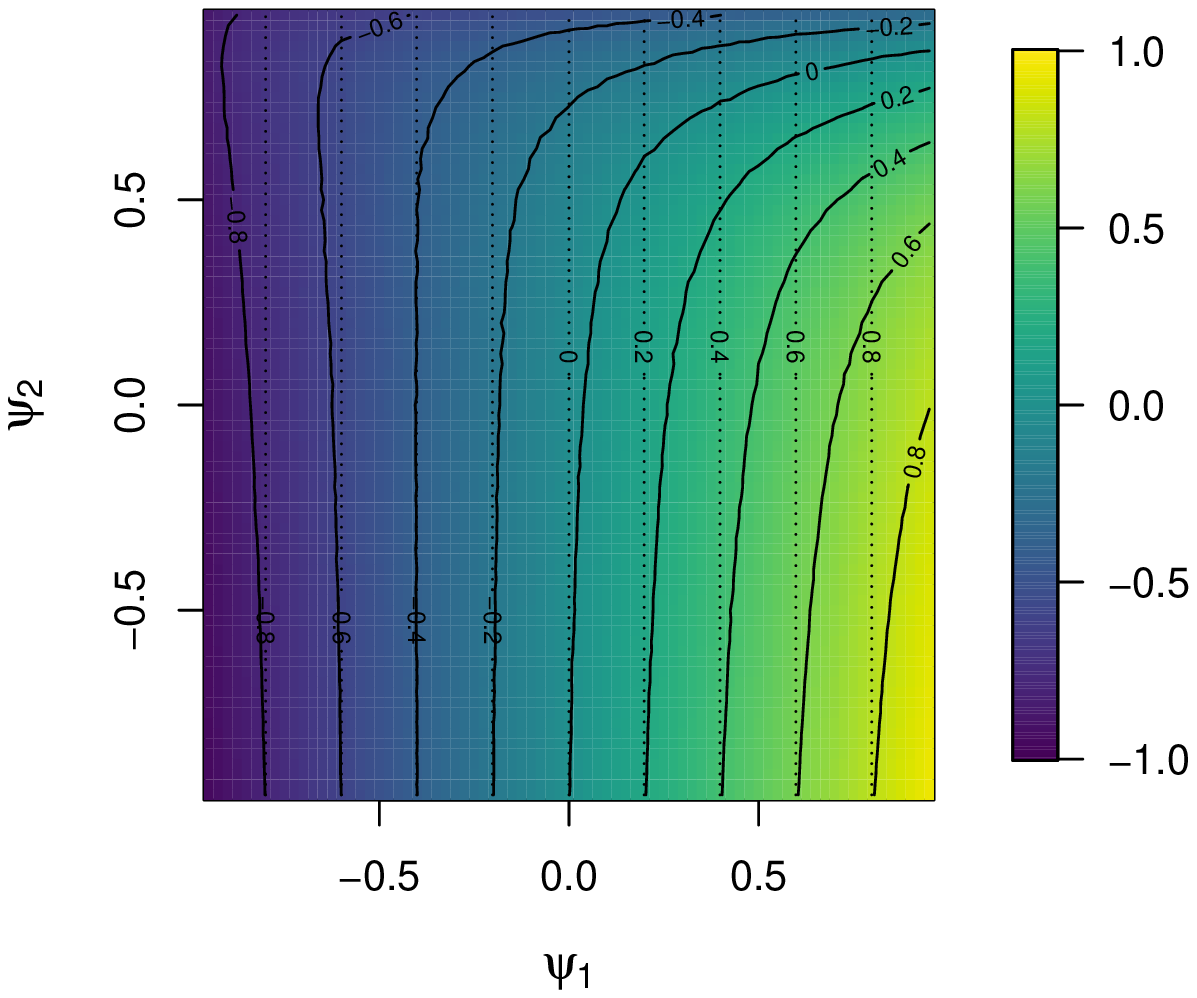}
\includegraphics[scale=0.38]{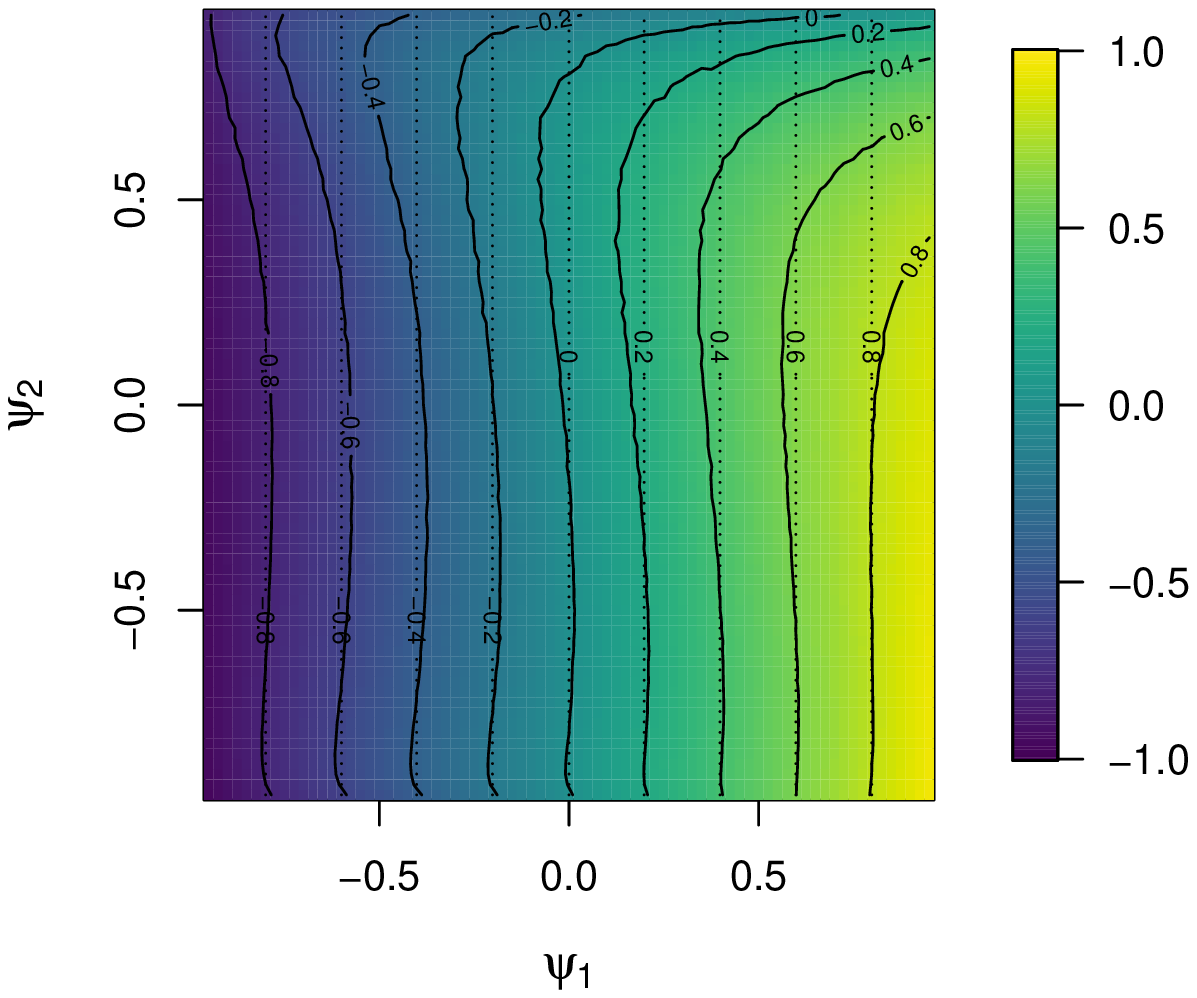}
\includegraphics[scale=0.38]{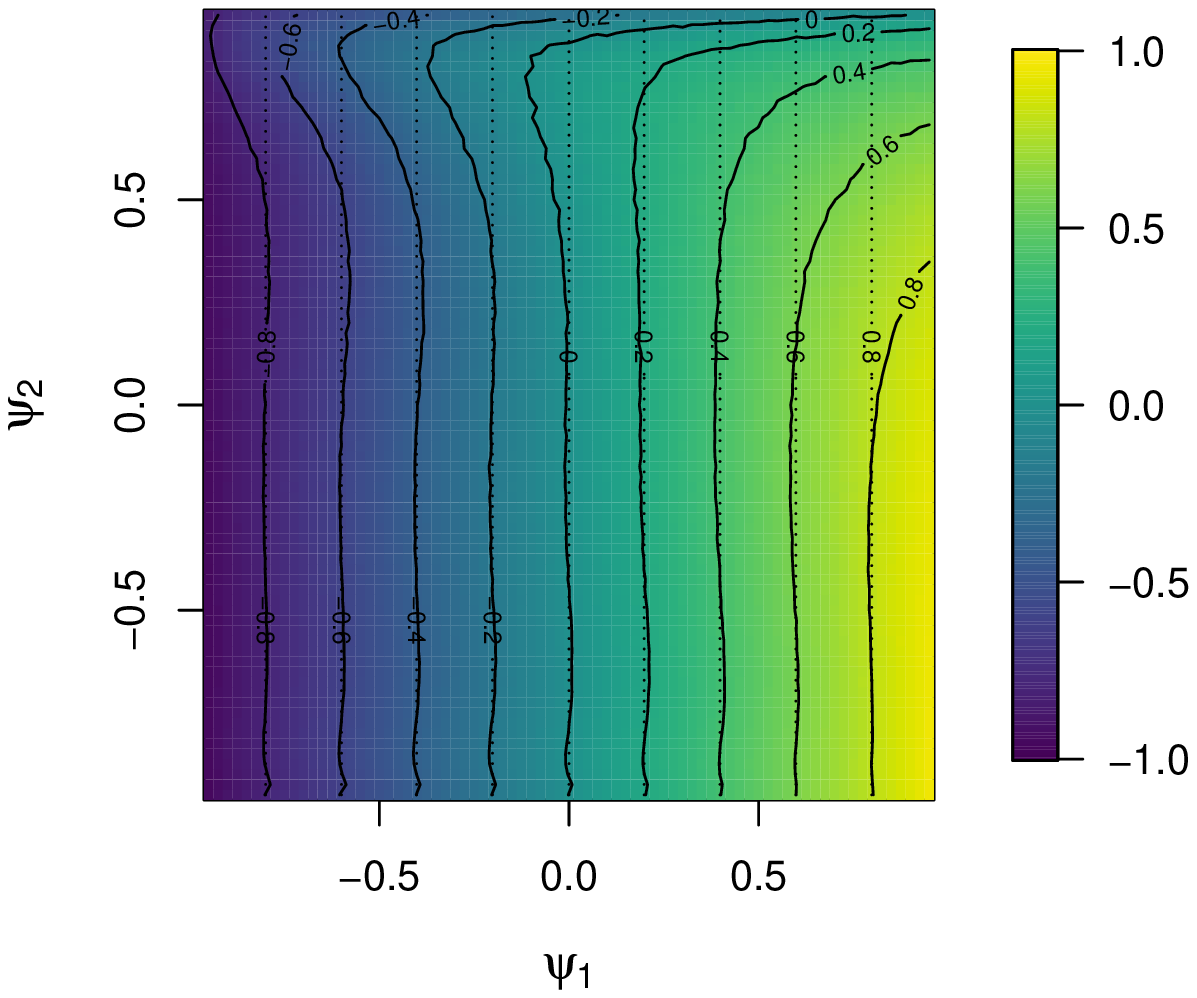}\\
\includegraphics[scale=0.38]{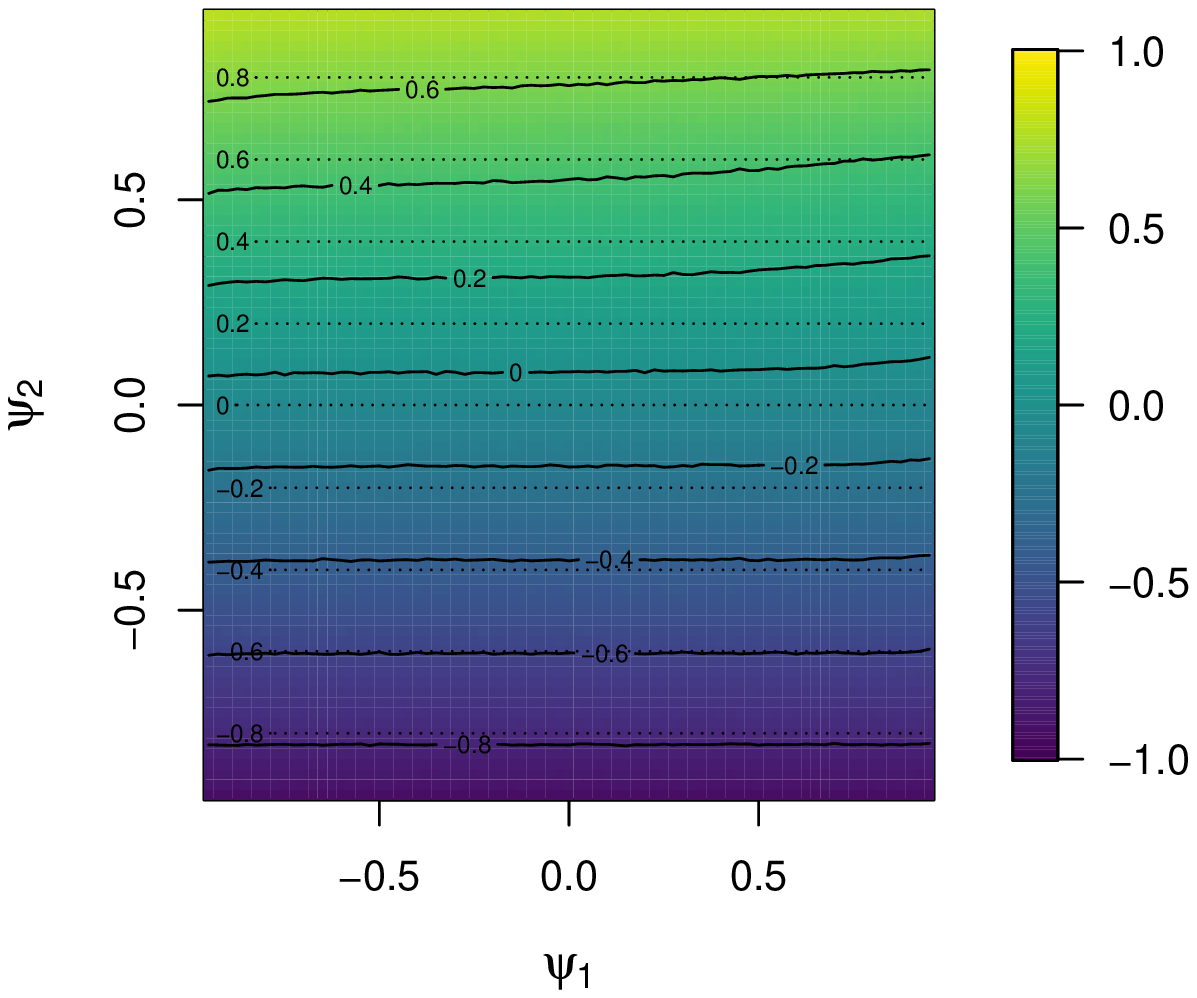}
\includegraphics[scale=0.38]{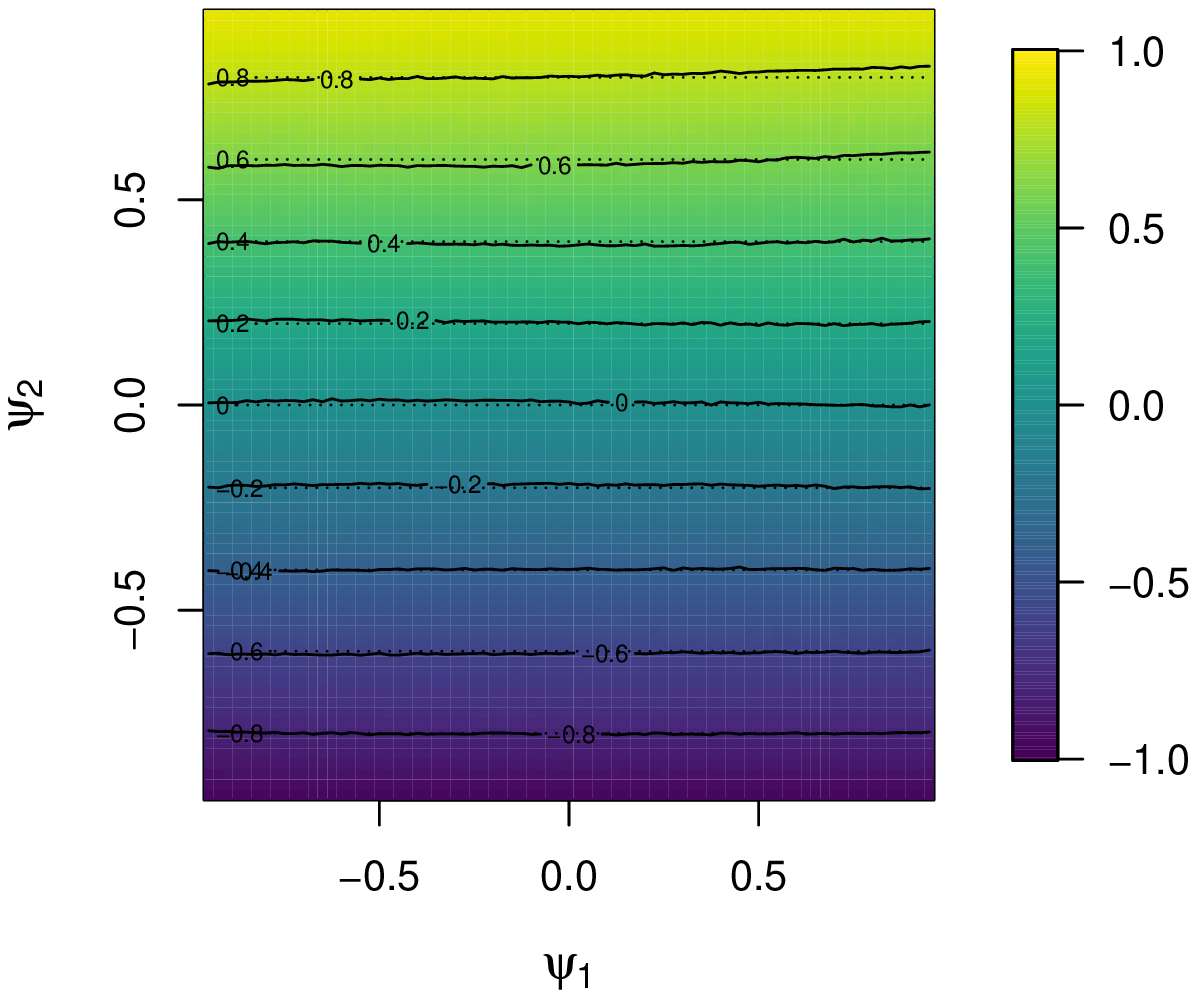}
\includegraphics[scale=0.38]{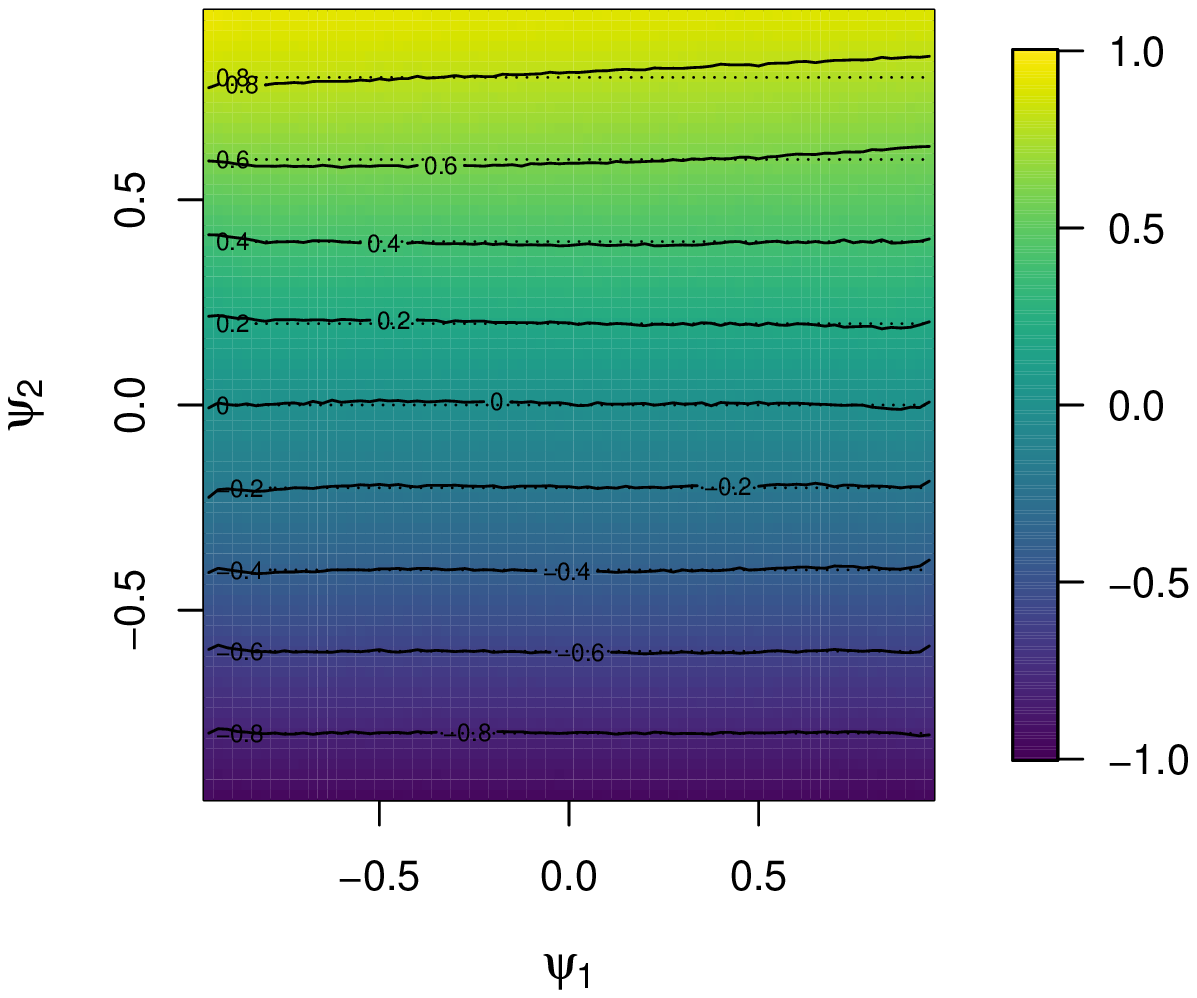}
\caption{Results for $n=30$. 
The mean estimated values of $\psi_1$ (upper) and $\psi_2$ (lower)  using the exact MLE (left), the bias-corrected estimator with $K=3$  (middle) and the bias-corrected estimator with $K=7$ (right).
}
\label{fig:ar2-image-n30}
\end{figure}

The estimated partial autocorrelations using the exact MLE and the corresponding bias-corrected estimates when $K=3$ and $K=7$ are shown for sample sizes $n=15$ and $n=30$ in Figure~\ref{fig:ar2-image-n15} -- ~\ref{fig:ar2-image-n30}, respectively. Visually, the bias properties using  $K=3$ and $K=7$ are quite similar. Both of these corrected estimators are very accurate in estimating $\psi_2$ but show some bias in estimating $\psi_1$, especially for the upper part of the square.  When $K=7$, the predictor in~\eqref{eq:optim-ar2} will have a total of 36 terms for each of the parameters $\psi_1$ and $\psi_2$. This reduces the overall bias slightly compared with using $K=3$, but it also creates some instability in the estimates and the variance increases.  Using $K=3$, the number of regression coefficients for each parameter is reduced to 10. 

To further study the properties of the given estimators, we have calculated the overall average bias, variance and the RMSE of these estimators when $K=3$ and $K=7$. The calculations are based on taking the averages for the $m$ simulations for each of the $l=5929$  combinations of ($\psi_1,\psi_2$), which are easily transformed to give estimates for the AR coefficients by \eqref{eq:psi2}. In calculating the overall averages we have added the results for both parameters. For example, RMSE for the bias-corrected estimator is given by 
\begin{eqnarray*}
\mbox{RMSE}((\hat {\phi}_{c,1},\hat{\phi}_{c,2})) & =  & \sqrt{\frac{1}{2l}\sum_{r=1}^l\sum_{i=1}^2\frac{1}{m}\sum_{j=1}^m (\hat\phi_{c,rji}-\phi_{ri})^2}.
\end{eqnarray*}
The bias and the variance are calculated correspondingly. 
Table~\ref{tab:ar2-rmse} summarizes the overall average bias, variance and the RMSE for the original and bias-corrected estimators using $K=3$. For comparison, we have also included RMSE when $K=7$. The bias-corrected estimators do have a smaller bias and larger variance than the original estimators, but the improved bias properties do not seem to appear at the expense of any significant increase in RMSE. When $n=30$, the RMSE is smaller for the bias-corrected versus the original estimators. When $n=15$, the increase in RMSE for the bias-corrected estimators is very small for the MLEs. When $K=7$,  the RMSE of the corrected estimators increase in all cases. We have also investigated properties of the bias-corrected estimator when $K=5$ but our results confirm that  $K=3$ is a reasonable choice.  

\begin{table}[h]
\begin{center}
\begin{tabular}{ll|rrr|rrrr}
& & \multicolumn{3}{c}{Original estimator} & \multicolumn{4}{c}{Bias-corrected estimator} \\
&Method & Bias & Variance & RMSE & Bias & Variance & RMSE & RMSE \\
 & & & & &  $K=3$ & $K=3$ &$K=3$ &$K=7$ \\\hline 
n=15 &Exact MLE& -0.127 & 0.068 & 0.311 & -0.017 & 0.099 & 0.317 & 0.332\\
&Conditional MLE & -0.126 & 0.068 & 0.312 & -0.017 & 0.099 & 0.318 &0.319\\
&Burg's method&  -0.119 & 0.064 & 0.306 & -0.018 & 0.102 & 0.323 &0.339\\
&Yule-Walker& -0.124 & 0.054 & 0.334 & -0.019 & 0.121 & 0.352 &0.401\\\hline
n=30&Exact MLE& -0.061 & 0.029 & 0.186 & -0.005 & 0.034 & 0.182 &0.188\\
&Conditional MLE & -0.061 & 0.029 & 0.186 & -0.005 & 0.034 & 0.183 &0.190\\
&Burg's method&  -0.061 & 0.029 & 0.188 & -0.005 & 0.035 & 0.184 &0.194\\
&Yule-Walker& -0.064 & 0.028 & 0.207 & -0.006 & 0.040 & 0.199 &0.208\\
 \end{tabular}
\caption{Bias, variance and root mean square error for the original estimators of $(\phi_1,\phi_2)$ and the corrected estimator for $n=15$ and $n=30$. These results are calculated using all the 10000 simulations for each of the selected pairs in $(\phi_1,\phi_2)$ within a fine grid of the stationary area.}
\label{tab:ar2-rmse}
\end{center}
\end{table}

\subsection{Sampling distributions using a Gaussian copula representation}
\begin{figure}[h]
\centering
\includegraphics[scale=0.28]{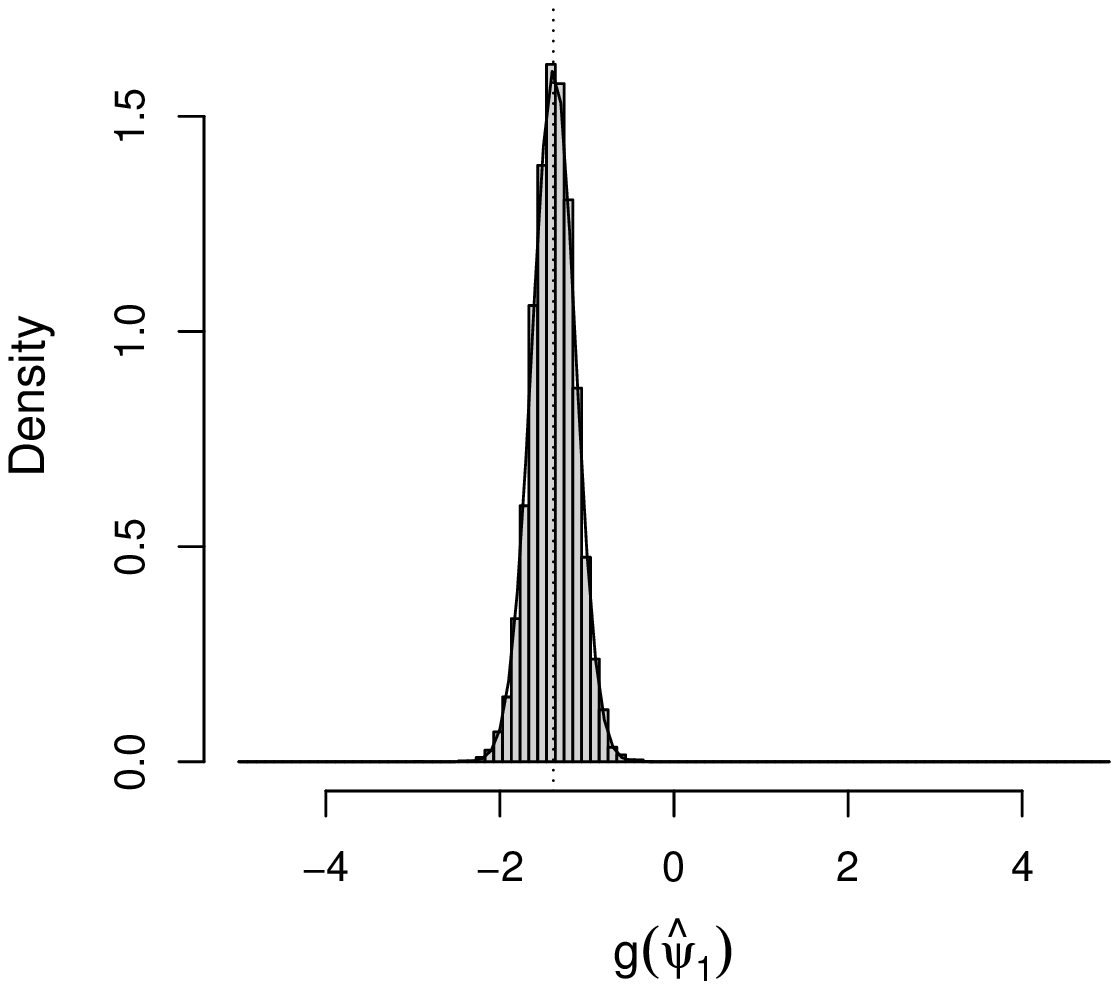}
\includegraphics[scale=0.28]{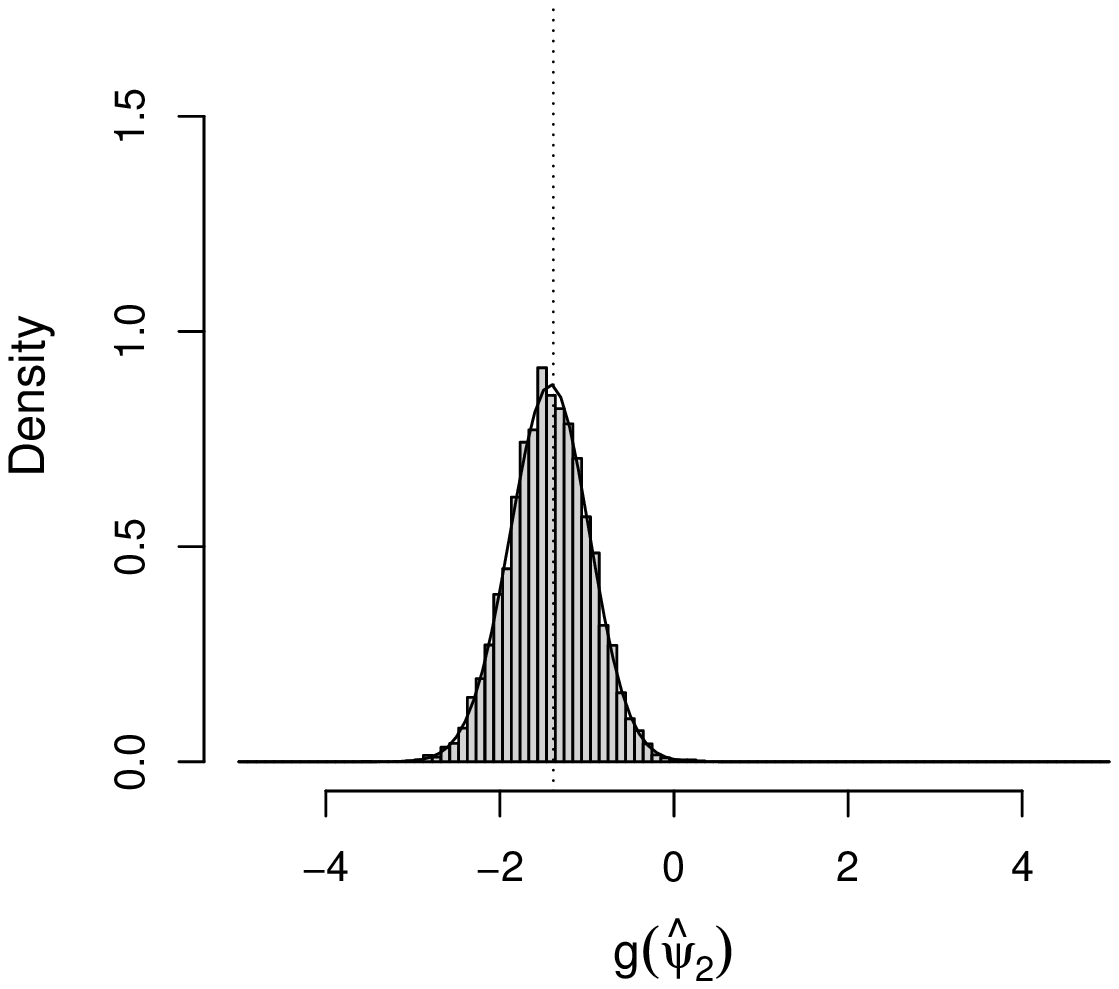}\hspace{0.2cm}
\includegraphics[scale=0.28]{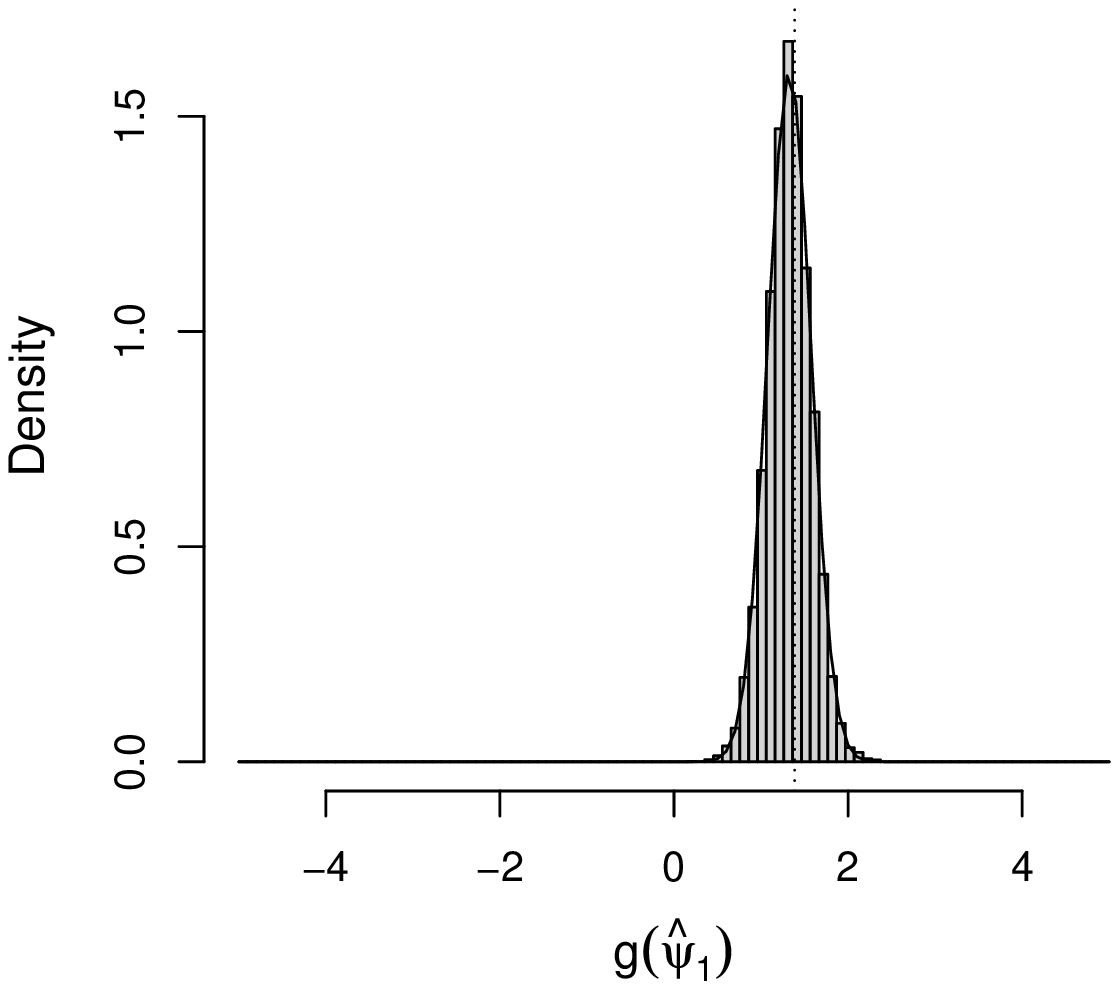}
\includegraphics[scale=0.28]{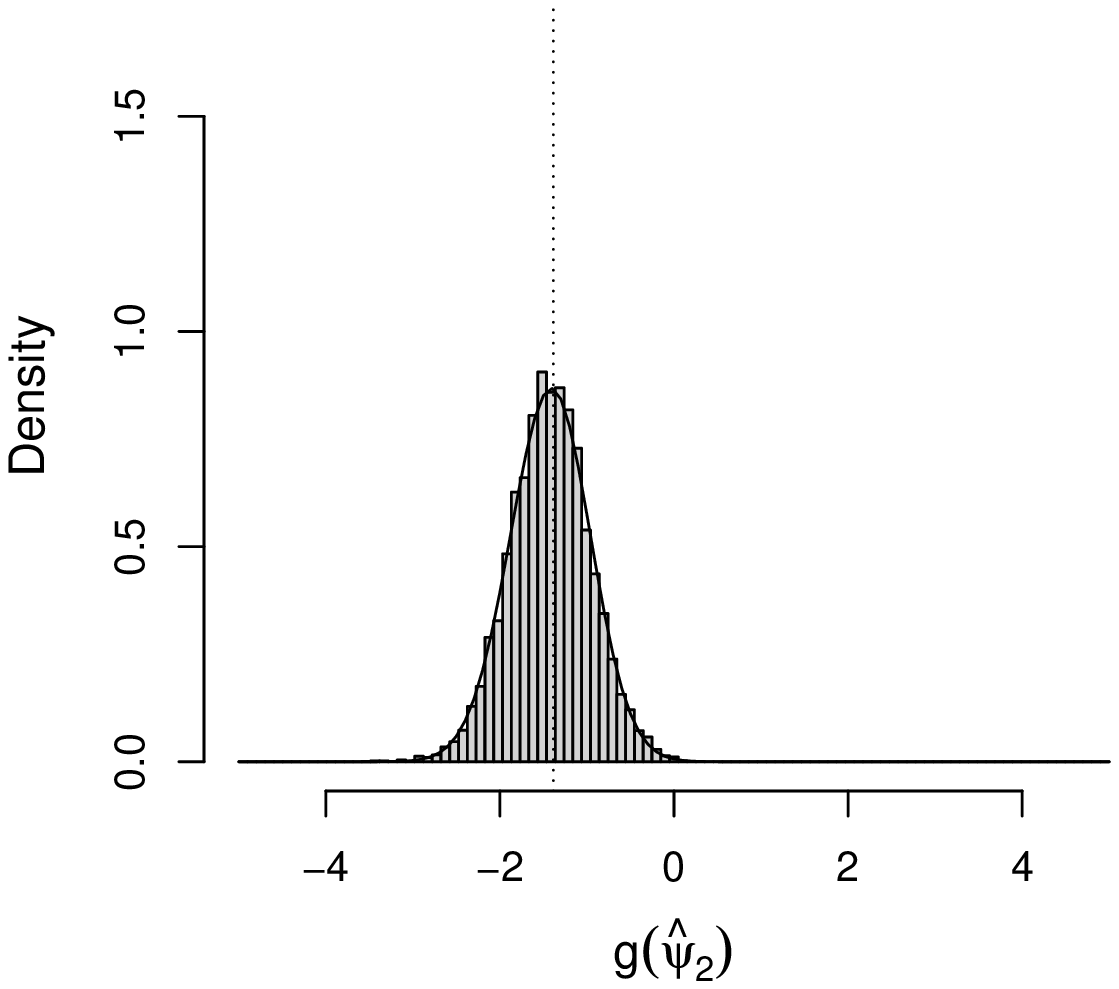}\\
\includegraphics[scale=0.28]{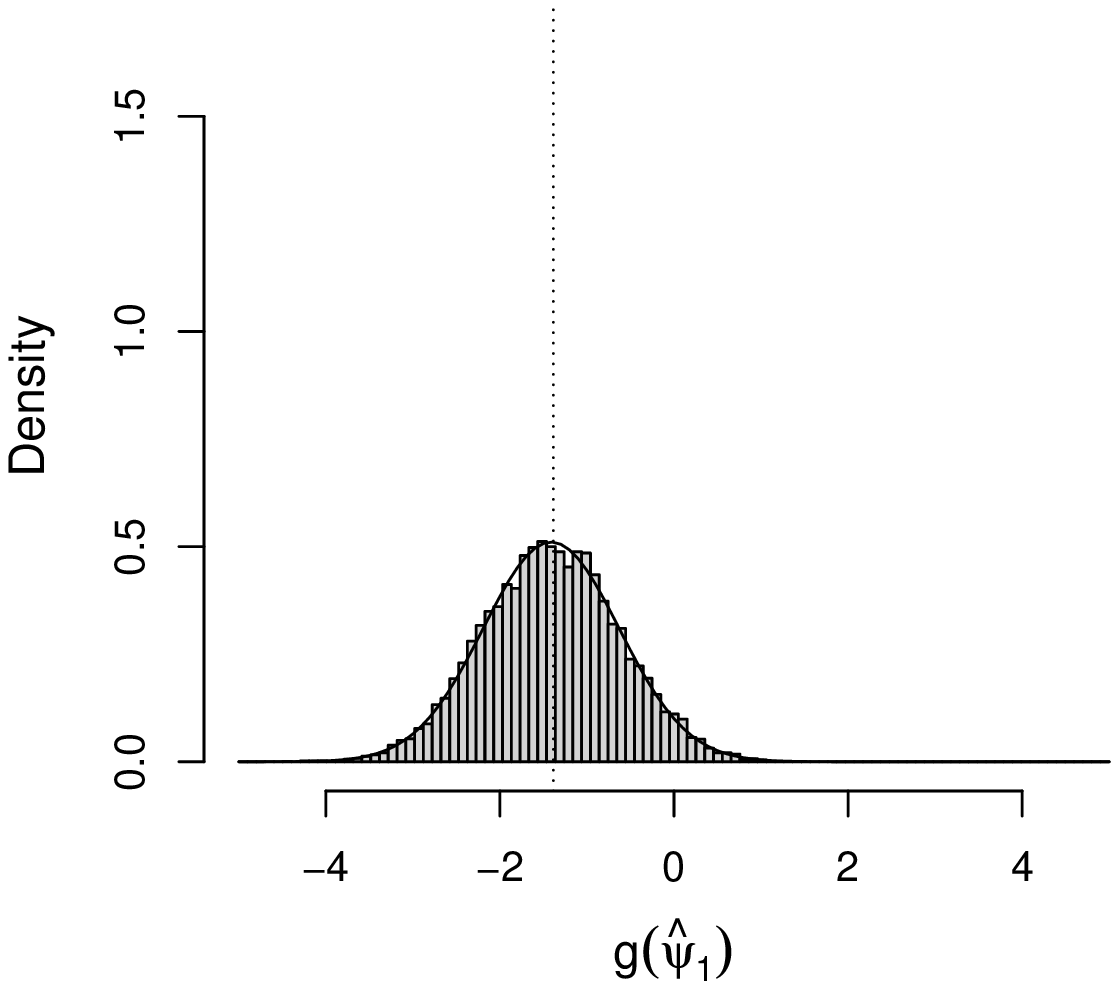}
\includegraphics[scale=0.28]{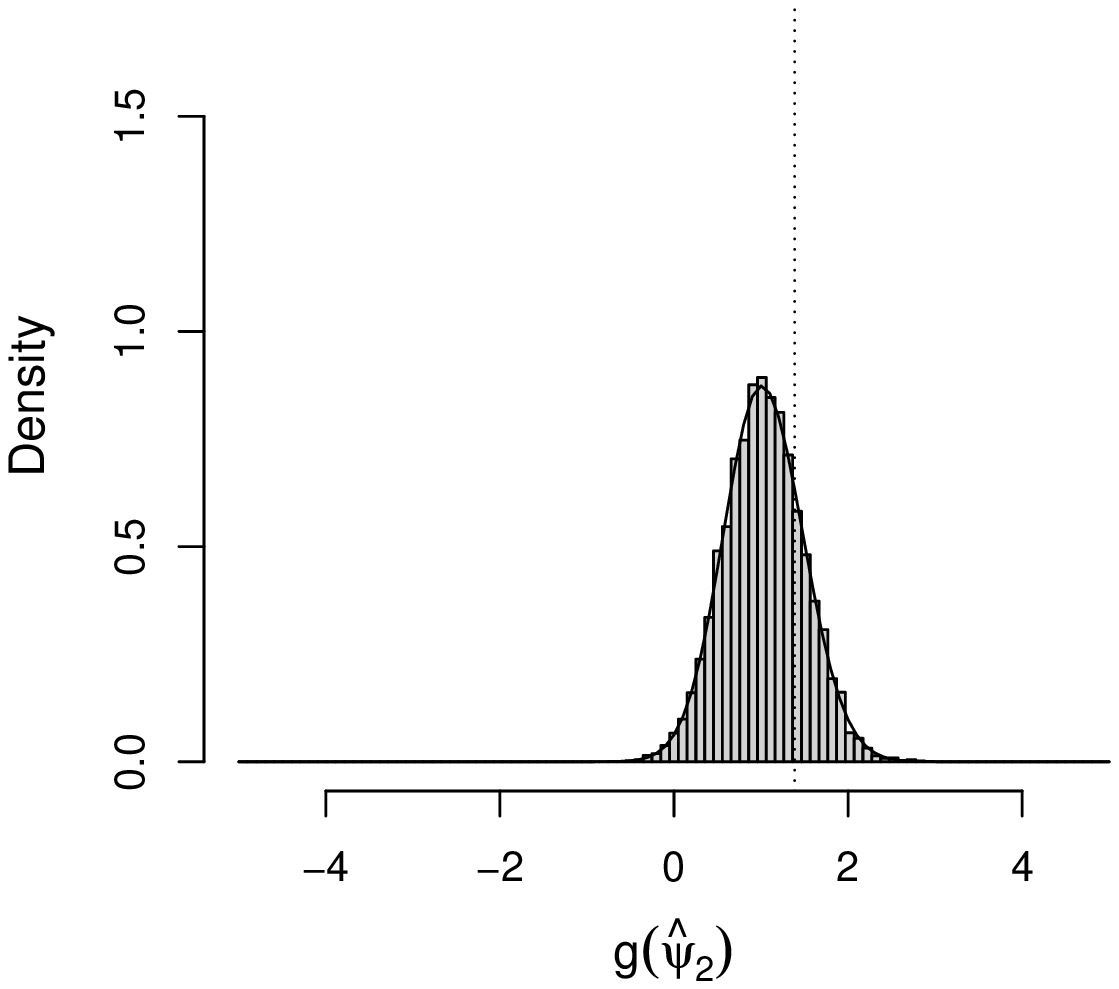}\hspace{0.2cm}
\includegraphics[scale=0.28]{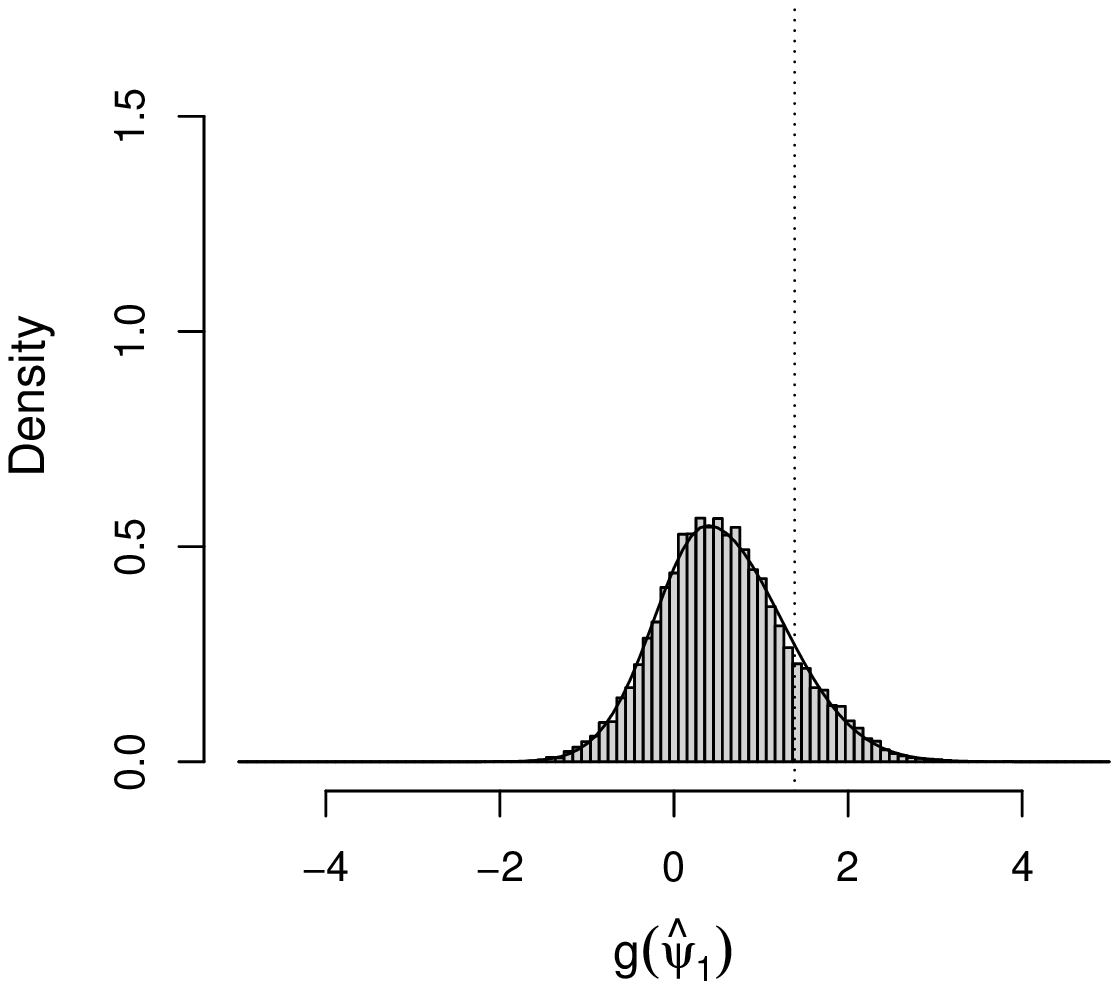}
\includegraphics[scale=0.28]{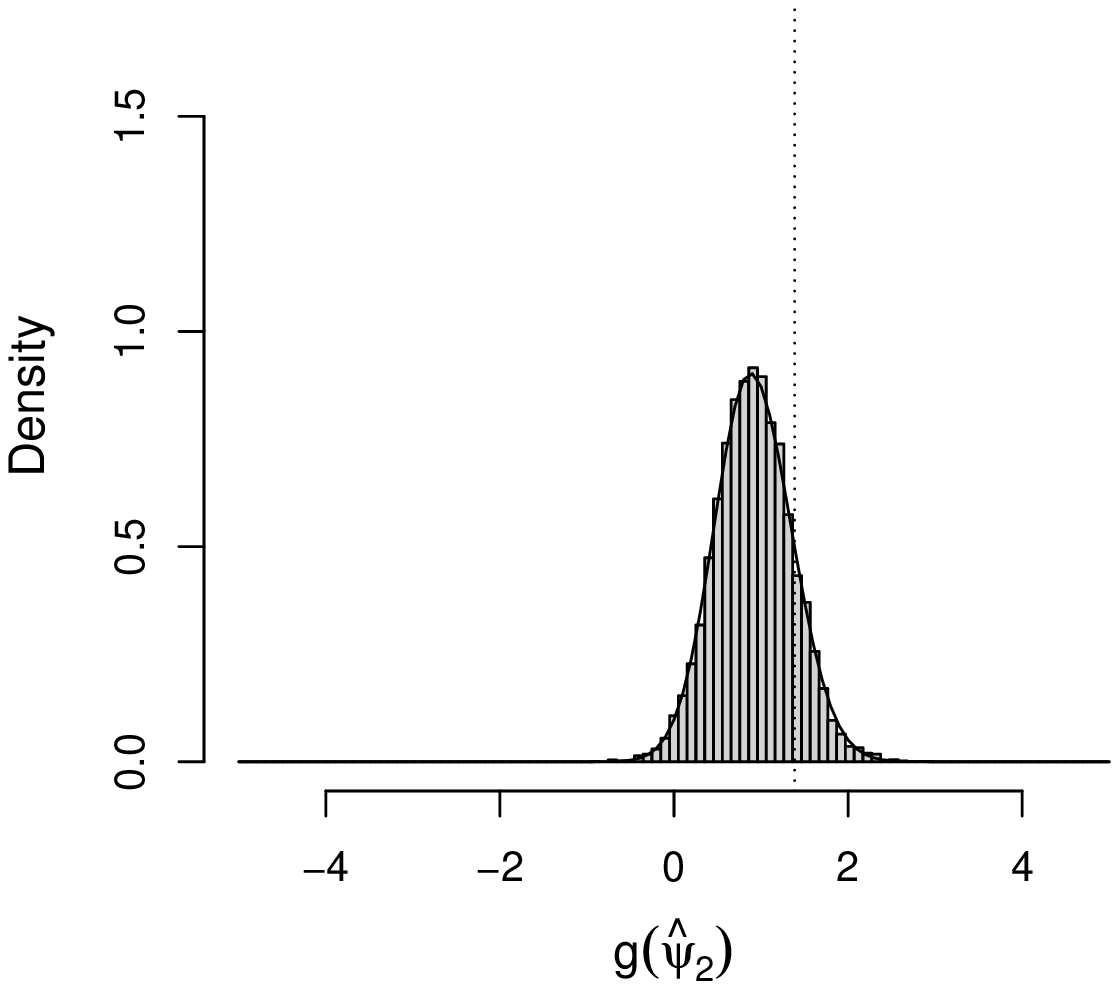}
\caption{The marginal sampling distributions for $g(\hat\psi_1)$ and $g(\hat\psi_2)$  for different combinations of $(\psi_1,\psi_2)$. These combinations include the values $(-0.6, -0.6)$ (upper left pair), $(-0.6, -0.6)$ (upper right pair), $(-0.6, 0.6)$ (bottom left pair) and  $(0.6, 0.6)$ (bottom right pair). The dotted vertical lines give the true values for $g(\psi_1)$  and $g(\psi_2)$.}
\label{fig:ar2-hist}
\end{figure}

Similar to the AR(1) case, we move on to study sampling distributions for the estimators of the parameters of the AR(2) model. For each method and each sample size, we have fitted skew-normal densities to all of the $r$ generated  samples of
$g(\hat\psi_{r,1})$ and $g(\hat\psi_{r,2})$ where $r=1,\ldots , l$.  Figure~\ref{fig:ar2-hist} illustrates the skew-normal approximation for a few selected pairs of the transformed  original estimators where the true underlying partial autocorrelations were set equal to $\psi_i = \pm 0.6$. The skew-normal densities seem to give  quite accurate approximations of the sampling distributions also in the AR(2) case. 

The next step is to find approximations of the sampling distributions for the original non-transformed estimators $(\hat\phi_1,\hat\phi_2)$ and the bias-corrected estimators $(\hat\phi_{c,1},\hat\phi_{c,2})$. This implies that for any pair  $(g(\hat\psi_{1}),g(\hat\psi_{2}))$, we need to predict the corresponding values for the parameters of the skew-normal densities.   We also need a predicted value of the correlation $\hat \rho = \mbox{Cor}(g(\hat\psi_{1}),g(\hat\psi_{2}))$. We choose to do this using an orthogonal polynomial regression model similar to \eqref{eq:polregr} where each of the seven parameters is modeled in terms of  $(\psi_1,\psi_2)$, in separate regression models.  Denote the parameters of the fitted skew-normal densities by  $\mm{\theta}_{i}=(\mu_{i},\sigma_{i},\xi_{i})$, $i=1,2$. Specifically, the predicted value of a parameter $\theta_s \in (\mm{\theta}_1,\mm{\theta}_2,\rho)$ is
\begin{eqnarray}
\hat \theta_{s} = \sum_{k=0}^K \sum_{q=0}^{K-k} \hat b_{k,q,s} h_{k,q}(g(\psi_{1}),g(\psi_{2})) \label{eq:snparam}
\end{eqnarray}
where the coefficients $\{\hat b_{k,q,s}\}$ are found by the ordinary least squares approach. Using $K=3$, this gives a set of 10 regression coefficients $\{b_{k,q,s}\}$ for each of the seven parameters, stored for all methods and sample sizes. 

In generating samples for the original and bias-corrected estimators, we need to preserve the correlation between the transformed partial autocorrelations. This can be done by constructing a two-dimensional Gaussian copula by 
$$C(u_1,u_2) = \Phi_{\hat{\rho}}(F_{sn}^{-1}(u_1;\hat{\mm{\theta}}_{1}),F_{sn}^{-1}(u_2;\hat{\mm{\theta}}_{2})).$$
The function $\Phi_{\hat \rho}(.)$ denotes the joint cumulative distribution of a bivariate standard normal vector with correlation between the components being equal to $\hat \rho$. The functions  $F_{sn}(.)$ represent  the skew-normal cumulative distribution functions (cdf) of $g(\hat \psi_1)$ and $g(\hat \psi_2)$. By the probability integral transform, 
$$\mm{x} =( F_{sn}^{-1}(u_1;\hat{\mm{\theta}}_1),F_{sn}^{-1}(u_2;\hat{\mm{\theta}}_2))$$
represents samples from the given skew-normal densities where the uniformly distributed variables $u_1$ and $u_2$  are generated from the inverse cdf of the standard normal marginals of the bivariate distribution.  

The resulting samples for $(\hat \phi_1,\hat \phi_2)$ and $(\hat \phi_{c,1},\hat \phi_{c,2})$ can be used to find confidence intervals for $(\phi_1,\phi_2)$. We performed a similar simulation study as in the AR(1) case,  generating 10000 AR(2) processes where the partial autocorrelation coefficients are drawn randomly from (-1,1). The coverage probabilities of the estimated $95\%$ confidence intervals are given in Table ~\ref{tab:coverage2} for sample sizes $n=15$ and $n=30$. We do notice that the coverage probabilities using the original estimators are smaller than the nominal level in all cases. In particular, the coverage is very low for $\phi_2$ giving values below 0.80 also when $n=30$. The confidence intervals using the bias-corrected estimators have coverage probabilities quite close to the nominal levels being within the interval $0.95\pm 0.03.$

\begin{table}[h]
\begin{center}
\begin{tabular}{ll|cc|cc}
 &  & \multicolumn{2}{c|}{Original estimator} & \multicolumn{2}{c}{Bias-corrected estimator} \\
n & Method  & $\phi_1$& $\phi_2$ & $\phi_1$ & $\phi_2$ \\\hline
15 &  Exact MLE & 0.8965  & 0.7454  & 0.9398 & 0.9637 \\
 &  Conditional MLE & 0.8967  &0.7474  & 0.9413   & 0.9656\\
 &  Burg's method & 0.8940  & 0.7268  & 0.9485  & 0.9687\\
 &  Yule-Walker & 0.7896 & 0.5485 & 0.9507  & 0.9778 \\\hline
30 &  Exact MLE & 0.9138 & 0.7815 & 0.9459 & 0.9443\\
 &  Conditional MLE &  0.9139 & 0.7819 & 0.9459  & 0.9459 \\
 &  Burg's method &  0.9100 & 0.7609 & 0.9500  & 0.9438 \\
 &  Yule-Walker &  0.8335& 0.6133  &0.9360  & 0.9217 \\
 \end{tabular}
\caption{Coverage for  $95\%$ confidence intervals for $\phi_1$ and $\phi_2$ found by sampling using a Gaussian copula and transformations of skew-normal densities.  The results are based on 10000 simulations where $\psi_1$ and $\psi_2$ are randomly generated from $(-1,1)$.}
\label{tab:coverage2}
\end{center}
\end{table}  

\section{Application in ecology}\label{sec:example}
Wildlife ecological research studies are often characterized by small sample sizes \citep{bissonette:99}.  This can be due to sparse distributions of the animal species of interest and also the research design in which data are collected by fieldwork. As expressed by  \cite{ives:10}:  {\it `` While a time series covering 40 years might represent an ecologist’s entire career, such time series are short for statistical purposes"}. Both AR(1) and AR(2) processes have commonly been used as models for ecological time series, for example modeling population cycles of different animal species like small rodents \citep{bjornstad:95,  hansen:99}. In such studies, the first and second-order coefficients are interpreted in terms of direct and delayed intra-specific dependence, respectively.  

To demonstrate the effect of bias correction for a real world example, we consider a dataset on gray-sided voles (\textit{Myodes rufocanus}), collected by the Japanese Forestry Agency at 85 different sites in Hokkaido, Japan \citep{Saitoh1998}. This dataset has been extensively studied and AR(2) model approximations have been used to assess density dependence, periodicities and synchrony of the vole populations \citep{Stenseth2003, hugueny:06, Cohen2016}. The dataset includes time series for the raw counts of the voles at each site, collected during both spring and fall for a total of 31 years (1962--1992). 
Here, we fit the AR(2) model to the log of the density estimates for the fall time series derived by \citet{Cohen2016} which account for  differences in sampling effort.

\begin{figure}[h]
\centering
\includegraphics[width=4.5cm]{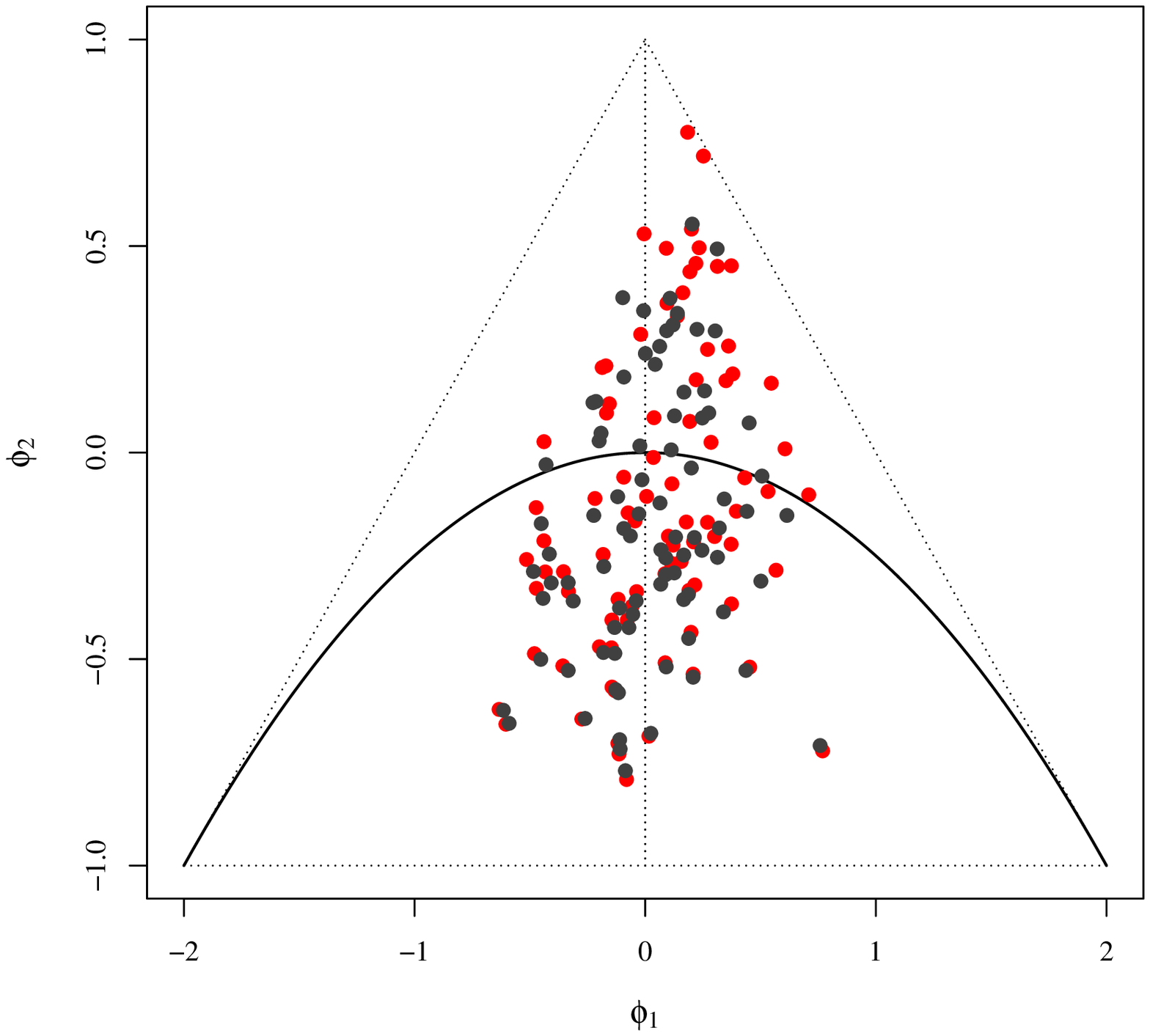}
\includegraphics[width=4.5cm]{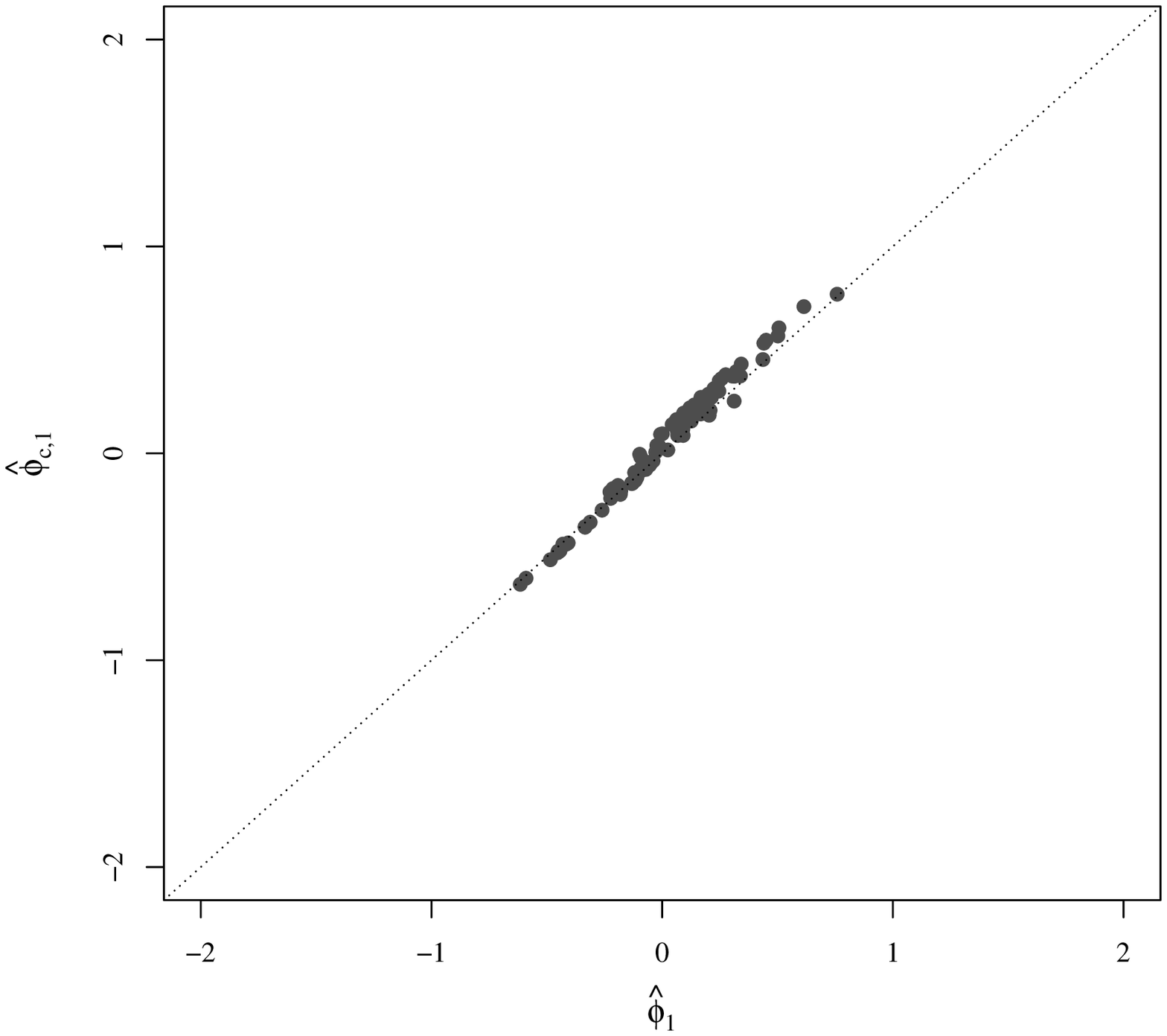}
\includegraphics[width=4.5cm]{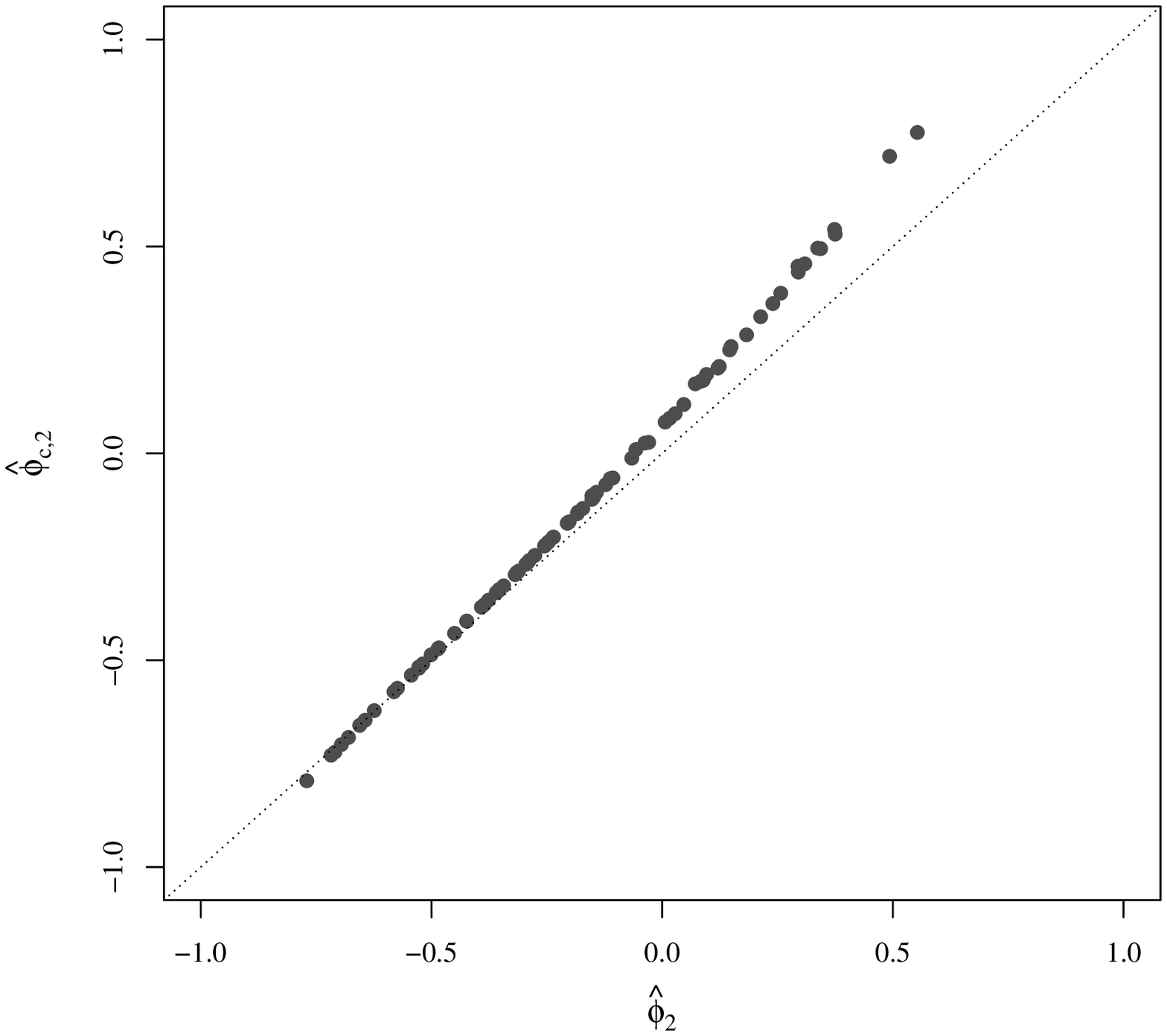}
\caption{Coefficient estimates fitting AR(2) processes to log-density estimates of Hokkaido vole populations observed at 85 different sites. The left panel shows the estimates of $(\phi_1,\phi_2)$ using the exact MLE (black) and the bias-corrected estimates (red). The middle and left panels show the individual scatter plots of the bias-corrected versus original estimates.}
\label{fig:voles}
\end{figure}

The left panel of Figure~\ref{fig:voles} illustrates the estimated AR(2) coefficients for the 85 time series using the exact MLE (black) and the corresponding bias corrected estimates (red). Two-third of the  original estimates are within the pseudo-periodic area which implies cyclic population dynamics where shorter periods imply stronger density dependence \citep{Stenseth2003}.  As expected, the difference between the original and bias-corrected estimates are not very large for this dataset as the time series length is  $n=31$. However, the bias correction is systematic in the sense that the estimates of $\phi_1$ and $\phi_2$ are mainly shifted right and upwards, respectively. This is further illustrated in the middle and left panels of Figure~\ref{fig:voles} showing scatterplots of the corrected versus the original estimates.  In correspondence with our simulation results, the bias correction of $\phi_1$ is quite small for estimates that are not too close to the borders of the triangle, while $\phi_2$ is increasingly underestimated for larger values of the parameter.  The given bias-correction implies slightly weaker estimated density dependence and some of the series can no longer be considered to be cyclic as the pairs of autoregressive coefficients fall outside of the pseudo-periodic area. Overestimation of the strength of density dependence have been associated with ignoring sampling variance \citep{Stenseth2003} and ignoring the estimation bias could thus add to this overestimation.   
\section{Concluding remarks}\label{sec:conclusions}

The simplicity and parsimonious parameterization of first and second-order AR processes make them attractive as plausible models for short time series. The AR(1) model reflects the Markov property, providing an important extension to  a temporal independence assumption. The AR(2) process is more flexible and is particularly useful in modelling pseudo-periodic dynamics.  
The bias involved in estimating the coefficients of short AR processes has been well-known for decades but this remains a problem, even for the simple first and second-order cases. The choice of estimator does make a difference for small sample sizes and incautious use of commonly applied estimators might give misleading results. The default choice using the \texttt{ar}-function in \texttt{R} gives the \texttt{}Yule-Walker estimates which are clearly not optimal, neither for short nor long time series. As stated by \cite{tjostheim:83}, ``{\it uncritical use of Yule-Walker estimates may be hazardous}". The \texttt{ar}-function also provides implementation of the popular least squares approach. We have not considered this method here as the coefficient estimates often fall outside the stationary area of AR processes.

The main goal of this paper was to provide a quick and easily available option to compute bias-corrected versions  of original coefficient estimates for AR(1) and AR(2) processes and assess estimation uncertainty by confidence intervals. This is achieved by modeling the true parameter values in terms of estimated values for a huge number of simulations, accounting for the sampling distribution of the chosen original estimators.  The model fitting  step is computationally expensive but needs to be done only once for each original estimator and each sample size.  
Similarly, we model the parameters of skew-normal distributions in terms of the estimated AR coefficients providing approximate finite sampling distributions for the original and bias-corrected estimators. The resulting approximate confidence intervals reflect the estimation variance. This is important as improved bias properties do come at the cost of increased variance and in practical use this inherent bias-variance trade-off has to be considered. 

\section*{Data availability statement}
The data on the Hokkaido vole population in Section~\ref{sec:example} were downloaded from the Supporting Information of \cite{Cohen2016}, available at \href{https://doi.org/10.1002/ecy.1575}{ https://doi.org/10.1002/ecy.1575}. We use the log of the density estimates given in the file {\em App3BayesCountsParameterEstimates.csv} in their Zip archive.

\bibliographystyle{apalike}
\bibliography{shs}

\begin{thebibliography}{}

\bibitem[Andrews, 1993]{andrews:93}
Andrews, D. W.~K. (1993).
\newblock Exactly median-unbiased estimation of first order autoregressive/unit
  root models.
\newblock {\em Econometrica}, 61:139--165.

\bibitem[Andrews and Chen, 1994]{andrews:94}
Andrews, D. W.~K. and Chen, H.-Y. (1994).
\newblock Approximately median-unbiased estimation of autoregressive models.
\newblock {\em Journal of Business and Economic Statistics}, 12:187--204.

\bibitem[Arnau and Bono, 2001]{arnau:01}
Arnau, J. and Bono, R. (2001).
\newblock Autocorrelation and bias in short time series: {A}n alternative
  estimator.
\newblock {\em Quality and Quantity}, 35:365--387.

\bibitem[Bissonette, 1999]{bissonette:99}
Bissonette, J.~A. (1999).
\newblock Small sample size problems in wildlife ecology: a contingent
  analytical approach.
\newblock {\em Wildlife Biology}, 5:65--71.

\bibitem[Bj{\o}rnstad et~al., 1995]{bjornstad:95}
Bj{\o}rnstad, O.~N., Falck, W., and Stenseth, N.~C. (1995).
\newblock A geographic gradient in small rodent density fluctuations: a
  statistical modelling approach.
\newblock {\em Proceedings of the Royal Society London B}, 262:127--133.

\bibitem[Box et~al., 2008]{box:08}
Box, G. E.~P., Jenkins, G.~M., and Reinsel, G.~C. (2008).
\newblock {\em Time Series Analysis: Forecasting and Control}.
\newblock John Wiley and Sons, Inc. Hoboken, New Jersey.

\bibitem[Box and Luceno, 1997]{box:97}
Box, G. E.~P. and Luceno, A. (1997).
\newblock {\em Time Series Analysis: Forecasting and Control}.
\newblock Wiley, New York.

\bibitem[Brockwell and Davis, 2002]{brockwell:02}
Brockwell, P.~J. and Davis, R.~A. (2002).
\newblock {\em Introduction to Time Series and Forecasting}.
\newblock Springer-Verlag, New Work, 2nd edition.

\bibitem[Burg, 1967]{burg:67}
Burg, J.~P. (1967).
\newblock Maximum entropy spectral analysis.
\newblock {\em Proceedings of 37th Meeting of the Society of Exploration
  Geophysics, Oklahoma City}.

\bibitem[Cheang and Reinsel, 2000]{cheang:00}
Cheang, W.~K. and Reinsel, G.~C. (2000).
\newblock Bias reduction of autoregressive estimates in time series regression
  model through restricted maximum likelihood.
\newblock {\em Journal of the American Statistical Association}, 95:1173--1184.

\bibitem[Cohen and Saitoh, 2016]{Cohen2016}
Cohen, J.~E. and Saitoh, T. (2016).
\newblock Population dynamics, synchrony, and environmental quality of
  {H}okkaido voles lead to temporal and spatial {T}aylor's laws.
\newblock {\em Ecology}, 97:3402--3413.

\bibitem[Cordeiro and Klein, 1994]{cordeiro:94}
Cordeiro, G.~M. and Klein, R. (1994).
\newblock Bias correction in {ARMA} models.
\newblock {\em Statistics and Probability Letters}, 19:169--176.

\bibitem[DeCarlo and Tryon, 1993]{decarlo:93}
DeCarlo, L.~T. and Tryon, W.~W. (1993).
\newblock Estimating and testing autocorrelation with small samples: {A}
  comparison of the c-statistic to a modified estimator.
\newblock {\em Behaviour Research and Therapy}, 31:781--788.

\bibitem[Fernandez and Steel, 1998]{fernandez:98}
Fernandez, C. and Steel, M. F.~J. (1998).
\newblock On {B}ayesian modeling of fat tails and skewness.
\newblock {\em Journal of the American Statistical Association}, 93:359--371.

\bibitem[Hannan, 1970]{hannan:70}
Hannan, E.~J. (1970).
\newblock {\em Multiple Time Series}.
\newblock Wiley, New York.

\bibitem[Hansen et~al., 1999]{hansen:99}
Hansen, T.~F., Stenseth, N.~C., Henttonen, H., and Tast, J. (1999).
\newblock Interspecific and intraspecific competition as causes of direct and
  delayed density dependence in a fluctuating vole population.
\newblock {\em Proceedings of the National Academy of Sciences of the United
  States of America}, 96:986--991.

\bibitem[Hugueny, 2006]{hugueny:06}
Hugueny, B. (2006).
\newblock Spatial synchrony in population fluctuations: extending the {M}oran
  theorem to cope with spatially heterogeneous dynamics.
\newblock {\em OIKOS}, 115:3--14.

\bibitem[Huitema and McKean, 1991]{huitema:91}
Huitema, B.~E. and McKean, J.~W. (1991).
\newblock Autocorrelation estimation and inference with small samples.
\newblock {\em Psycological Bulletin}, 110:291--304.

\bibitem[Ives et~al., 2010]{ives:10}
Ives, A.~R., Abbott, K.~C., and Ziebarth, N.~L. (2010).
\newblock Analysis of ecological time series with {ARMA}(p,q) models.
\newblock {\em Ecology}, 91:858--871.

\bibitem[Kendall, 1954]{kendall:54}
Kendall, M.~G. (1954).
\newblock Note on bias in the estimation of autocorrelation.
\newblock {\em Biometrika}, 41:403--404.

\bibitem[Kim, 2003]{kim:03}
Kim, J.~H. (2003).
\newblock Forecasting autoregressive time series with bias-corrected parameter
  estimators.
\newblock {\em International Journal of Forecasting}, 19:493--502.

\bibitem[Kim, 2014]{bootpr:14}
Kim, J.~H. (2014).
\newblock {\em BootPR: Bootstrap Prediction Intervals and Bias-Corrected
  Forecasting}.
\newblock R package version 0.60.

\bibitem[Krone et~al., 2017]{krone:17}
Krone, T., Albers, C.~J., and Timmerman, M.~E. (2017).
\newblock A comparative simulation study of {AR}(1) estimators in short time
  series.
\newblock {\em Quality and Quantity}, 51:1--21.

\bibitem[Marriott and Pope, 1954]{marriott:54}
Marriott, F. H.~C. and Pope, J.~A. (1954).
\newblock Bias in the estimation of autocorrelations.
\newblock {\em Biometrika}, 41:390--402.

\bibitem[McLeod and Zhang, 2006]{mcleod:06}
McLeod, A.~I. and Zhang, Y. (2006).
\newblock Partial autocorrelation parameterization for subset autoregression.
\newblock {\em Journal of Time Series Analysis}, 27:599--612.

\bibitem[McLeod and Zhang, 2008]{mcleod:08b}
McLeod, A.~I. and Zhang, Y. (2008).
\newblock Improved subset autoregression: With {R} package.
\newblock {\em Journal of Statistical Software}, 28:1--28.

\bibitem[{R Core Team}, 2020]{rteam:20}
{R Core Team} (2020).
\newblock {\em R: A Language and Environment for Statistical Computing}.
\newblock R Foundation for Statistical Computing, Vienna, Austria.

\bibitem[Roy and Fuller, 2001]{roy:01}
Roy, A. and Fuller, W.~A. (2001).
\newblock Estimation for autoregressive time series with a root near one.
\newblock {\em Journal of Business and Economic Statistics}, 19:482--493.

\bibitem[Saitoh et~al., 1998]{Saitoh1998}
Saitoh, T., Stenseth, N.~C., and Bj{\o}rnstad, O.~N. (1998).
\newblock The population dynamics of the vole {\em {c}lethrionomys rufocanus}
  in {H}okkaido, {J}apan.
\newblock {\em Researches on Population Ecology}, 40:61--76.

\bibitem[Shaman and Stine, 1988]{shaman:88}
Shaman, P. and Stine, R.~A. (1988).
\newblock The bias of autoregressive coefficient estimators.
\newblock {\em Journal of the American Statistical Association}, 83:842--848.

\bibitem[Shumway and Stoffer, 2017]{shumway:17}
Shumway, R.~H. and Stoffer, D.~S. (2017).
\newblock {\em Time Series Analysis and its Applications with R Examples}.
\newblock Springer, New York.

\bibitem[Stenseth et~al., 2003]{Stenseth2003}
Stenseth, N.~C., Viljugrein, H., Saitoh, T., Hansen, T.~F., Kittilsen, M.~O.,
  B{\o}lviken, E., and Glockner, F. (2003).
\newblock Seasonality, density dependence, and population cycles in {H}okkaido
  voles.
\newblock {\em Proceedings of the National Academy of Sciences},
  100:11478--11483.

\bibitem[Tanaka, 1984]{tanaka:84}
Tanaka, K. (1984).
\newblock An asymptotic expansion associated with the maximum likelihood
  estimators in {ARMA} models.
\newblock {\em Journal of the Royal Statistical Society, Series B}, 46:58--67.

\bibitem[Thombs and Schucany, 1990]{thombs:90}
Thombs, L.~A. and Schucany, W.~R. (1990).
\newblock Bootstrap prediction intervals for autoregression.
\newblock {\em Journal of the American Statistical Association}, 85:486--492.

\bibitem[Tj{\o}stheim and Paulsen, 1983]{tjostheim:83}
Tj{\o}stheim, D. and Paulsen, J. (1983).
\newblock Bias of some commonly-used time series estimates.
\newblock {\em Biometrika}, 70:389--399.

\bibitem[Walker, 1931]{walker:31}
Walker, G. (1931).
\newblock On periodicity in series of related terms.
\newblock {\em Proceedings of the Royal Society, Series A}, 131:518--532.

\bibitem[Wuertz et~al., 2020]{fgarch:20}
Wuertz, D., Setz, T., Chalabi, Y., Boudt, C., Chausse, P., and Miklovac, M.
  (2020).
\newblock {\em fGarch: Rmetrics - Autoregressive Conditional Heteroskedastic
  Modelling}.
\newblock R package version 3042.83.2.

\bibitem[Yule, 1927]{yule:27}
Yule, G.~U. (1927).
\newblock On a method of investigating periodicities in disturbed series, with
  special reference to {W}olfer's sunspot numbers.
\newblock {\em Philosophical Transactions of the Royal Society A: Mathematical,
  Physical and Engineering Sciences}, 226:267--298.

\end{thebibliography}

\section{Appendix: R-package}\label{sec:appendix}
To make the presented methodology easily available, we include the  \texttt{R}-package \texttt{ARbiascorrect} which can be used to obtain the bias-corrected estimates and their approximate $95\%$ confidence intervals, in addition to $95\%$ confidence intervals for the original estimates.  The package consists of only one function as defined below. The user can either use the relevant time series as input or the parameter estimates found by one of the original methods including the exact and conditional MLE, Burg's algorithm or the Yule-Walker solution. The \texttt{R}-package can be downloaded and installed from \url{https://github.com/pedrognicolau/ARbiascorrect}, or installed directly in \texttt{R} through the \texttt{devtools} package by running:
\begin{verbatim} 
	devtools::install_packages("pedrognicolau/ARbiascorrect")
\end{verbatim}

\noindent The specifications of this package are as follows:

\begin{description}
\item[Description] \

Gives bias-corrected estimates and $95\%$ confidence intervals for autoregressive coefficients of  AR(1) and AR(2) processes, for sample sizes $n=10,11, \ldots, 50$. 
\item[Usage]\

\begin{verbatim} 
biascorrect(phi = NULL, n = NULL,  
            method = c("mle",  "cmle", "burg", "yw"), 
            order = NULL, x = NULL)
\end{verbatim}

\item[Arguments]\

\begin{enumerate}
\item[\texttt{phi}]   Single value (AR1) or two-dimensional vector (AR2) containing the coefficient estimates subject to bias correction.  Not required if the time series \texttt{x} is used as input.

\item[\texttt{n}]   
Integer for the length of the time series. Needs to be between 10 and 50. Not required if the time series \texttt{x} is used as input.

\item[\texttt{method}]  Character string specifying the method used to estimate the autoregressive coefficients. Needs to be either \texttt{mle}, \texttt{cmle}, \texttt{burg} or \texttt{yw}, specifying the exact MLE, the conditional MLE, Burg's method and the Yule-Walker solution, respectively.

\item[\texttt{order}]   Order of the estimated autoregressive process. Needs to be provided if  \texttt{x} is used as input.

\item[\texttt{x}]    The time series to be fitted by AR(1) or AR(2) in which \texttt{order} needs to be specified. The default estimation method is the exact MLE if none is specified.

\end{enumerate}

\item[\texttt{Value}]\

\begin{enumerate}
\item[\texttt{phi.hat}] The original estimates of the AR coefficients
\item[\texttt{phi.correct}] The bias-corrected estimates
\item[\texttt{ci.hat}] The $95\%$ confidence interval for the original estimates
\item[\texttt{ci.correct}] The $95\%$ confidence interval for the bias-corrected estimates
\end{enumerate}

\end{description}
\end{document}